# Self-avoiding Tethered Membranes at the Tricritical Point

**Kay Jörg Wiese*** **and François David**†‡
CEA, Service de Physique Théorique, CE-Saclay
F-91191 Gif-sur-Yvette Cedex, FRANCE

**Abstract**

The scaling properties of self-avoiding tethered membranes at the tricritical point ($\Theta$-point) are studied by perturbative renormalization group methods. To treat the 3-body repulsive interaction (known to be relevant for polymers), new analytical and numerical tools are developped and applied to 1-loop calculations. These technics are a prerequisite to higher order calculations for self-avoiding membranes. The cross-over between the 3-body interaction and the modified 2-body interaction, attractive at long range, is studied through a new double $\varepsilon$-expansion. It is shown that the latter interaction is relevant for 2-dimensional membranes at the $\Theta$-point.

*Email: wiese@amoco.saclay.cea.fr
†Email: david@amoco.saclay.cea.fr
‡Physique Théorique CNRS

# Contents

# Appendix

# 1 Introduction

Two-dimensional tethered surfaces, which model polymerized flexible membranes, offer interesting problems of statistical mechanics (for a general introduction see [1]). One problem, which is now reasonnably well understood, is the influence of bending rigidity: for high bending rigidity (or low temperature), such a membrane is flat, with fractal dimension $d_f = 2$, while for low bending rigidity (or high temperature) it is crumpled, with infinite $d_f$. The existence of a crumpling transition, separating these two phases, has been established by numerical simulations and by renormalization group calculations for "phantom surfaces", where self-avoidance is ignored.

The effect of self-avoidance and its interplay with bending rigidity is more difficult to study and it is not so well understood. Numerical simulations of self-avoiding flexible polymerized surfaces in 3-dimensional space favour the idea that self-avoidance has drastic consequences and flattens the surfaces at any temperature [2, 3, 4]. There is however no fully convincing analytical argument for such a behavior. Another question is how the behavior of self-avoiding surfaces depends on the dimension of bulk space or on the details of the contact interaction.

The standard model for theoretical studies of self-avoiding surfaces has been first discussed in [5, 6] and is inspired by the Edwards model [7] for polymers (for a general presentation see e.g. [8]): it consists in an extension of this model from a line (the polymer) to $D$-dimensional manifolds. The case $D = 2$ corresponds to surfaces. The internal points of the manifold, which belong to the nodes of a $D$-dimensional network, are labelled by continuous coordinates $x \in \mathbb{R}^D$. The position of these points in the external $d$-dimensional bulk space is described by the vector field $x \to r(x) \in \mathbb{R}^d$. The continuum Hamiltonian is

$$\mathcal{H}[r] = \int_x \frac{1}{2}(\nabla r(x))^2 + t \int_x \int_y \delta^d(r(x) - r(y)) \ . \tag{1.1}$$

The first term is the Gaussian elastic term, which describes the crumpled phase of "phantom" surfaces. The second term is a "weak" self-avoidance 2-body $\delta$-potential, which models the contact interaction in bulk space. $t > 0$ is the coupling constant.

Dimensional analysis shows that the contact interaction is relevant at large distances if $d < d^\star$, where $d^\star = 4D/(2 - D)$ is the upper critical dimension. As in the case of polymers, it is natural to perform an $\varepsilon$-expansion about $d^\star$ to evaluate scaling properties such as the fractal dimension $d_f$. Since for $D = 2$, the upper critical dimension $d^\star = \infty$, it is in fact better to perform both an expansion in $d$ and $D$, starting from $(d^\star, D^\star)$ with $D^\star < 2$, aiming e.g. at $d = 3$, $D = 2$.

The first calculations for the model (1.1) [5, 6] used the direct renormalization method, adopted from polymer theory [9, 10]. It has been developed by several authors to perform calculations at 1-loop order [11, 12, 13]. In this method the theory is reexpressed in terms of dimensionless physical quantities, determined for finite surfaces with internal extent $L$. This length $L$ provides a renormalization scale and allows to calculate the renormalization group flow for the model at 1-loop order. The method has been checked to be valid at 1-loop order [14].

Recently a different and more general formalism has been introduced by B. Duplantier, E. Guitter and one of the authors in [15]. (It partly relies on previous studies of mem-branes interacting with a fixed element [17, 18, 19]). Although the interaction term in the Hamiltonian (1.1) is a non-local and singular function of the field $r(x)$, it is shown that the short distance behavior of the model can be encoded in a multilocal operator product expansion (hereafter abbreviated as MOPE), which generalizes the Wilson operator product expansion valid for local field theories. This allows a systematic analysis of the short distance ultra-violet (UV) singularities of the model and shows that the theory (1.1) is renormalizable at the critical dimension $d^{\star}$. This means that the model is rendered UV finite in perturbation theory at $d^{\star}$ by a renormalization of the coupling $t$ and the field $r$ ("wave-function" renormalization). In parallel with the derivation of the renormalization group equations for the $\Phi^4$-theory in dimension $d = 4-\varepsilon < 4$, which gives the scaling laws for a large class of critical phenomena, renormalization group equations for the model of self-avoiding tethered surfaces are derived in a similar expansion for $d < d^{\star}$. The 1-loop results obtained through this method confirm the previous calculations of [5, 6]. The consistency of the direct renormalization method follows from the validity of finite size scaling laws for finite manifolds, also established in [15].

In this paper we apply the renormalization group approach of [15] to a different problem: that of tethered surfaces at the tricritical point, the so-called $\Theta$-point. Our motivation is twofold:

- Physical: the study of the $\Theta$-point for membranes is physically interesting in its own: For polymers it exists due to a competition between 2-body attractive interactions (like long-range Van der Waals forces) and hard-core repulsive interactions. At high temperature the repulsive interactions dominate and the polymer is swollen. At low temperature attractive interactions dominate and the polymer is in a collapsed compact state. For a single long polymer, the transition between these two states occurs at the $\Theta$-point. This point represents a different multi-critical state for the polymer [20] (we refer to [8] for a general presentation). One expects a similar transition to occur for membranes and an interesting fact was first pointed out in [15]: for polymers and for membranes with internal dimension $D$ small enough, one expects an effective 3-body repulsive interaction to be relevant to describe the $\Theta$-point close to the upper critical dimension $d_c = 3D/(2-D)$. For higher $D$, it is a modified 2-body interaction, repulsive at short range, but attractive at larger range, which is relevant to describe the $\Theta$-point close to the upper-critical dimension, now given by $d'_c = 2(3D-2)/(2-D)$. The crossover between the two interactions occurs at $D = 4/3$, $d = 6$. While the modified 2-body interaction leads, at 1-loop order, to calculations for the critical exponents which are analytically computable and quite similar to those for self-avoiding membranes, the 3-body interaction has not been considered up to now – except of course for polymers ($D = 1$) which have been extensively studied – and it is not known which theory should describe "physical membranes" with $D = 2$ and $d = 3$. As we shall show here, the interplay between the 2-body and the 3-body interaction can be studied via a "double $\varepsilon$-expansion" around the critical point $D = 4/3$, $d = 6$.
- Mathematical: 1-loop calculations for the model of membranes with the repulsive 3-body interaction are already non-trivial and the first order term of the $\varepsilon$-expansion appears not to be computable analytically, i.e. cannot be expressed in terms of standard special functions except for $D = 1$. In fact, at 1-loop order, one encounters problems similar to those occuring in the evaluation of 2-loop corrections for self-avoiding surfaces but without the additional difficulty of double poles. Thus, this model can be considered as a (not so enjoyable) toy-model to develop analytical as well as numerical technics which should apply to higher order calculations for self-avoiding surfaces.

This paper is organized as follows: In section 2 we define the model with the 3-body interaction, recall how its perturbative expansion is obtained and how the short distance UV divergences are organized according to the MOPE. We then show explicitly how these divergences can be subtracted in order to construct a renormalized theory and how the scaling laws are obtained.

Sections 3–6 are devoted to the explicit calculation of the 1-loop counterterms for the 3-body interaction, valid a priori for $D < 4/3$.

The counterterm associated to the "wave-function" renormalization, i.e. to the renormalization of the elastic energy term in the Hamiltonian, is treated in full details in section 3. The integral representation of the counterterm in terms of the MOPE coefficient is derived. Various technical problems are discussed: extraction of the singular part (residue of the pole in $\varepsilon$), subtraction of divergences associated to the fine-tuning of the 2-body interaction needed to reach the $\Theta$-point and the definition of the measure in non-integer dimension $D$. Using these methods, the counterterm is evaluated numerically in the range $1 < D < 4/3$. Finally the result for $D \to 1$ is compared with the value for $D = 1$, already known analytically [21].

In section 4 we introduce a useful change of variables for the integral representation of the counterterms, based on conformal transformations.

In section 5 the first two diagrams for the coupling constant counterterm are discussed. The first one is calculated analytically. The second is evaluated by methods similar to those used in section 3 for the wave-function counterterm.

Section 6 is devoted to the evaluation of the last diagram needed for the coupling constant counterterm. It can only be evaluated numerically. A full use of the technical tricks developped previously (analytic continuation of the integration measure, conformal mappings) is required as well as the implementation of an original adaptative Monte-Carlo integration method.

After this rather technical part we gather in section 7 the various counterterms for the model with 3-body interaction. We further give the results for the anomalous correction at 1-loop order to the exponent $\nu$ in the range $1 \leq D < 4/3$. This exponent is related to the fractal dimension of the membrane at the $\Theta$-point.

In section 8 we study the crossover between the 3-body interaction and the modified 2-body interaction. First we show that the 1-loop corrections obtained previously from the 3-body interaction have a smooth limit for $D \to 4/3$. Then we recall the 1-loop corrections obtained from the modified 2-body interaction (valid for $D > 4/3$) and show that the limit $D \to 4/3$ exists and is close to but different from the previous one. Finally we show that the interplay between the two interactions can be studied for $D$ and $d$ close to the critical values $D = 4/3$, $d = 6$, by a new "double $\varepsilon$-expansion". At 1-loop order this expansion is analytically computable and the renormalization group flow can be studied explicitly. As a result we show that depending on $D$ and $d$, the $\Theta$-point is described either by: (a) a Gaussian fixed point, (b) the 3-body repulsive interaction, (c) the modified 2-body interaction. The three corresponding domains in the 2-dimensional $(d, D)$ plane are depicted in figure 1.1. The fat lines are the separatices between these domains. The fat dashed line is a linear extrapolation of the 1-loop result. This line separates the domains (b) and (c). As discussed in section 8, this result indicates that the modified 2-body interaction should be relevant to describe 2-dimensional membranes at the $\Theta$-point, independently of the dimension $d$.

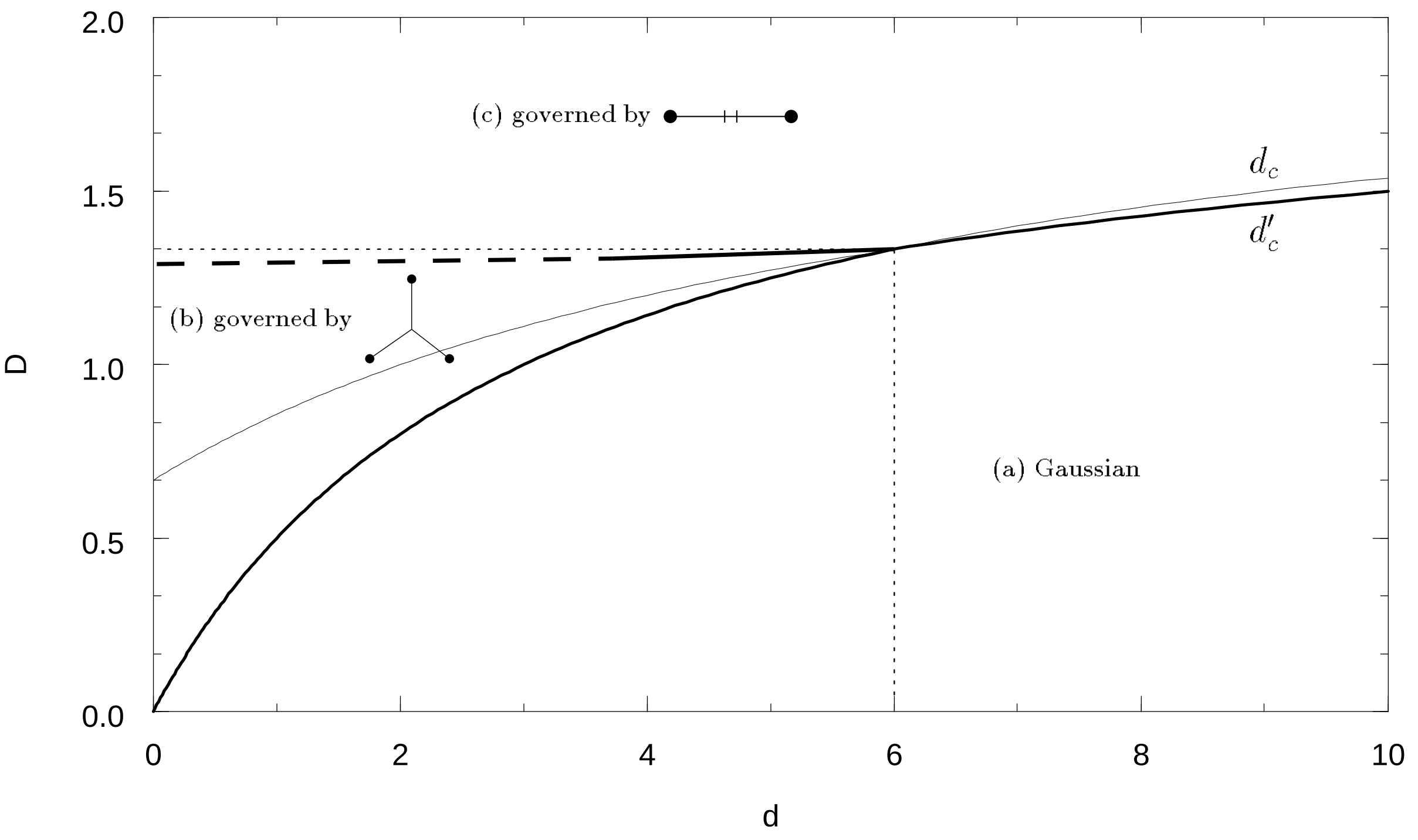

Figure 1.1: Critical dimensions for various operators (solid lines) and phase diagram. The phase separatrices are the fat lines.

The results are summarized in the conclusions. More technical points are discussed in the appendices.

Appendix A treats problems associated with the finite part prescription which we use to subtract the relevant UV divergences. It is shown that different prescriptions may be adopted but lead to the same 1-loop results.

In appendix B diagrams not calculated in the main text are given.

Appendix C briefly discusses the anomalous dimension of the 2-body self-avoiding interaction. This allows to describe the model in the neighborhood of the tricritical point.

# 2 The Model

## 2.1 The 3-body Hamiltonian

The field $r(x)$, $r \in \mathbb{R}^d$, $x \in \mathbb{R}^D$ describes the configuration of the $D$-dimensional polymerized flexible membrane in $d$-dimensional space. The Hamiltonian for this membrane with 3-body repulsive interaction is

$$\mathcal{H}_g^0[r] = \int_x \frac{1}{2}\big(\nabla r(x)\big)^2 + g \int_x \int_y \int_z \delta^d(r(x) - r(y))\delta^d(r(x) - r(z)) \tag{2.1}$$

and $\int_x = \int d^D x$. The calculations of physical observables are performed as an expansion in $g$ and an analytical continuation in the internal dimension $D$ and the external dimension $d$ (dimensional regularization), along the line of [15]. Dimensional regularization allows to deal with the short distance (ultraviolet) divergences, which appear as poles in the complex $D$ or $d$ planes. Otherwise a physical short distance regulator has to be used. One can for instance add a term proportional to the curvature to the Hamiltonian, which in the spirit of the Pauli-Villars regularization contributes higher derivatives to the elastic energy term $(\nabla r)^2$, thus modifying the free propagator at short distances. Large distance (infrared) divergences also occur. They can be cured by using an IR regulator, for instance by considering a finite membrane. However the calculations become technically more difficult and one has to keep track of curvature, boundary and finite-size effects. Alternatively observables invariant under global translations ($r(x) \to r(x) + r_0$) in bulk space may be considered, since these observables are expected to be IR finite even for an infinite membrane. For more details cf. [16].

Dimensional analysis shows that the dimension of $r$ and of the coupling constant $g$ are (in internal momentum units such that $[x] = -1$)

$$[r] = -\nu = \frac{D-2}{2}\,, \quad [g] = \varepsilon = 3D - 2\nu d \tag{2.2}$$

The interaction is relevant in the sense of Wilson if $\varepsilon > 0$ and the perturbation theory is expected to be UV finite up to subtractions associated with relevant perturbation terms, which we shall discuss later. The interaction is irrelevant if $\varepsilon < 0$. The short distance divergences will occur as poles in $\varepsilon$ at $\varepsilon = 0$. As for standard Landau-Ginzburg-Wilson $\Phi^4$ models, which describe critical phenomena in the $\varepsilon = 4 - D$ expansion, these poles have to be subtracted in order to define a renormalized field theory UV finite at $\varepsilon = 0$. This theory will give the scaling behavior of the model for $\varepsilon > 0$.

For clarity we shall graphically represent the different interaction terms which have to be considered. The local operators are

$1 =$  (2.3)

$\frac{1}{2}(\nabla r(x))^2 =$ 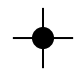 . (2.4)

The bi-local operators are

$\delta^d(r(x) - r(y)) =$  (2.5)

$(-\Delta_r)\delta^d(r(x) - r(y)) =$  . (2.6)

The tri-local operator in the Hamiltonian (2.1) is

$$\delta^d(r(x)-r(y))\delta^d(r(x)-r(y)) \;=$$

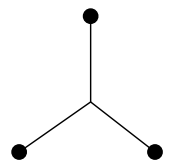

. (2.7)

The first terms of the perturbative expansion in $g$ of the expectation value of an observable $\mathcal{O}$ are

$$\langle \mathcal{O} \rangle_g \;=\; \langle \mathcal{O} \rangle_0 \;-\; g \iiint \langle \mathcal{O}$$

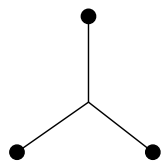

$$\rangle_0$$

$$+\;\frac{1}{2} g^2 \iiint \iiint \langle \mathcal{O}$$

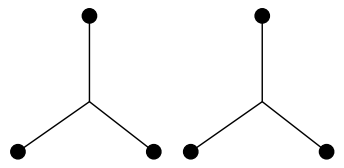

$$\rangle_0 \;+\; \ldots \qquad (2.8)$$

where $\langle \mathcal{O} \rangle_0$ will denote from now on the expectation value of an operator $\mathcal{O}$ for the free theory with $g=0$

$$\langle \mathcal{O} \rangle_0 \;=\; \frac{\int \mathcal{D}[r] e^{-\int_x \,\dotplus}\, \mathcal{O}}{\int \mathcal{D}[r] e^{-\int_x \,\dotplus}} \;. \qquad (2.9)$$

The perturbation expansion (2.8) will suffer from short distance singularities, which occur when subsets of points coalesce. This gives rise to the renormalization discussed in the following. Examples of IR finite observables are provided by "neutral" products of vertex operators

$$\mathcal{O} \;=\; \prod_{a=1}^{N} : e^{i k_a r(x_a)} : \qquad \sum_{a=1}^{N} k_a \;=\; 0 \;. \qquad (2.10)$$

To compute the $\langle \ldots \rangle_0$ in equation (2.8), one writes the $\delta$-functions of the interaction operator as the Fourier transform of vertex operators, performs the free average, then inverses the Fourier transformation and ends up with an integral over the positions of the internal points belonging to the interaction operators. The integrand is a function of all the distances between the internal points and the external points (the $x_a$'s). We will show in 3.1 how this works.

## 2.2 The MOPE: mixing of 3-, 2- and 1-body operators

Short distance singularities may arise when the distances between internal points vanish. In fact, for observables of the form (2.10), no additional singularity occurs when distances between internal and external points vanish as long as the distances between external points stay non-zero. As shown in [15, 16], the short distance behavior of expectation values of operators in the free theory is given by a multilocal operator product expansion. This implies that the short distance divergences can be absorbed by adding to the Hamiltonian (2.1) counterterms proportional to multi-local operators.

At first order in $g$, these operators are generated by contracting points in the single 3-body operator. If the three points $(x_1, x_2, x_3)$ are contracted towards their center of mass $o$, 1-body operators are obtained:

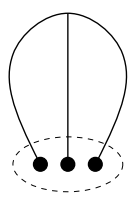

$$=\; A(\{x_1,x_2,x_3\}) \;\bullet\; +\; B(\{x_1,x_2,x_3\}) \;\dotplus\; +\; \ldots \qquad (2.11)$$

where $\{x_1, x_2, x_3\}$ means the relative distances between the 3 points $x_i$. By power counting the coefficients $A$ and $B$ are homogeneous functions of the difference between the positions $(x_1, x_2, x_3)$ of the points, with degree given respectively by

$$\deg(A) \;=\; -2\nu d\ , \qquad \deg(B) \;=\; D - 2\nu d\ . \tag{2.12}$$

In equation (2.11) we made use of the equation of motion of the theory and neglected total derivatives such as $\nabla(r\nabla r)$, since they do not give UV divergences. The dots ... represent operators of higher dimension.

If only two points are contracted (e.g. $x_1, x_2$), 2-body operators are generated:

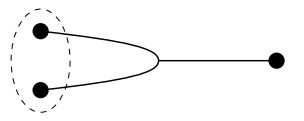 $= C(\{x_1, x_2\})$  $+ \; D(\{x_1, x_2\})$  $+ \; \ldots$ (2.13)

with

$$\deg(C) \;=\; -\nu d\ , \qquad \deg(D) \;=\; (2-d)\nu\ . \tag{2.14}$$

At order $g^2$, one can contract subsets of points of a pair of 3-body operators. Since we are interested in evaluating anomalous dimensions at 1-loop order, we only need to know the divergences which arise when two 3-body operators coalesce into a single 3-body operator. Three contractions are possible:

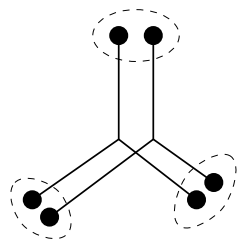 $= E(\{x_1, y_1\}, \{x_2, y_2\}, \{x_3, y_3\})$ 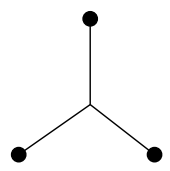 $+ \; \ldots$ (2.15)

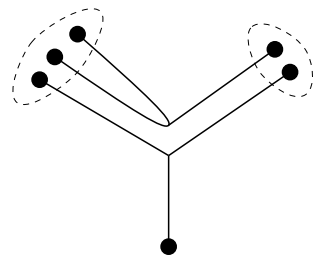 $= F(\{x_1, y_1, y_2\}, \{x_2, y_3\})$ 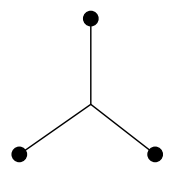 $+ \; \ldots$ (2.16)

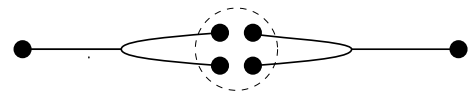 $= G(\{x_2, x_3, y_2, y_3\})$ 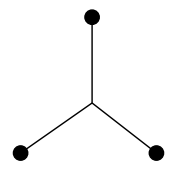 $+ \; \ldots$ (2.17)

with

$$\deg(E) \;=\; \deg(F) \;=\; \deg(G) \;=\; -2\nu d\ . \tag{2.18}$$

The coefficients $A$ in equation (2.11) and $C$ in equation (2.13) give strong short distance divergences. We call a divergence strong or relevant if it is not integrable and thus has to be treated by a finite part prescription. The relevant divergence here is expected, since the corresponding operators 1 and $\delta^d(r(x) - r(y))$ are *relevant* for $\varepsilon = 0$. The divergence proportional to the unity operator 1 is just a "vacuum energy" term. It does not occur for expectation values of physical observables, but will be present in the partition function of finite membranes. The divergence proportional to the 2-body operator  has to be cancelled by adding a 2-body counterterm

$$\Delta\mathcal{H}^0[r] \;=\; \Delta t \int_x \int_y \delta^d(r(x) - r(y)) \tag{2.19}$$

and by fine tuning $\Delta t$ so that the renormalized 2-body interaction vanishes. This situation is known from the $\Theta$-point of polymers: the fine tuning of the 2-body interaction is required in order to reach the $\Theta$-point which separates the swollen phase, where self-avoidance is relevant, from the collapsed phase, where attraction and short-distance repulsion dominate. It also arises in standard scalar field theories: Quadratic divergences, associated to a mass renormalization, have to be subtracted in order to stay at the critical point.

When using dimensional regularization to define the theory, these operators do not give rise to logarithmic divergences at $\varepsilon = 0$ and thus can be subtracted unambiguously by a finite part prescription. This amounts to analytically continue beyond the poles caused by these operators. With this prescription, the renormalized coupling of the 2-body interaction is automaticaly zero if the bare coupling $t$ is set to zero, i.e. if one starts with the 3-body Hamiltonian (2.1).

The subleading term $D(\{x_1, x_2\})$ in equation (2.13), proportional to the modified 2-body interaction (2.6), gives a divergence for $\varepsilon = 0$ if the internal dimension $D$ of the membrane is larger than or equal to 4/3. This reflects the fact, first outlined in [15], that at the $\Theta$-point, the modified 2-body interaction is more relevant than the 3-body interaction if $D > 4/3$, while for $D < 4/3$, including the case of polymers, the 3-body interaction is the most relevant one. The Hamiltonian (2.1) thus describes the $\Theta$-point for $D < 4/3$ and $\varepsilon$ small. For $D > 4/3$ it describes a multicritical point reached by fine tuning both the 2-body and the modified 2-body couplings. A more serious investigation of the relative relevance of these two operators for finite $\varepsilon$ requires a study of the model with both couplings,

$$\mathcal{H}^0_{g,b} = \int_x \; [\text{diagram 1}] \; + g \int_x \int_y \int_z \; [\text{diagram 2}] \; + b \int_x \int_y \; [\text{diagram 3}] \; , \qquad (2.20)$$

around the point $D = 4/3$ and $d = 6$. This topic is discussed in section 8.

## 2.3 UV divergences and 1-loop renormalization

Let us now concentrate on the Hamiltonian (2.1) and on the divergences at $\varepsilon = 0$. The MOPE structure of the UV singularities implies that the theory can be made UV finite for $\varepsilon \to 0$ by considering the Hamiltonian

$$\mathcal{H}^R_g[r] = Z \int_x \frac{1}{2}\big(\nabla r(x)\big)^2 + g Z_g \mu^\varepsilon \int_x \int_y \int_z \delta^d(r(x) - r(y))\delta^d(r(x) - r(z)) \, , \qquad (2.21)$$

where $\mu$ is a renormalization momentum scale. $r$ and $g$ are the renormalized field and the renormalized coupling constant. As in [15, 16], the counterterms $Z$ (wave-function renormalization) and $Z_g$ (coupling constant renormalization) subtract the poles at $\varepsilon = 0$.

The counterterms are evaluated as follows. The divergence proportional to the operator 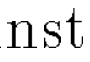 occurs because of the integration over a global length scale in the MOPE (2.11). For instance, if we integrate over the three points $(x_1, x_2, x_3)$ in a domain $\mathcal{D}_L(o)$ of size $L$ around their center of mass $o$

$$\mathcal{D}_L(o) = \{x_1 + x_2 + x_3 = 3o \text{ and } |x_a - x_b| < L, \; a, b = 1, 2, 3\} \, , \qquad (2.22)$$

using equation (2.11) gives a pole of the form

$$\iint_{\mathcal{D}_L(x)}$$ 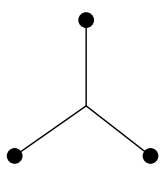 $$= \frac{L^{\varepsilon}}{\varepsilon}$$ 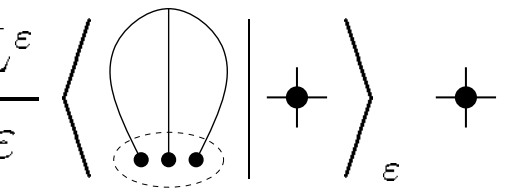 $$+ \; O(\varepsilon^0) \qquad (2.23)$$

with the residue determined by the coefficient $B$ of the MOPE and abbreviated graphically as

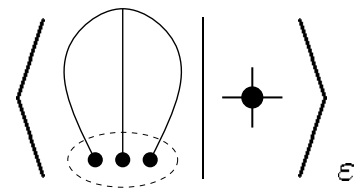 $$= \iint_{\substack{x_1+x_2+x_3=0 \\ \sup(|x_a-x_b|)=L}} B(\{x_1,x_2,x_3\})\Big|_{\varepsilon=0} \;. \qquad (2.24)$$

We will explain that in more detail in section 3.3.

The other residues come from a similar integration at a "typical" distance $L$ between the points in the clusters of the coefficients $E$, $F$ and $G$ in equations (2.15), (2.16) and (2.17). They are abbreviated similarly as

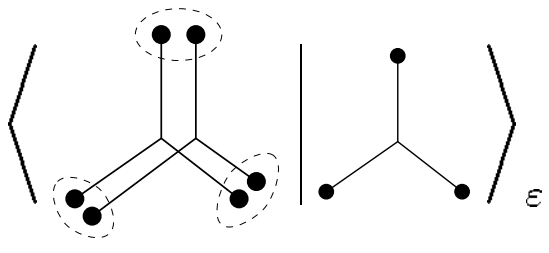 $$= \iiint_{\substack{\text{sup of distances} \\ \text{in clusters } = \; L}} E(\{x_1,y_1\},\{x_2,y_2\},\{x_3,y_3\})\Big|_{\varepsilon=0} \;, \qquad (2.25)$$

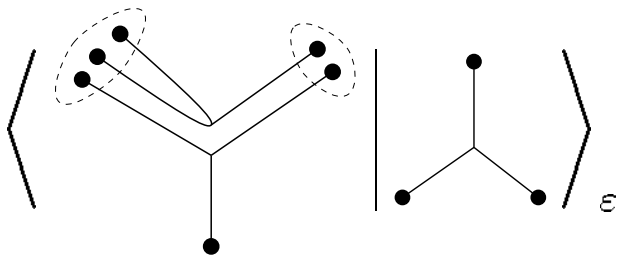 $$= \iiint_{\substack{\text{sup of distances} \\ \text{in clusters } = \; L}} F(\{x_1,y_1,y_2\},\{x_2,y_3\})\Big|_{\varepsilon=0} \;, \qquad (2.26)$$

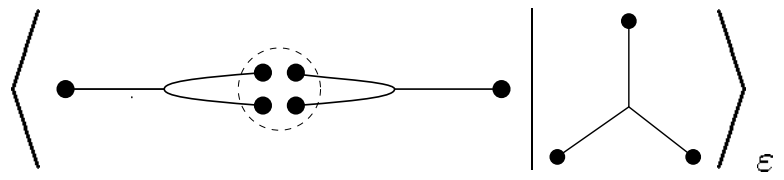 $$= \iiint_{\substack{\text{sup of distances} \\ \text{in cluster } = \; L}} G(\{x_2,x_3,y_2,y_3\})\Big|_{\varepsilon=0} \;. \qquad (2.27)$$

With these notations and using equation (2.8), the counterterms which make the theory finite at 1-loop order are

$$Z = 1 + g\,\frac{a}{\varepsilon} + \mathcal{O}(g^2)\;, \qquad a = -$$ 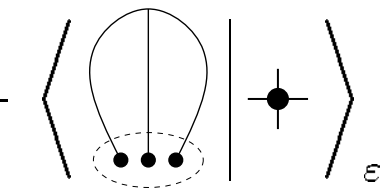 $$(2.28)$$

$$Z_g = 1 + g\,\frac{b}{\varepsilon} + \mathcal{O}(g^2)\;,$$

$$b = 3$$ 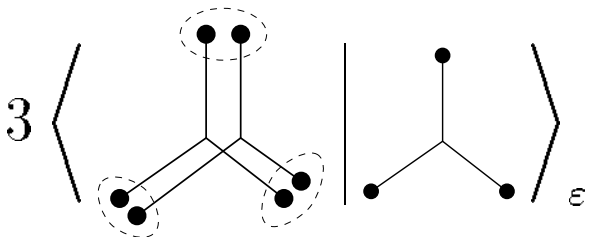 $$+ \; 18$$ 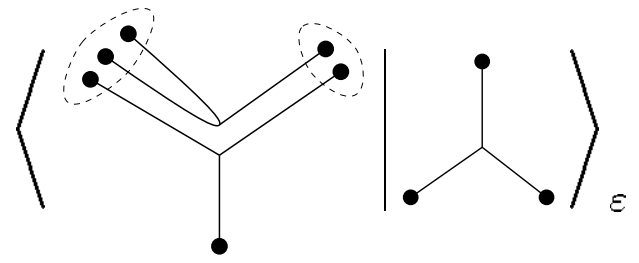

$$+ \frac{9}{2}$$ 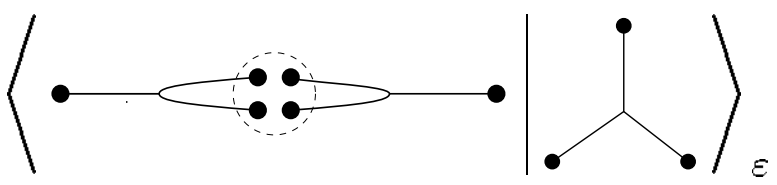 $$. \qquad (2.29)$$

The coefficients of the counterterms, $a$ and $b$, depend of course on the point $(D_c, d_c)$ on the curve $\varepsilon = 3D - 2\nu d = 0$, where the renormalization is performed.

The picture associated with this renormalization prescription at 1-loop order is to cut out domains of size $L = \frac{1}{\mu}$ around collapsing points in the perturbation expansion (2.8). The freedom in the choice of a renormalization scale $\mu$ is represented as the freedom to choose the size $L = \frac{1}{\mu}$ of these domains.

## 2.4 RG equations and scaling relations

From the existence of counterterms and of a UV finite perturbation theory for $\varepsilon \geq 0$ one deduces renormalization group (Callan-Symanzik) equations for the renormalized theory and scaling laws for the model for $\varepsilon > 0$ in the standard way. The renormalized Hamiltonian (2.21) can be rewritten as a bare Hamiltonian (2.1) through the change of variables to bare field $r_0(x)$ and bare coupling constant $g_0$

$$r_0(x) \;=\; Z^{1/2}\, r(x)\;, \qquad g_0 \;=\; g Z^d Z_g \mu^\varepsilon\;. \tag{2.30}$$

The Callan-Symanzik equations, which give the scale dependence of the renormalized theory, are obtained via the $\mu$ dependence of the renormalized couplings keeping the bare couplings fixed. One thus obtains the renormalization group $\beta$-function for the coupling constant

$$\beta_g(g) \;=\; \mu \frac{\partial}{\partial \mu}\bigg|_{g_0} g \;=\; -\,\varepsilon g \,+\, (ad+b)g^2 \,+\, \mathcal{O}(g^3) \tag{2.31}$$

and the scaling dimension $\nu$ of the field $r$

$$\begin{aligned} \nu(g) \;&=\; \nu \,-\, \frac{1}{2}\mu\frac{\partial}{\partial\mu}\bigg|_{g_0} \ln Z \\ &=\; \frac{2-D}{2} \,+\, \frac{1}{2} a g \,+\, \mathcal{O}(g^2) \end{aligned} \tag{2.32}$$

As we shall see, both $a$ and $b$ are positive and the renormalization group flow has an IR stable fixed point for positive $g = g^\star$. $\nu^\star = \nu(g^\star)$ is related to the fractal dimension $d_F^\star$ of the membrane at the $\Theta$-point via

$$d_F^\star \;=\; \frac{D}{\nu^\star}\;. \tag{2.33}$$

## 2.5 A change in normalizations

Before describing the details of the calculations for the counterterms, let us change the normalizations used in the definition of the theory. This is done for technical purpose only, but appears convenient to avoid a lot of factors $\pi$ and $\Gamma$ functions in intermediate expressions. We rewrite the bare Hamiltonian as

$\mathcal{H}_g^0[r] \;=\; (2-D)^{-1} \int_x$ 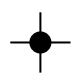 $\;+\; g \int_x\int_y\int_z$ 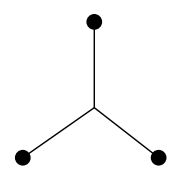 (2.34)

with as before

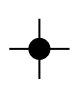 $\;=\; \frac{1}{2}\big(\nabla r(x)\big)^2$ (2.35)

but with a modified integration measure

$$\int_x \;=\; \frac{1}{S_D}\int d^D x\;, \qquad S_D \;=\; \frac{2\pi^{D/2}}{\Gamma(D/2)} \tag{2.36}$$

being the volume of the $(D-1)$-dimensional unit sphere. With this normalization the free propagator becomes simply

$$\langle r(x) r(y) \rangle_0 \;=\; -\,|x-y|^{2\nu} \quad . \tag{2.37}$$

Similarly, we normalize the measure in Fourier space $\mathbb{R}^d$ as

$$\int_p \;=\; \pi^{-d/2} \int d^d p \tag{2.38}$$

and the $\delta^d$-distribution in $\mathbb{R}^d$ is modified to

$$\tilde{\delta}^d(r-r') \;=\; (4\pi)^{d/2}\,\delta^d(r-r') \;=\; \int_p \mathrm{e}^{ip(r-r')} \quad , \tag{2.39}$$

so that the 2-and 3-body operators, which have to be considered, are

 $= \tilde{\delta}^d(r(x)-r(y))$ , (2.40)

 $= (-\Delta_r)\tilde{\delta}^d(r(x)-r(y))$ , (2.41)

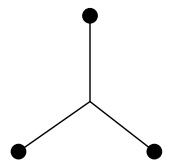 $= \tilde{\delta}^d(r(x)-r(y))\tilde{\delta}^d(r(x)-r(z))$ . (2.42)

The only change in the expressions for the 1-loop counterterms (2.28) and (2.29) is

$a \;=\; -\,(2-D)$ 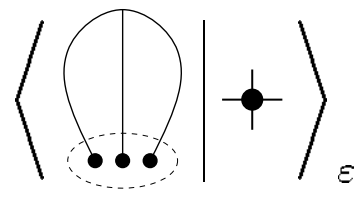 . (2.43)

# 3 Elastic Term Renormalization

## 3.1 Explicit evaluation of the 3-point MOPE coefficient

Let us first derive explicitly the MOPE (2.11) for the 3-body operator (2.42). The 3-body operator has as Fourier integral

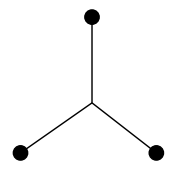 $= \int_p \int_q :\mathrm{e}^{ipr(x)} :: \mathrm{e}^{iqr(y)} :: \mathrm{e}^{-i(p+q)r(z)} :$ (3.1)

$: :$ denotes the usual normal product for free fields. We use the OPE for products of vertex operators

$$\begin{aligned} &:\mathrm{e}^{ipr(x)} : \; : \mathrm{e}^{iqr(y)} : \; : \mathrm{e}^{-i(p+q)r(z)} : \\ &\quad = \; : \mathrm{e}^{ipr(x)} \mathrm{e}^{iqr(y)} \mathrm{e}^{-i(p+q)r(z)} : \langle \mathrm{e}^{ipr(x)+iqr(y)-i(p+q)r(z)} \rangle_0 \end{aligned} \tag{3.2}$$

and since $r$ is a free field we have explicitly, using (2.37)

$$\begin{aligned} &\langle \mathrm{e}^{ipr(x)+iqr(y)-i(p+q)r(z)} \rangle_0 \\ &\quad = \; \mathrm{e}^{-\frac{1}{2}\langle (p(r(x)-r(z))+q(r(y)-r(z)))^2 \rangle_0} \\ &\quad = \; \mathrm{e}^{-p^2|x-z|^{2\nu}+(pq)\left(|x-z|^{2\nu}+|y-z|^{2\nu}-|x-y|^{2\nu}\right)-q^2|y-z|^{2\nu}} \end{aligned} \tag{3.3}$$

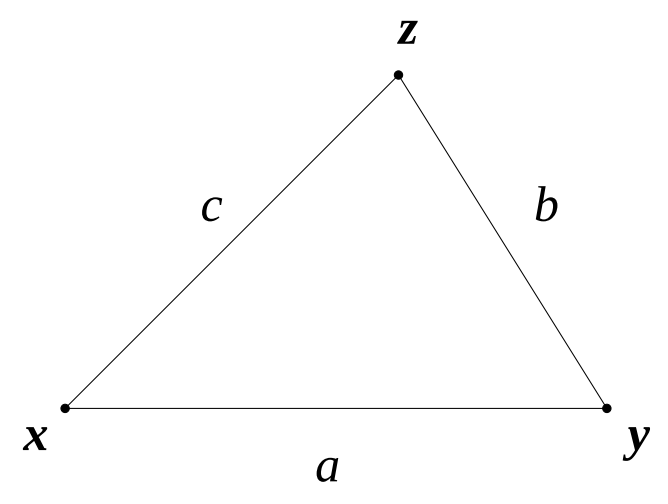

Figure 3.1: The triangle $\Delta(a,b,c)$

Let $o$ be the center-of-mass of the 3 points $x$, $y$ and $z$ and let us denote

$$x = o + \tilde{x} \ , \ y = o + \tilde{y} \ , \ z = o + \tilde{z} \ , \ \tilde{x} + \tilde{y} + \tilde{z} = 0 \tag{3.4}$$

The MOPE is obtained by performing a Taylor expansion around the center-of-mass $o$

$$: \mathrm{e}^{ipr(x)} \mathrm{e}^{iqr(y)} \mathrm{e}^{-i(p+q)r(z)} := \ \mathrm{e}^{\tilde{x}\partial_x} \mathrm{e}^{\tilde{y}\partial_y} \mathrm{e}^{\tilde{z}\partial_z} : \mathrm{e}^{ipr(x)} \mathrm{e}^{iqr(y)} \mathrm{e}^{-i(p+q)r(z)} : \Big|_{x=y=z=o} \tag{3.5}$$

Expanding in powers of $\partial$, we find up to order 2 in $\partial$

$$\begin{aligned} & \mathrm{e}^{\tilde{x}\partial_x} \mathrm{e}^{\tilde{y}\partial_y} \mathrm{e}^{\tilde{z}\partial_z} : \mathrm{e}^{ipr(x) + iqr(y) - i(p+q)r(z)} : \Big|_{x=y=z=o} \\ & = \ 1(o) \ + \ \frac{-|x-z|^2 \, p^2 + 2(y-z)(z-x)\,(pq) - |y-z|^2 \, q^2}{dD} \, \frac{1}{2} : (\nabla r)^2(o) : + \ldots \end{aligned} \tag{3.6}$$

The factor $1/dD$ comes from the contractions between internal space indices (in the partial derivatives $\partial_x$ and $\tilde{x}$) and between external space indices (in $r$ and the external momenta $p$).

Using (3.3) and (3.6) and performing the explicit integration over $p$ and $q$, the term of order 0 will give the coefficient $A$ in (2.11), whereas the term of order 2 yields $B$. Denoting by $a$, $b$, $c$ the respective distances between the three points $x$, $y$ and $z$ (see figure 3.1)

$$a = |y-x| \ , \qquad b = |z-y| \ , \qquad c = |x-z| \ , \tag{3.7}$$

we find

$$\begin{aligned} A(\{x,y,z\}) \ &= \int_p \int_q \langle \mathrm{e}^{ipr(x) + iqr(y) - i(p+q)r(z)} \rangle_0 \\ &= \left[ \frac{4}{(a^\nu + b^\nu + c^\nu)(a^\nu + b^\nu - c^\nu)(b^\nu + c^\nu - a^\nu)(c^\nu + a^\nu - b^\nu)} \right]^{\frac{d}{2}} \end{aligned} \tag{3.8}$$

and

$$\begin{aligned} B(\{x,y,z\}) \ &= \int_p \int_q \frac{-|x-z|^2 \, p^2 + 2(y-z)(z-x)\,(pq) - |\tilde{y}-\tilde{z}|^2 \, q^2}{dD} \\ & \qquad \langle \mathrm{e}^{ipr(x) + iqr(y) - i(p+q)r(z)} \rangle_0 \\ &= -\frac{2^d}{D} \, \frac{(c^2 + a^2 - b^2) b^{2\nu} + (a^2 + b^2 - c^2) c^{2\nu} + (b^2 + c^2 - a^2) a^{2\nu}}{\left[ (a^\nu + b^\nu + c^\nu)(a^\nu + b^\nu - c^\nu)(b^\nu + c^\nu - a^\nu)(c^\nu + a^\nu - b^\nu) \right]^{1+\frac{d}{2}}} \end{aligned} \tag{3.9}$$

## 3.2 MOPE and relevant subdivergences

If two out of the three points in 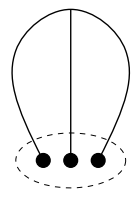 are contracted, a dominant subdivergence occurs. For instance, if $x \to y$, the coefficient $B$ given by equation (3.9) behaves as

$$B(\{x,y,z\}) \ \simeq \ -\frac{1}{2D}\,|x-y|^{-\nu d}\,|x-z|^{D-\nu d} \tag{3.10}$$

As explained in section 2.2, this follows from the MOPE (2.13). Using similar techniques as for calculating the coefficients of the MOPE (2.11), the leading coefficient $C$ in equation (2.13) is found to be

$$C(\{x,y\}) \ = \ |x-y|^{-\nu d} \tag{3.11}$$

To subtract this divergence with a finite part prescription, we must add a 2-body counterterm (2.19) which is at first order in $g$

$$\Delta\mathcal{H}^0 \ = \ -g\,\frac{1}{2}\int_x\int_y\int_z \left\{|x-z|^{-\nu d}\tilde{\delta}^d(r(x)-r(y)) + \begin{matrix}\text{permutations}\\ \text{of } x,\, y,\, z\end{matrix}\right\} \tag{3.12}$$

or equivalently we can redefine the 3-body operator (2.42) as

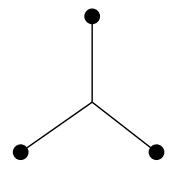

$$\begin{aligned} &= \ \tilde{\delta}^d(r(x)-r(y))\tilde{\delta}^d(r(x)-r(z)) \\ &\quad -\frac{1}{2}\left\{|x-z|^{-\nu d}\tilde{\delta}^d(r(x)-r(y))\,1(z) + \begin{matrix}\text{permutations}\\ \text{of } x,\, y,\, z\end{matrix}\right\} \end{aligned} \tag{3.13}$$

With this subtraction prescription the coefficient $B$ in the MOPE (2.11), given originally by equation (3.9), becomes

$$\begin{aligned} &B(\{x,y,z\}) \ = \\ &-\frac{2^d}{D}\,\frac{(c^2+a^2-b^2)b^{2\nu}+(a^2+b^2-c^2)c^{2\nu}+(b^2+c^2-a^2)a^{2\nu}}{\left[(a^\nu+b^\nu+c^\nu)(a^\nu+b^\nu-c^\nu)(b^\nu+c^\nu-a^\nu)(c^\nu+a^\nu-b^\nu)\right]^{1+\frac{d}{2}}} \\ &+\frac{1}{4D}\left\{\frac{b^{D-\nu d}+c^{D-\nu d}}{a^{\nu d}}+\frac{c^{D-\nu d}+a^{D-\nu d}}{b^{\nu d}}+\frac{a^{D-\nu d}+b^{D-\nu d}}{c^{\nu d}}\right\} \end{aligned} \tag{3.14}$$

This can be checked by an explicit calculation along the lines of section 3.1.

## 3.3 IR regularization and extraction of the residue

As discussed in [18], integrals over points $x$, $y$, ... on the membrane are defined in non-integer dimension $D$ by switching to distance variables, that is by integrating over the relative Euclidean distances $|x-y|$, ... between these points. Through this change of variables the usual measure $d^Dx \wedge d^Dy \wedge \ldots$ becomes a distribution over the distance space, which depends analytically on the dimension $D$. This distribution will be shown more explicitly in section 3.4.

We already discussed that UV-divergences have to be treated by dimensional regularization. The problem arising in this context is that our Hamiltonian has no intrinsic length scale. Choosing the dimensions $D$ and $d$ so that $\varepsilon > 0$ in order to make Feynman diagrams UV-finite thus necessarily involves IR-divergences. To extract the counterterms an IR-regulator has to be introduced. This regulator will set the renormalization scale. We want to extract the UV divergence associated to the global contraction of subsets of points. An example of such a contraction, involving four subsets indicated by dotted lines (one subset is reduced to one point), is depicted in figure 3.2. According to [15] this divergence is proportional to the marginal multilocal operators appearing in the MOPE for this contraction. The divergence is obtained by integrating the corresponding MOPE coefficients for small distances inside each subset and by keeping the logarithmically divergent part. By construction, the MOPE coefficient $C$ depends only on the relative distances between points inside each subset and not on the relative distance between the different subsets. (This is why in figure 3.2 the coefficient $C$ is noted as an independant function of each subset; in fact the last subset, reduced to one point, does not play any role in $C$.)

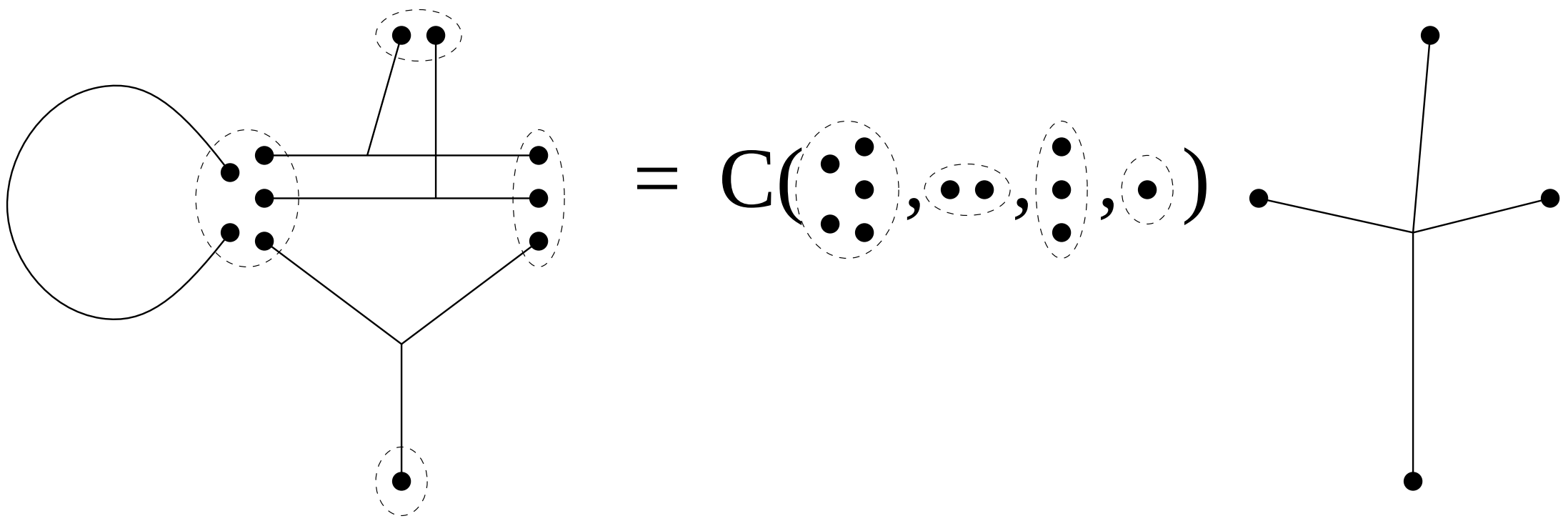

Figure 3.2: A contraction of points into 4 subsets and the corresponding MOPE coefficient for 4-body operators.

As schematically discussed in section 2.3, we choose the following IR-regularization prescription: The distances in any contracted subset have to be smaller than $L$, the IR-cutoff. Compared to other prescriptions like considering closed membranes with the topology of a hypercube or a hypersphere of diameter $L$, this prescription has a great calculatory advantage. The integrands and the MOPE coefficients are not modified and the interesting pole term can be extracted simply by using homogenity.

Let us apply this regularization prescription to the graph 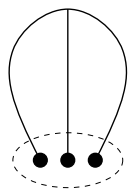, that we consider here. The MOPE (2.11) gives with the conventions of figure 3.2:

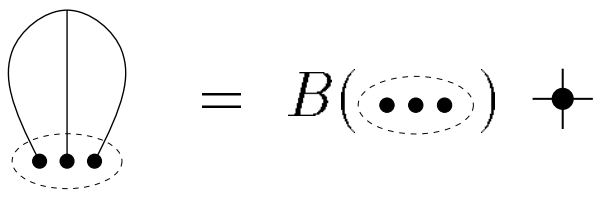 (3.15)

When integrating $B$ over the three points $x$, $y$, $z$ with fixed barycenter $o$ and with the IR regulator $L$, i.e. in the domain

$$\mathcal{D}_L(o) \;=\; \{x+y+z=3o \text{ and } |x-y|,\, |y-z|,\, |z-x| \;\leq\; L\} \tag{3.16}$$

we obtain the UV divergent term, which is denoted

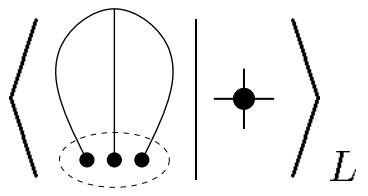

$$= \int_{\mathcal{D}_L(o)} B(\;\cdot\;) \; . \tag{3.17}$$

Since the integrand (3.14) is homogenous, we deduce that

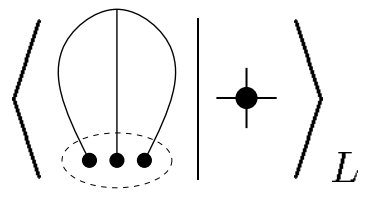

$$= c(\varepsilon)\frac{L^\varepsilon}{\varepsilon} \; . \tag{3.18}$$

For $\varepsilon = 3D - 2\nu d > 0$ the integral (3.17) is UV and IR convergent and the residue $c(\varepsilon)$ is analytic in $\varepsilon$. We are interested in the residue $c(0)$. Applying $L\frac{\partial}{\partial L}$ on both sides yields:

$$L\frac{\partial}{\partial L}$$

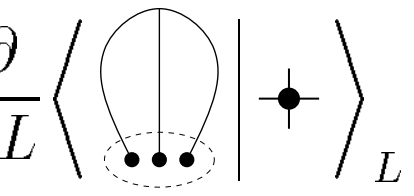

$$= c(\varepsilon)L^\varepsilon \tag{3.19}$$

Acting on the r.h.s. of equation (3.17) the operation $L\frac{\partial}{\partial L}$ extracts the boundary $\partial\mathcal{D}_L(o)$ of $\mathcal{D}_L(o)$. There the largest distance between the three points equals exactly $L$.

$$c(0) \; =$$

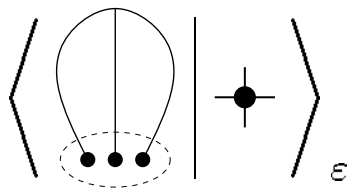

$$= \int_{\partial\mathcal{D}_L(o)} B(\;\cdot\;)\Big|_{\varepsilon=0} \tag{3.20}$$

The normalization introduced in equation (2.36) was chosen in order to eliminate the additional factor $S_D$ one might expect here.

We can consider separately the contribution of each so-called sector, where $a = |x-y|$, $b = |y-z|$ or $c = |z-x|$ respectively is the largest distance (the set where two distances have exactly the same length is a set of measure zero and thus can be neglected). By symmetry each sector gives the same contribution. For $c(\varepsilon)$ the prescription yields:

$$\begin{aligned} c(\varepsilon) \;&= 3 \int_{L=a>b,c} \\ &- \frac{2^d}{D} \frac{(c^2+a^2-b^2)b^{2\nu} + (a^2+b^2-c^2)c^{2\nu} + (b^2+c^2-a^2)a^{2\nu}}{\left[(a^\nu+b^\nu+c^\nu)(a^\nu+b^\nu-c^\nu)(b^\nu+c^\nu-a^\nu)(c^\nu+a^\nu-b^\nu)\right]^{1+\frac{d}{2}}} \\ &+ \frac{1}{4D}\left\{\frac{b^{D-\nu d}+c^{D-\nu d}}{a^{\nu d}} + \frac{c^{D-\nu d}+a^{D-\nu d}}{b^{\nu d}} + \frac{a^{D-\nu d}+b^{D-\nu d}}{c^{\nu d}}\right\} \end{aligned} \tag{3.21}$$

The final expression for the residue $c(0)$ is obtained by replacing $d$ by $d_c = 3D/(2-D)$ in equation (3.21). That this replacement is justified will be discussed in appendix A.

## 3.4 Analytic continuation of the measure

We now define the explicit form for the integration measure in non-integer dimension $D$, that will be used in the calculations. For other but equivalent formulations we refer to [18].

The general problem is to integrate some function $f$ invariant under translations and Euclidean rotations over all configurations of $N$ points $x_1, \dots, x_N$ imbedded in $D$ dimensions. This implies that $f$ depends only on the $N(N-1)/2$ relative distances $|x_i - x_j|$ between these points. In the following the integral over the center of mass is therefore always excluded. In order to be able to define the integration, let us take $D \geq N-1$ and integer. For $i < N$ denote by $y_i = x_i - x_N$ the $i$'th distance-vector and by $y_i^a$ its $a$'th component ($a = 1, \dots, D$).

The integral over $y_1$ is simple: Using rotation invariance, we fix $y_1$ to have only the $a = 1$ component non-zero. The measure becomes

$$\int d^D y_1 = S_D \int_0^\infty dy_1^1 \, (y_1^1)^{D-1} \ , \qquad y_1 = (y_1^1, 0, \dots, 0) \tag{3.22}$$

where $S_D$ is the volume of the unit-sphere in $\mathbb{R}^D$, defined in (2.36).

We now fix $y_2$ to have only $a = 1$ and $a = 2$ as non-zero components. The integral over $y_2$ consists of the integration along the direction fixed by $y_1$ and the integration in the orthogonal space $\mathbb{R}^{D-1}$:

$$\int d^D y_2 = S_{D-1} \int_{-\infty}^\infty dy_2^1 \int_0^\infty dy_2^2 \, (y_2^2)^{D-2} \ , \qquad y_2 = (y_2^1, y_2^2, 0, \dots, 0) \tag{3.23}$$

For the $j$-th point, one proceeds recursively to integrate first over the hyperplane defined by $y_1, \dots, y_{j-1}$ and then the orthogonal complement:

$$\int d^D y_j = S_{D-j+1} \prod_{a<j} \int_{-\infty}^\infty dy_j^a \int_0^\infty dy_j^j \, (y_j^j)^{D-j} \ , \qquad y_j = (y_j^1, \dots, y_j^j, 0, \dots, 0) \tag{3.24}$$

The final result for an integral over all configurations of $N$ points is

$$\int \prod_{j=1}^{N-1} d^D y_j = S_D S_{D-1} \dots S_{D-N+2} \prod_{j=1}^{N-1} \left( \prod_{a=1}^{j-1} \int_{-\infty}^\infty dy_j^a \int_0^\infty dy_j^j \, (y_j^j)^{D-j} \right) \tag{3.25}$$

This provides a well defined analytic continuation of the measure to dimensions $D$ non-integer, even for $D \leq N-1$. It is equivalent to the measures defined in [18].

As we first want to contract 3 points with coordinates ($x_1$, $x_2$, $x_3$), let us look what the general expression results in: By translation invariance we fix one point, $x_3$ and integrate only over $y = x_1 - x_3$ and $z = x_2 - x_3$ as discussed above. We thus have to evaluate an integral of the form

$$\int d^D y \, d^D z \, f(y,z) = S_D S_{D-1} \int_0^\infty dy_1 y_1^{D-1} \int_{-\infty}^\infty dz_1 \int_0^\infty dz_2 z_2^{D-2} f(y_1, z_1, z_2) \tag{3.26}$$

This expression is well defined and integrable for $D > 1$. For $D \to 1$ it should reduce to the integral over a line. Let us take $D = 1 + \varepsilon$ then we get

$$2\varepsilon \int_0^\infty dy_1 \int_{-\infty}^\infty dz_1 \int_0^\infty dz_2 \, z_2^{\varepsilon - 1} f(y_1, z_1, z_2) \ , \tag{3.27}$$

where subleading terms in $\varepsilon$ from the expansion of $S_D S_{D-1}$ are neglected. As $\varepsilon z_2^{\varepsilon-1}$ is a representation of the $\delta$-distribution for $\varepsilon \to 0$, (3.27) reduces in this limit to

$$2 \int_0^\infty dy_1 \int_{-\infty}^\infty dz_1 \, f(y_1, z_1, 0) = \int_{-\infty}^\infty dy_1 \int_{-\infty}^\infty dz_1 \, f(y_1, z_1, 0) \tag{3.28}$$

as expected.

## 3.5 Explicit integration

From (3.21) the explicit expression for the pole term proportional to $\frac{1}{\varepsilon}L^{\varepsilon}$ of the diagram contributing to the wave function renormalization is:

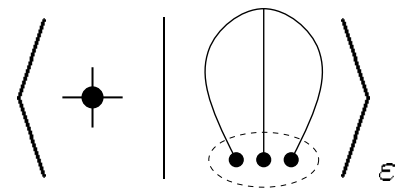

$$= -\frac{6}{4D}\int_{1=a>c>b}\left[(a^{2\nu}+c^{2\nu}-b^{2\nu})b^2+(c^{2\nu}+b^{2\nu}-a^{2\nu})a^2+(b^{2\nu}+a^{2\nu}-c^{2\nu})c^2\right]$$
$$\left(\frac{4}{(a^\nu+c^\nu+b^\nu)(a^\nu+c^\nu-b^\nu)(a^\nu-c^\nu+b^\nu)(c^\nu+b^\nu-a^\nu)}\right)^{d_c/2+1}$$
$$-a^{-\nu d_c}(b^{D-\nu d_c}+c^{D-\nu d_c})-b^{-\nu d_c}(a^{D-\nu d_c}+c^{D-\nu d_c})$$
$$-c^{-\nu d_c}(a^{D-\nu d_c}+b^{D-\nu d_c}) \qquad (3.29)$$

The factor 6 is due to the explicit ordering of the distances $a$, $b$ and $c$. This expression has to be integrated numerically. From (3.23) and with the change of normalization in (2.36) we know that the measure is

$$\int_b f(a,b,c) = \frac{S_{D-1}}{S_D}\int_{-\infty}^{\infty} db_1 \int_0^{\infty} db_2\, b_2^{D-2} f(a,b,c) \qquad (3.30)$$
$$a=1\ , \quad b=\sqrt{b_1^2+b_2^2}\ , \quad c=\sqrt{(1-b_1)^2+b_2^2}\ ,$$

restricted to the domain, where $b<c<a$:

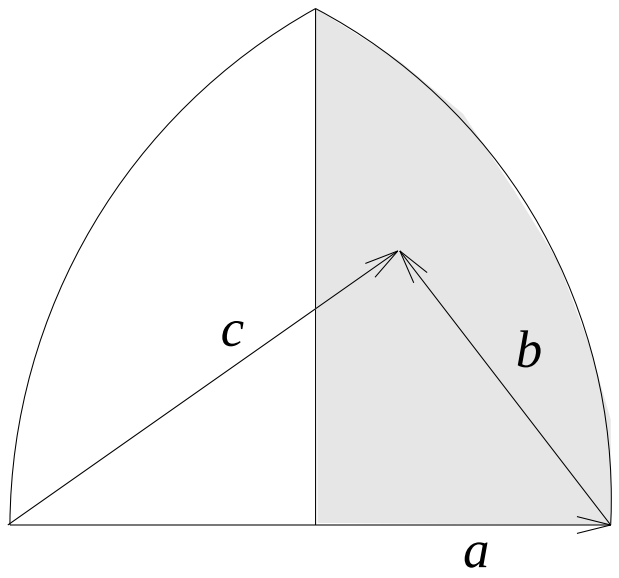

Figure 3.3: The half-sector used for the numerical integration

For $1<D<4/3$, the expression (3.29) is integrable everywhere. For $D\le 1$, the measure has a non-integrable singularity. For $D\ge 4/3$, an additional non-integrable singularity appears for small $b$.

Various numerical problems exist: The first is due to the integrable singularitiy for $b_2\to 0$ from the analytical continuation of the measure to $D<2$. The second is also an integrable singularity for $b\to 0$. They are handled by the following variable transformations:

$$\int_b f(1,b,c) = \frac{1}{D-\gamma}\frac{1}{D-1}\frac{\pi}{2}\frac{S_{D-1}}{S_D}\int_0^1 d\alpha\, \alpha^{\frac{2-D}{D-1}}\sin(\beta)^{D-2}$$
$$\int_0^1 du\, b^\gamma f(1,b,c)\Theta(1-b)\Theta(c-b)\ , \qquad (3.31)$$

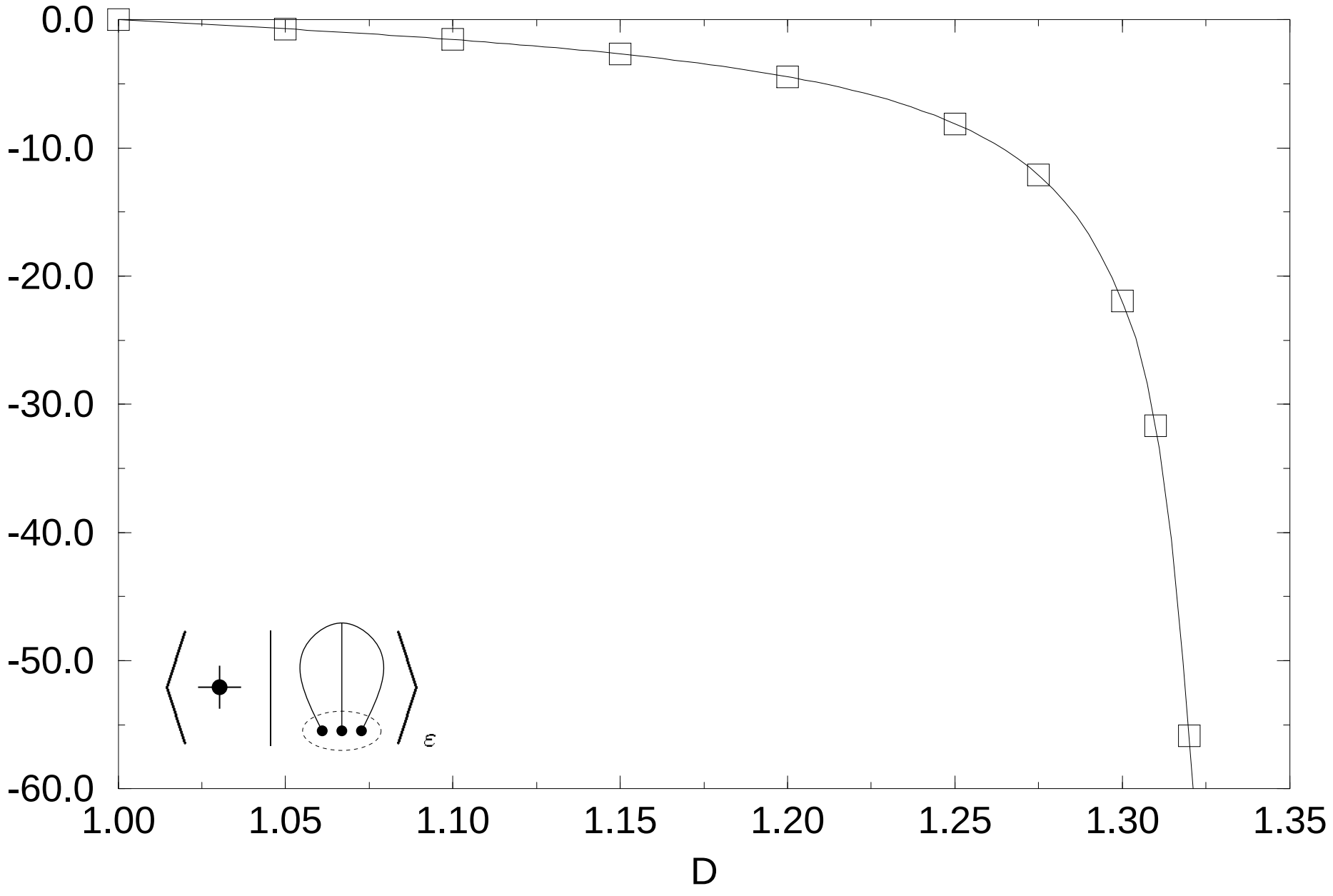

| $D$ | residue |
|---|---|
| 1.00 | 0 |
| 1.05 | -0.6993 |
| 1.10 | -1.5336 |
| 1.15 | -2.6599 |
| 1.20 | -4.4451 |
| 1.25 | -8.0963 |
| 1.275 | -12.111 |
| 1.30 | -21.9321 |
| 1.31 | -31.6537 |
| 1.32 | -55.845 |

Figure 3.4: Numerical results for the diagram (3.29)

where

$$\beta = \frac{\pi}{2}\alpha^{\frac{1}{D-1}} \tag{3.32}$$

$$b = u^{\frac{1}{D-\gamma}} \tag{3.33}$$

$$\gamma = \nu(d_c - 2) = \frac{5D}{2} - 2 \tag{3.34}$$

$$c = \sqrt{b^2 + 1 - 2b\cos(\beta)} \tag{3.35}$$

The transformations are constructed in order to generate factors which compensate the singularities of the integrand. The factor $b^\gamma$ in (3.31) e.g. exactly cancels the singularity $b^{-\gamma}$ in (3.29) as can be seen from the small $b$ expansion ($a = 1$) of the integrand:

$$-\frac{d_c + 2}{16D} b^{-\nu(d_c-2)} \left(1 + \mathcal{O}(b^{1-2\nu}) + \mathcal{O}(b^{2\nu})\right) \tag{3.36}$$

The parametrization with the angle $\beta$ is convenient to disentangle the divergence for $b_2 \to 0$ from the divergence for $b \to 0$.

An additional difficulty arises as for small $b$ (3.29) is the difference of two diverging terms. One can use the small $b$ expansion, equation (3.36), in a domain determined by the program of approximately $b < 10^{-7}$.

This integral is now performed by a simple numerical integration routine using Simpson's rule. On a workstation, this integration takes some minutes for a precision of $10^{-4}$.

The result of the calculation can be found in figure 3.4. For $D \to 0$, the diagram vanishes, whereas for $D \to 4/3$ it diverges. This is due to the fact that for $D \geq 4/3$ a new subdivergence for $b \to 0$ occurs.

### 3.6 The limit $D \to 1$

The case $D = 1$ can be treated analytically. For $D = 1$ and omitting the counterterms, (3.29) reduces to

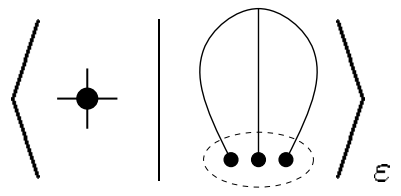

$$
\begin{aligned}
\left\langle \cdots \right\rangle_{\varepsilon} &= -\frac{3}{4}\int_0^1 db\,(b-b^2)^{-d_c/2} \\
&= -\frac{3}{4}\frac{\Gamma^2(1-d_c/2)}{\Gamma(2-d_c)} \\
&= 0 \quad \text{for } d_c = 3
\end{aligned}
\tag{3.37}
$$

The factor 3 in (3.37) is due to symmetry. The integral runs over one of three possible and equivalent sectors only. For simplicity the counterterms for the relevant subdivergences were omitted, as the integrand can by calculated analytically without counterterms and since they were constructed from a finite part prescription and thus give no contribution. This is indeed the case, as the integral of the counterterms is:

$$
\begin{aligned}
-\frac{3}{4}\int_0^1 db\,&\left(b^{-d_c/2} + (1-b)^{1-d_c/2} + b^{-d_c/2}(1-b)^{1-d_c/2}\right) \\
&= -\frac{3}{4}\left(\frac{2}{2-d_c} + \frac{2}{4-d_c} + \frac{\Gamma(1-d_c/2)\Gamma(2-d_c/2)}{\Gamma(3-d_c)}\right) \\
&= 0 \quad \text{for } d_c = 3
\end{aligned}
\tag{3.38}
$$

These calculations show that the limit $D \to 1$ is correctly reproduced [21].

# 4 Conformal Mapping of the Sectors

In the previous section, we had to calculate some integral over distances, restricted to various sectors, such that one of the distances is larger than all the others and is set to $L$. In that case, the integrand was symmetric with respect to the three distances $a$, $b$ and $c$ and each of the three sectors gave the same contribution. In the following we shall deal with non-symmetric integrands and with integrals over more distances. We now introduce an extremely useful tool, the mapping of sectors, which will reveal its full power in the analysis of the 4-point divergences. Besides calculational convenience it also shows that the simple pole at 1-loop order is an universal quantity. We shall discuss this later. First we show how different sectors, arising in the analysis of the 3-point divergences, can be mapped onto each other. Let us remind that, with the measure (3.23) and the normalization (2.36), we had to compute an integral over some domain in the upper half plane, with the measure

$$
\int_y = \frac{S_{D-1}}{S_D}\int_{-\infty}^{\infty} dy_1 \int_0^{\infty} dy_2\,(y_2)^{D-2} , \tag{4.1}
$$

of a function $f$ of the three distances $a$, $b$ and $c$ between the points $\mathrm{F} = (0,0)$, $\mathrm{E} = (-L,0)$ and $\mathrm{G} = (y_1, y_2)$, given explicitly by (cf. figure 4.1)

$$
a = L \text{ fixed}, \quad b = \sqrt{(y_1)^2 + (y_2)^2}\ , \quad c = \sqrt{(L-y_1)^2 + (y_2)^2}\ . \tag{4.2}
$$

Let us consider that $f$ is homogenous, with degree $\lambda$, but not necessarily symmetric

$$f(\kappa a, \kappa b, \kappa c) = \kappa^{-\lambda} f(a, b, c)\ . \tag{4.3}$$

The exponent $\lambda$ is called the conformal weight of the integrand $f$.

The upper half-plane in $y$ can be divided into three sectors $\mathcal{A}$, $\mathcal{B}$ and $\mathcal{C}$. The sector $\mathcal{A}$

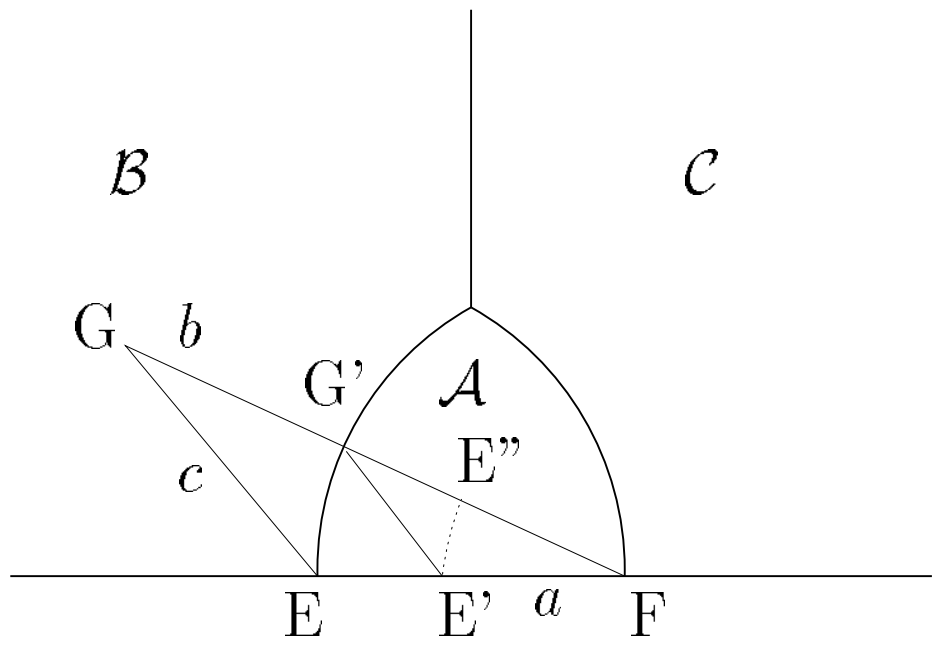

Figure 4.1: The sectors $\mathcal{A}$, $\mathcal{B}$ and $\mathcal{C}$

consists of all triangles $\Delta(a, b, c)$ with $a = L$ and $b, c < a$ as indicated in figure 4.1. This is the sector over which the integration was performed in section 3.3. The sectors $\mathcal{B}$ and $\mathcal{C}$ are the domains, where $b$ and $c$ respectively are the largest distances.

We can map $\mathcal{B}$ onto $\mathcal{A}$. This mapping consists of 2 steps:

- the rescaling with respect to F by a factor $L/|b|$ which maps G onto G' and E onto E'. (4.3) implies that $f$ is changed by a factor $\left(\frac{L}{|b|}\right)^{-\lambda}$.
- a mirror operation, which maps G' onto E and E' onto E", leaving invariant F. This operation is a permutation of the first two arguments of $f$.

The mapping G→E" is a special conformal transformation, the inversion with respect to the circle $S_L$(F). In complex coordinates it is

$$y = y_1 + \mathrm{i} y_2 \longrightarrow \tilde{y} = \tilde{y}_1 + \mathrm{i}\tilde{y}_2 = \frac{L^2}{\bar{y}} \tag{4.4}$$

One easily checks that this mapping $\mathcal{B} \rightarrow \mathcal{A}$ is one to one. The measure in (4.1) transforms as

$$d^D y \ = \ d^D \tilde{y} \left(\frac{L}{|\tilde{y}|}\right)^{2D} \ . \tag{4.5}$$

The final result is:

$$\int_{y \in \mathcal{B}} f(a = L, b, c) = \int_{y \in \mathcal{A}} \left(\frac{b}{L}\right)^{\lambda - 2D} f(b, a = L, c) \tag{4.6}$$

For the integrals which give the residue at the critical dimension, (4.2) is dimensionless and the degree $\lambda$ in (4.3) is $\lambda = 2D$ so that we have

$$\int_{y \in \mathcal{B}} f(a = L, b, c) = \int_{y \in \mathcal{A}} f(b, a = L, c)\ . \tag{4.7}$$

An analogous transformation is valid for the mapping of $\mathcal{C}$ onto $\mathcal{A}$:

$$\int_{y\in\mathcal{C}} f(a=L,b,c) = \int_{y\in\mathcal{A}} f(c,b,a=L) \tag{4.8}$$

We call $\kappa = \lambda - 2D$ the conformal dimension of the integral. For integrals with conformal dimension zero, a conformal change of coordinates to map the various sectors simply permutes the vertices of the triangle.

One word should be said about the conformal mapping for integrals which have not conformal dimension 0. As $a$ is fixed to equal $L$ one can always multiply the integrand by a power of $a/L$, by this way adjusting the conformal dimension to 0. Then the conformal mapping again consists in a pure permutation of the arguments.

A more general method to look at the mapping of sectors consists in using the measure (3.7) of [18] over the distances, considered as independent variables (we set $L = 1$). The integral of $f$ over (4.1) is

$$\int_{y\in\mathcal{A}} f(a,b,c) \ = \ \int d\mu_D(a,b,c)\, \chi_{\mathcal{A}}(a,b,c)\, \delta(a-1)\, f(a,b,c) \tag{4.9}$$

with the measure $d\mu_D$ defined as

$$d\mu_D(a,b,c) = \frac{1}{8}\frac{S_{D-1}}{S_D} da^2 db^2 dc^2 \ (2\Delta(a,b,c))^{D-3} \ . \tag{4.10}$$

$\Delta(a,b,c)$ is the area of the triangle with edge lengths $a$, $b$ and $c$ and $\chi_{\mathcal{A}}$ the characteristic function of the sector $\mathcal{A}$:

$$\chi_{\mathcal{A}}(a,b,c) \ = \ \theta(a-b)\theta(a-c) \tag{4.11}$$

Substituting $a = \tilde{a}b$, $c = \tilde{c}b$ and using the homogeneity of $f$ and of the measure, the r.h.s. of (4.9) becomes

$$\frac{1}{8}\frac{S_{D-1}}{S_D} \int d\tilde{a}^2\, db^2\, d\tilde{c}^2\, \delta(\tilde{a}b-1)\, b^{2D-2-\lambda}\, (2\Delta(\tilde{a},1,\tilde{c}))^{D-3}\, f(\tilde{a},1,\tilde{c})\ \chi_{\mathcal{A}}(\tilde{a},1,\tilde{c}) \tag{4.12}$$

Substituting further $b = \tilde{b}/\tilde{a}$ and using the fact that the $\delta$-distribution restricts $\tilde{b}$ to equal 1 yields:

$$\frac{1}{8}\frac{S_{D-1}}{S_D} \int d\tilde{a}^2\, d\tilde{b}^2\, d\tilde{c}^2\, \delta(\tilde{b}-1)\, \tilde{a}^{-2D+\lambda} (2\Delta(\tilde{a},\tilde{b},\tilde{c}))^{D-3} f(\tilde{a},\tilde{b},\tilde{c}) \chi_{\mathcal{A}}(\tilde{a},\tilde{b},\tilde{c}) \tag{4.13}$$

In order to get the same formula as (4.7), one considers functions $f$ with conformal weight $\lambda = 2D$ and renames $b = \tilde{a}$, $a = \tilde{b}$ and $c = \tilde{c}$, to obtain:

$$\int d\mu_D(a,b,c)\, \delta(a-1)\, f(a,b,c)\, \chi_{\mathcal{A}}(a,b,c) = \int d\mu_D(a,b,c)\, \delta(a-1)\, f(b,a,c)\, \chi_{\mathcal{B}}(a,b,c) \tag{4.14}$$

The result is equivalent to (4.8).

The method extends to the contraction of four or more points. We give the general result for $N$ points connected by the distances $x = z_1$, $y = z_2$, $z_3, \ldots, z_{N(N-1)/2}$ in a slightly different formulation:

$$\begin{aligned} &\int d\mu_D(x, y, z_3, \ldots, z_{N(N-1)/2}) f(x, y, z_3, \ldots, z_{N(N-1)/2})\, \delta(x-1) \\ &\quad = \int d\mu_D(x, y, z_3, \ldots, z_{N(N-1)/2}) f(x, y, z_3, \ldots, z_{N(N-1)/2})\, \delta(y-1) x^{-(N-1)D+\lambda} \end{aligned} \tag{4.15}$$

The ordering of the distances on both sides of (4.15) has to coincide.

In the case of two or more contraction domains the same rules apply: one adjusts the conformal dimension of the integral to equal 0 and then can freely choose the length which has to be fixed. The ordering of the distances of course has to be respected.

The mapping of sectors immediately proves the universality of the pole term in $\varepsilon$ at 1-loop order, i.e. as long as no double pole appears, once the general normalization has been fixed: Demanding the longest, the shortest or any intermediate distance to equal $L$ results in the same pole term. From these regularization prescription every other prescription can be constructed. This observation completes the proof. (The question whether different prescriptions to subtract the relevant divergences change the pole term is discussed in appendix A).

# 5 3-body Interaction Renormalization: Easy Graphs

## 5.1 Graph reducible to 2-point integrals

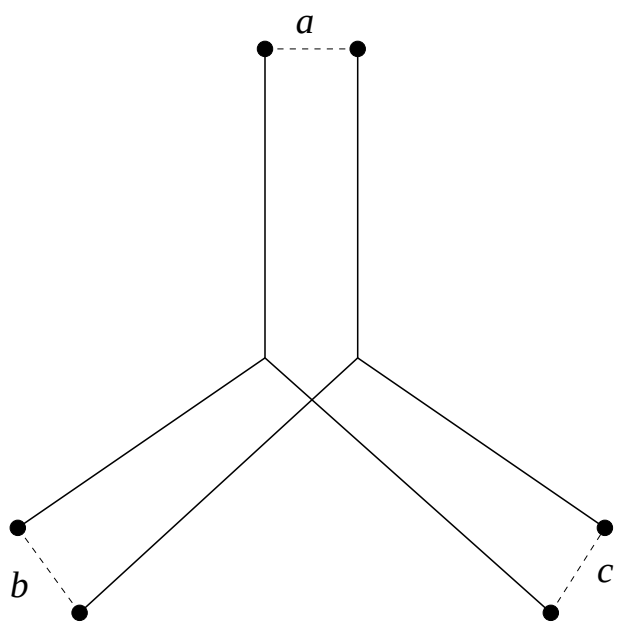

Figure 5.1: Distances in (5.1)

The first contribution to the renormalization of the 3-point interaction comes from the MOPE (2.15). The divergent integral is:

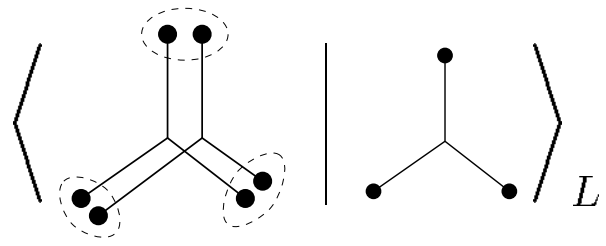

$$= \int_{a<L} \int_{b<L} \int_{c<L} \left( \frac{1}{a^{2\nu}c^{2\nu} + c^{2\nu}b^{2\nu} + b^{2\nu}a^{2\nu}} \right)^{d/2} \tag{5.1}$$

By the same procedure as in section 3.5, the pole term is extracted by fixing one by one the distances $a$, $b$ and $c$ to equal $L$ and to be the largest. Using the mapping of section

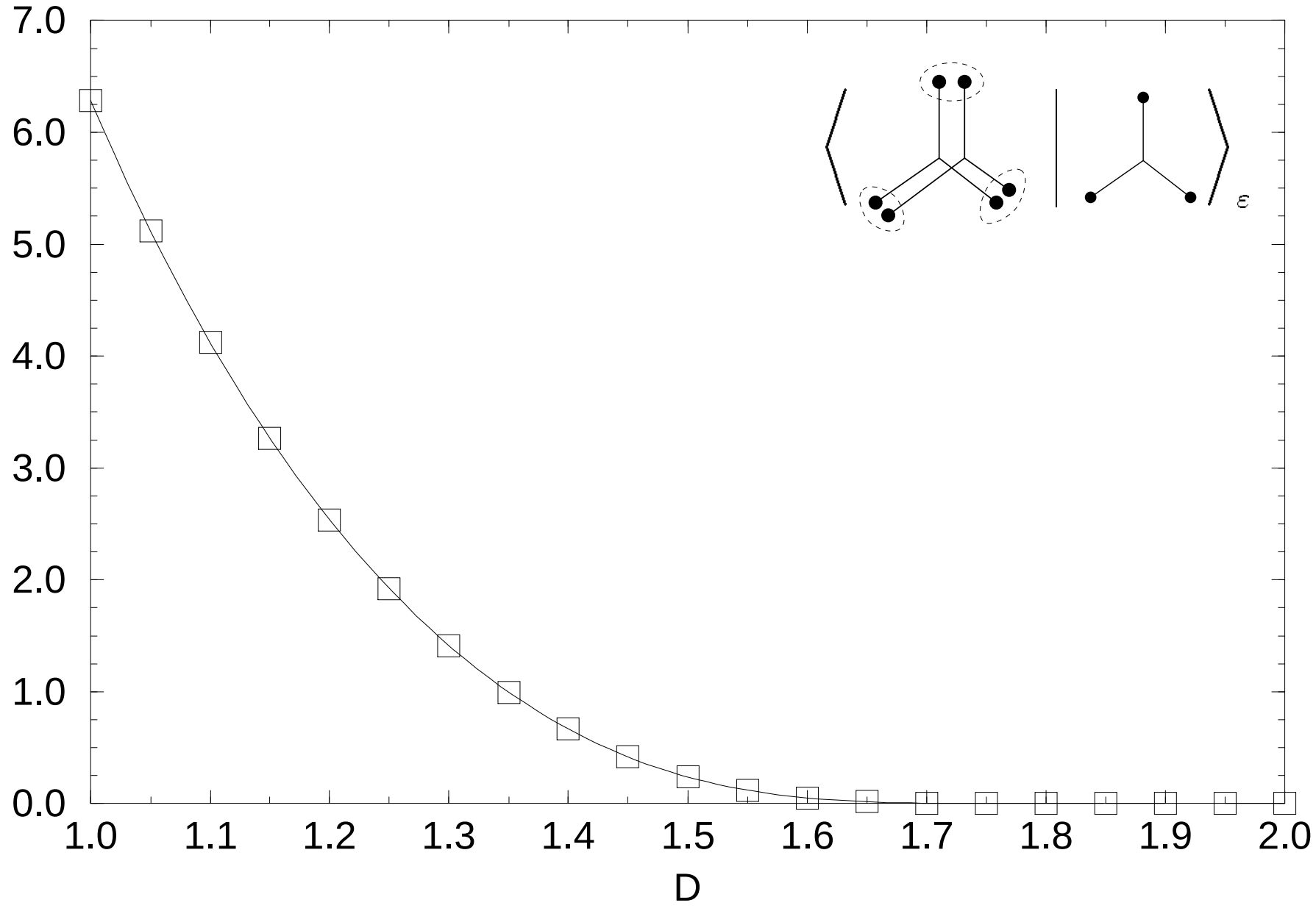

| $D$ | | $D$ | | $D$ | | $D$ | |
|---|---|---|---|---|---|---|---|
| 1.00 | $2\pi$ | 1.25 | 1.923434 | 1.50 | 0.239359 | 1.75 | 0.000518 |
| 1.05 | 5.118675 | 1.30 | 1.412483 | 1.55 | 0.121477 | 1.80 | 0.000023 |
| 1.10 | 4.120247 | 1.35 | 0.996341 | 1.60 | 0.052083 | 1.85 | 0.000000 |
| 1.15 | 3.265599 | 1.40 | 0.667377 | 1.65 | 0.017497 | 1.90 | 0.000000 |
| 1.20 | 2.537700 | 1.45 | 0.417915 | 1.70 | 0.004061 | 2.00 | 0 |

Figure 5.2: Numerical results for the diagram (5.3)

4 this integral is converted to an integral, where $a = L = 1$ fixed, $b$ and $c$ now running from 0 to $\infty$. The pole term is obtained by setting $d = d_c$:

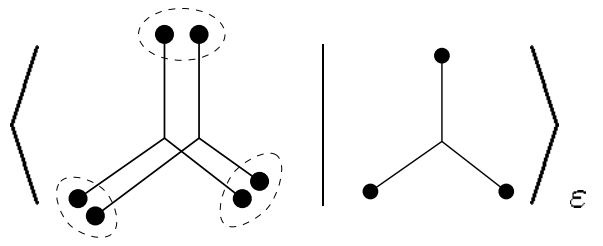

$$= \int_b \int_c \left( \frac{1}{c^{2\nu} + c^{2\nu} b^{2\nu} + b^{2\nu}} \right)^{d_c/2} \tag{5.2}$$

The integration can easily be done as the integrals over both $b$ and $c$ simply yield Beta-functions:

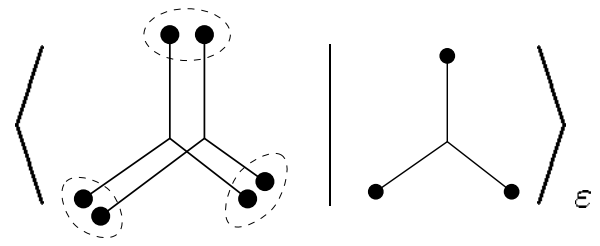

$$= \frac{1}{(2-D)^2} \frac{\Gamma\left(\frac{1}{2}\frac{D}{2-D}\right)^3}{\Gamma\left(\frac{3}{2}\frac{D}{2-D}\right)} \tag{5.3}$$

For $D = 1$ the integral is $2\pi$ and decays rapidly for $D \to 2$, where it vanishes. A plot and some numerical values can be found in figure 5.2.

## 5.2 Graph reducible to a 3-point integral

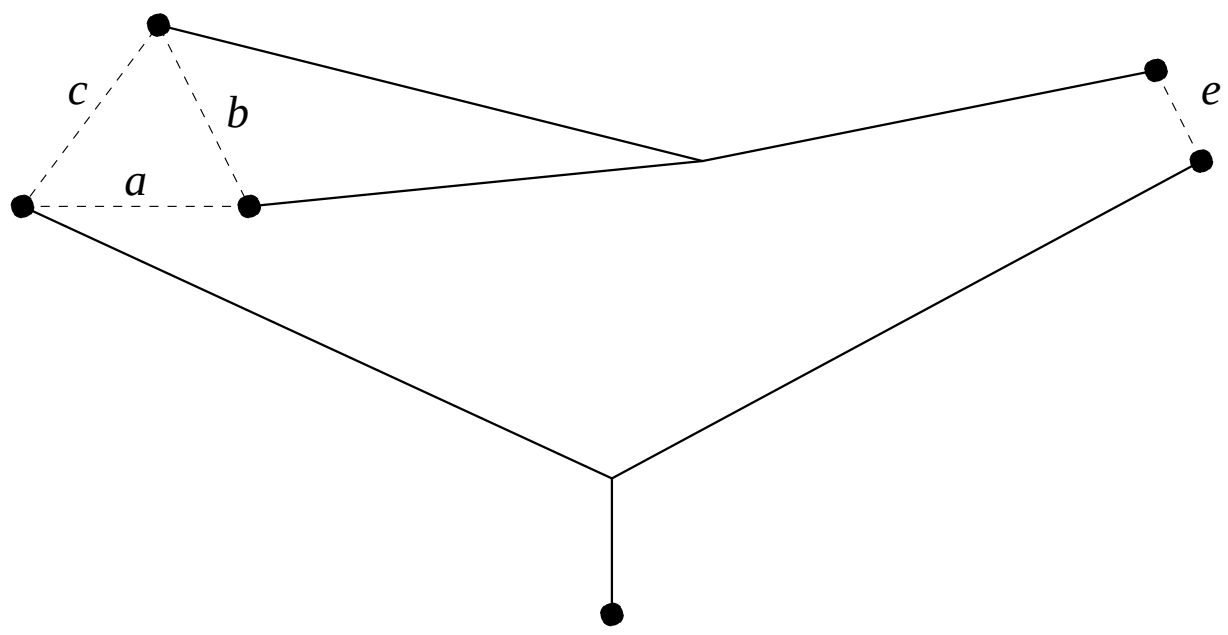

Figure 5.3: The distances in (5.4)

The next diagram, which has to be calculated, comes from the MOPE (2.16). The divergent integral is:

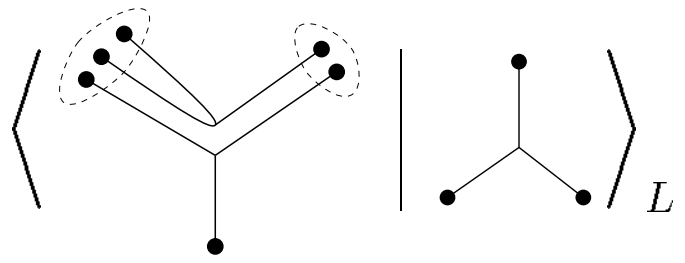

$$= \iint_{a,b,c<L} \int_{e<L}$$

$$\left( \frac{4}{4b^{2\nu}e^{2\nu} + (a^\nu + c^\nu + b^\nu)(a^\nu + c^\nu - b^\nu)(a^\nu - c^\nu + b^\nu)(c^\nu + b^\nu - a^\nu)} \right)^{d/2} \tag{5.4}$$

The pole term is again derived by applying the operator $L\frac{\partial}{\partial L}|_{d=d_c}$ to the diagram and leads to look at the sectors where $a$, $b$, $c$ or $e$ equals $L$ and all other distances are smaller. By mapping the sectors, it can be transformed to an integral over the sectors, where the largest of the distances $a$, $b$ or $c$ equals $L$, $e$ now running from 0 to $\infty$.

For the relevant divergence for $b \to 0$, the counterterms are found by applying a finite part prescription. This gives:

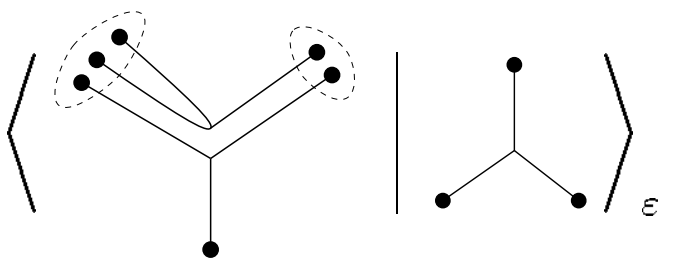

$$= \left[ \int_{L=a>b,c} + \int_{L=b>a,c} + \int_{L=c>a,b} \right] \int_e$$

$$\left( \frac{4}{4b^{2\nu}e^{2\nu} + (a^\nu + c^\nu + b^\nu)(a^\nu + c^\nu - b^\nu)(a^\nu - c^\nu + b^\nu)(c^\nu + b^\nu - a^\nu)} \right)^{d_c/2}$$
$$-\frac{1}{2} b^{-\nu d_c} \left[ \left( a^{2\nu} + e^{2\nu} \right)^{-d_c/2} + \left( c^{2\nu} + e^{2\nu} \right)^{-d_c/2} \right] \tag{5.5}$$

The integral over $e$ can still be performed analytically:

$$\frac{1}{2-D} \frac{\Gamma\left(\frac{D}{2-D}\right) \Gamma\left(\frac{D}{2(2-D)}\right)}{\Gamma\left(\frac{3D}{2(2-D)}\right)} \left[ \int_{L=a>b,c} + \int_{L=b>a,c} + \int_{L=c>a,b} \right]$$
$$b^{-D} \left( \frac{4}{(a^\nu + c^\nu + b^\nu)(a^\nu + c^\nu - b^\nu)(a^\nu - c^\nu + b^\nu)(c^\nu + b^\nu - a^\nu)} \right)^{\frac{D}{2(2-D)}}$$
$$-\frac{1}{2} b^{-3D/2} \left( a^{-D/2} + c^{-D/2} \right) \tag{5.6}$$

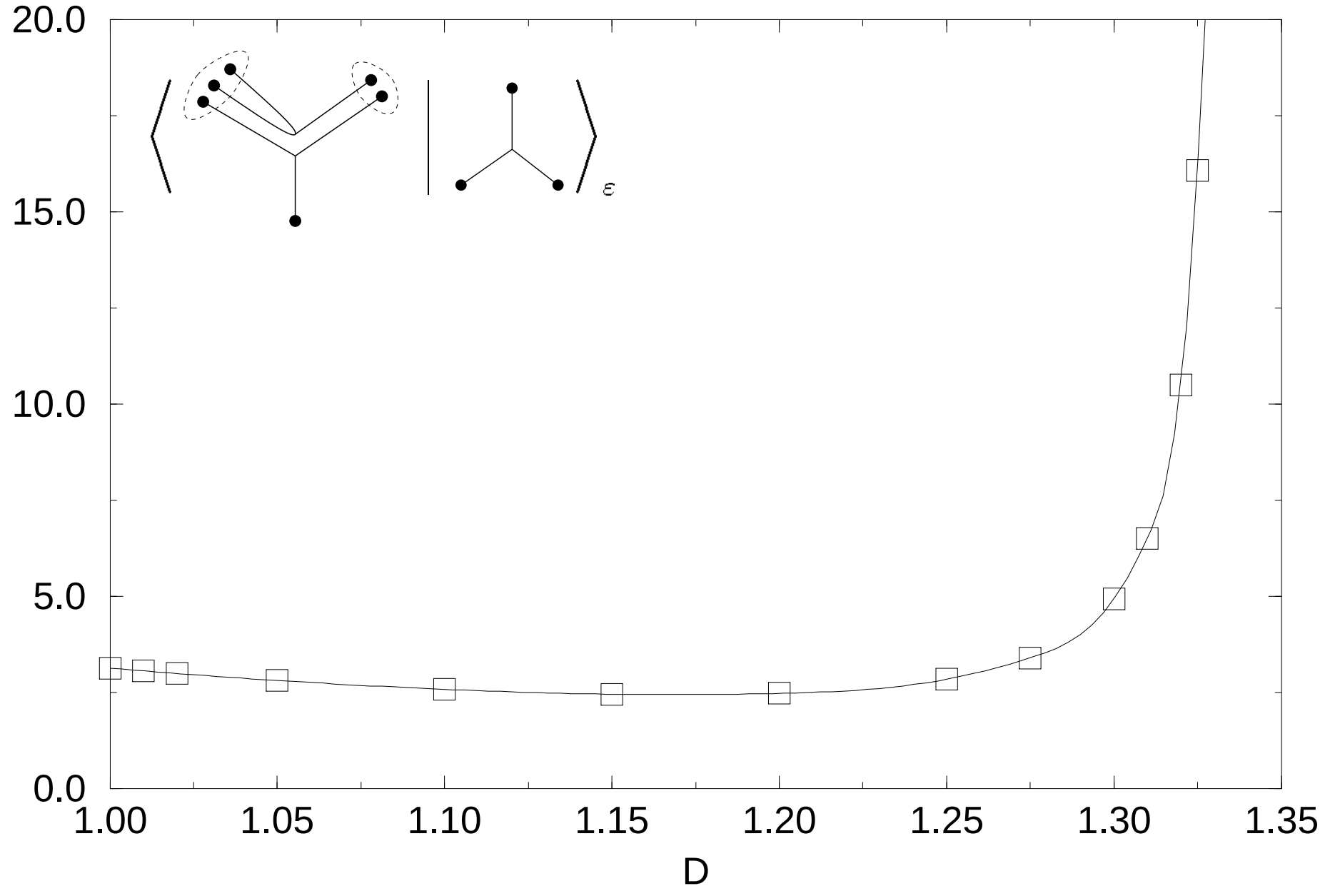

| $D$ | |
|---|---|
| 1.00 | $\pi$ |
| 1.01 | 3.07 |
| 1.02 | 3.0027 |
| 1.05 | 2.8210 |
| 1.10 | 2.5903 |
| 1.15 | 2.4618 |
| 1.20 | 2.4854 |
| 1.25 | 2.8518 |
| 1.275 | 3.4084 |
| 1.30 | 4.9365 |
| 1.31 | 6.5140 |
| 1.32 | 10.5032 |
| 1.325 | 16.09 |

Figure 5.4: Numerical results for the diagram (5.7)

This integral is numerically integrated in its symmetrized version:

$$\frac{1}{2-D}\frac{\Gamma\left(\frac{D}{2-D}\right)\Gamma\left(\frac{D}{2(2-D)}\right)}{\Gamma\left(\frac{3D}{2(2-D)}\right)}\int_{L=a>c>b} 2\left(a^{-D}+b^{-D}+c^{-D}\right)$$
$$\left(\frac{4}{(a^\nu+c^\nu+b^\nu)(a^\nu+c^\nu-b^\nu)(a^\nu-c^\nu+b^\nu)(c^\nu+b^\nu-a^\nu)}\right)^{\frac{D}{2(2-D)}}$$
$$-a^{-3D/2}\left(c^{-D/2}+b^{-D/2}\right)-b^{-3D/2}\left(a^{-D/2}+c^{-D/2}\right)-c^{-3D/2}\left(b^{-D/2}+a^{-D/2}\right) \quad (5.7)$$

The numerical problems and their solution are similar to those discussed in section 3.5. The singularity for $a=1$ and $b\to 0$ is:

$$\frac{D}{4(2-D)^2}\frac{\Gamma\left(\frac{D}{2-D}\right)\Gamma\left(\frac{D}{2(2-D)}\right)}{\Gamma\left(\frac{3D}{2(2-D)}\right)}b^{2-5D/2}\left(1+\mathcal{O}(b^{1-2\nu})+\mathcal{O}(b^{2\nu})\right) \quad (5.8)$$

This equation determines the exponent $\gamma$ for the transformation of the measure (3.31)

$$\gamma=\frac{5D}{2}-2 \quad (5.9)$$

as before. It also serves to eliminate the small $b$ difficulties as discussed at the end of section 3.5.

The explicit numerical results are given in figure 5.4. For $D\to 1$ the numerical value approaches $\pi$, which is the result obtained by analytical calculation. For $D\to 4/3$ it diverges, again reflecting the fact that for $D\geq 4/3$ a new non-integrable divergence appears for small $b$.

# 6 3-body Interaction Renormalization: Hard Graph

## 6.1 The 4-point diagram

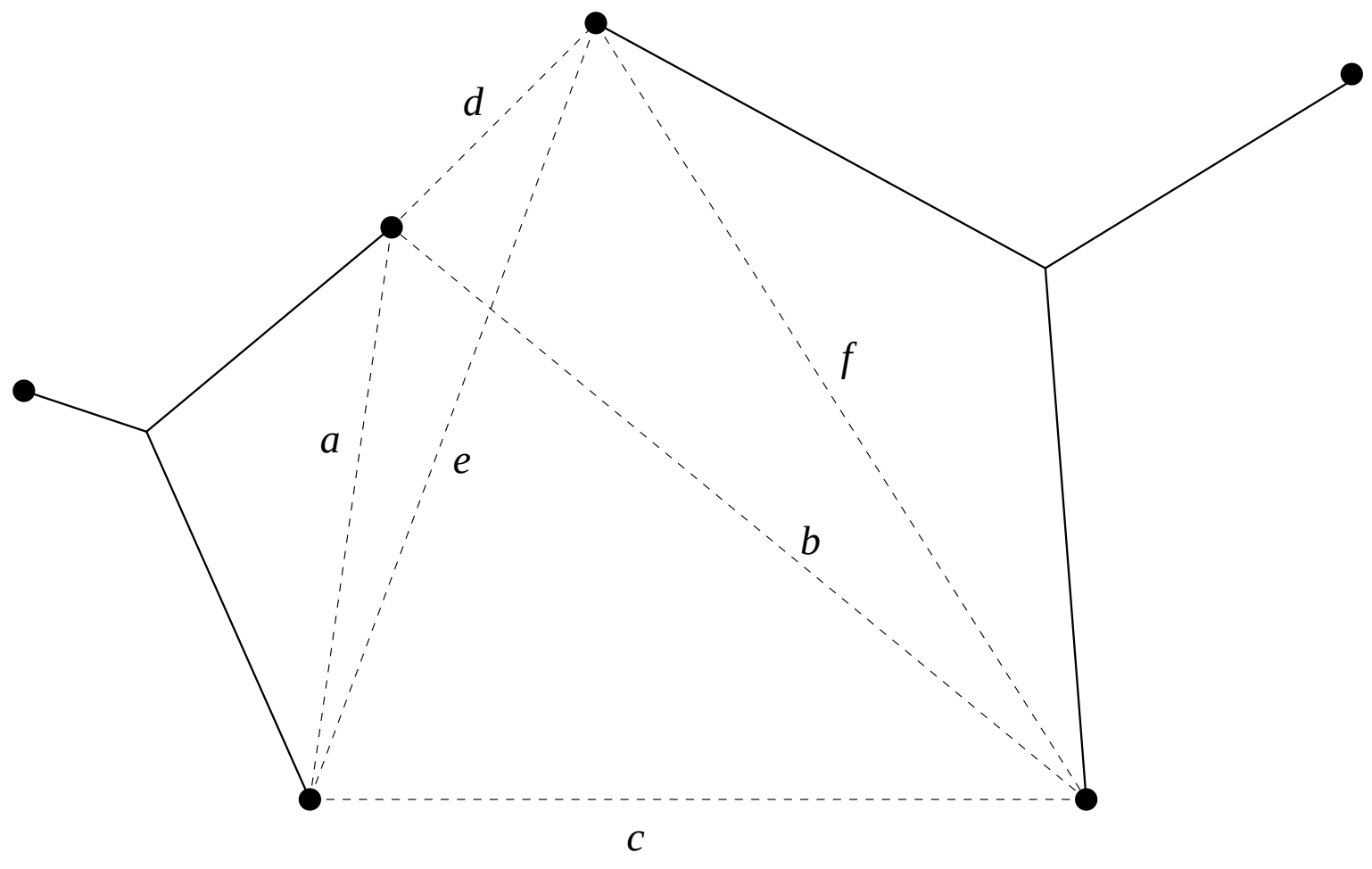

Figure 6.1: The distances in (6.1)

The last diagram, which contributes to the renormalization of the 3-point interaction, is depicted in figure 6.1. The associated divergent integral is:

$$\left\langle \;\cdot\; \middle| \;\cdot\; \right\rangle_L = \iiiint\!\!\iint_{a,b,c,d,e,f<L} \left[ a^{2\nu} f^{2\nu} - \frac{1}{4}\left( c^{2\nu} - e^{2\nu} - b^{2\nu} + d^{2\nu} \right)^2 \right]^{-d/2} \tag{6.1}$$

Applying the operator $L\frac{\partial}{\partial L}$ from section 3.3 to extract the pole term gives 6 different sectors: Each of the distances $a$, $b$, $c$, $d$, $e$ or $f$ might be the largest. In order to simplify the calculations, the mapping of section 4 shall be used. We want to end up with an integral, where $c = L = 1$ is fixed, whereas the other distances may vary freely:

$$\left\langle \;\cdot\; \middle| \;\cdot\; \right\rangle_\varepsilon = \iiiint\!\!\int_{c=1,a,b,d,e,f} \left[ a^{2\nu} f^{2\nu} - \frac{1}{4}\left( c^{2\nu} - e^{2\nu} - b^{2\nu} + d^{2\nu} \right)^2 \right]^{-d_c/2} \tag{6.2}$$

For $c = 1$ the measure, given by (3.25) and (2.36), simplifies to an integral over the vectors $a$ and $f$:

$$\frac{S_{D-1}S_{D-2}}{S_D^2} \int_{-\infty}^{\infty} da_1 \int_0^{\infty} da_2\, a_2^{D-2} \int_{-\infty}^{\infty} df_1 \int_{-\infty}^{\infty} df_2 \int_0^{\infty} df_3\, f_3^{D-3} F(a,b,c,d,e,f) \tag{6.3}$$

We herein abbreviated the integrand by $F(a,b,c,d,e,f)$.

For small $a$ or $f$, $F$ possesses relevant subdivergences. They are eliminated via a finite part prescription, if one uses the modified 3-body interaction (3.13):

$$\iiiint\!\!\int_{c=1,a,b,d,e,f} \left[ a^{2\nu} f^{2\nu} - \frac{1}{4}\left( c^{2\nu} - e^{2\nu} - b^{2\nu} + d^{2\nu} \right)^2 \right]^{-d_c/2} - (af)^{-\nu d_c} \tag{6.4}$$

The integrand now is integrable.

## 6.2 Improvement of the measure

For $D < 2$ the measure defined in (6.3) is a distribution and suffers from a relevant divergence for $f_3 \to 0$. Geometrically these are configurations, where the tetrahedron spanned by $a, b, \ldots, f$ has volume 0, i.e. is restricted to a plane. A finite part prescription has to be applied in order to make the measure finite.

One may think of implementing this prescription by subtracting the singularity. This method however imposes at least numerical difficulties. It is better to eliminate the singularity by a partial integration with respect to $f_3$, which is mathematically equivalent. As only $d$, $e$ and $f$ depend on $f_3$, the integral

$$\int_0^\infty df_3\, f_3^{D-3} F(a,b,c,d,e,f) \tag{6.5}$$

can be converted to

$$\begin{aligned} &\frac{1}{2-D}\int_0^\infty df_3\, f_3^{D-2}\frac{\partial}{\partial f_3}F(a,b,c,d,e,f) \\ &= \frac{1}{2-D}\int_0^\infty df_3\, f_3^{D-1}\mathbf{R}F(a,b,c,d,e,f) \end{aligned} \tag{6.6}$$

where $\mathbf{R}$ is defined via

$$\mathbf{R} = \frac{1}{d}\frac{\partial}{\partial d} + \frac{1}{e}\frac{\partial}{\partial e} + \frac{1}{f}\frac{\partial}{\partial f} \tag{6.7}$$

The strength of the divergences for $a \to 0$ or $f \to 0$ is unchanged. It is important to remark that this trick can not be used to eliminate the relevant divergences when $a \to 0$ or $f \to 0$. It works for the divergence in $f_3$, because the integrand does not directly depend on $f_3$ but on $d$, $e$ and $f$, which themselves depend on $f_3$. So the derivation of $F$ with respect to $f_3$ does not produce a factor $1/f_3$ but factors $1/d$, $1/e$ and $1/f$, which are not singular for $f_3 \to 0$. Explicitly:

$$\begin{aligned} \mathbf{R}F(a,b,c,d,e,f) = \frac{\nu d}{f^2}\Bigg[ &\left(\frac{1}{2}f^2\left(d^{-D}-e^{-D}\right)\left(c^{2\nu}-e^{2\nu}-b^{2\nu}+d^{2\nu}\right)-a^{2\nu}f^{2\nu}\right) \\ &\left(a^{2\nu}f^{2\nu}-\frac{1}{4}\left(c^{2\nu}-e^{2\nu}-b^{2\nu}+d^{2\nu}\right)^2\right)^{-d/2-1} + a^{-\nu d}f^{-\nu d}\Bigg] \end{aligned} \tag{6.8}$$

## 6.3 Parametrization of the measure

The main singularities for small distances appear for $a$ or $f$ small. As in section 3.5 we therefore want to parametrize the measure with the help of these distances. The divergences for small volume of the tetrahedron spanned by $a, \ldots, f$, $a_2 \to 0$ or $f_3 \to 0$ shall be treated by a parametrization in angles as by this way small distance and small volume singularities are best disentangled. We have chosen the parametrization indicated in the figure. One triangle is spanned by $c$ and $a$ with an angle $\beta$ between, another by $c$ and $f$, where the corresponding angle is $\sigma$. The angle between the planes spanned by these two triangles is $\tau$. The distances as functions of $a$, $f$ and $\beta$, $\sigma$, $\tau$ are:

$$b = \sqrt{a^2 + 1 - 2a\cos\beta}$$

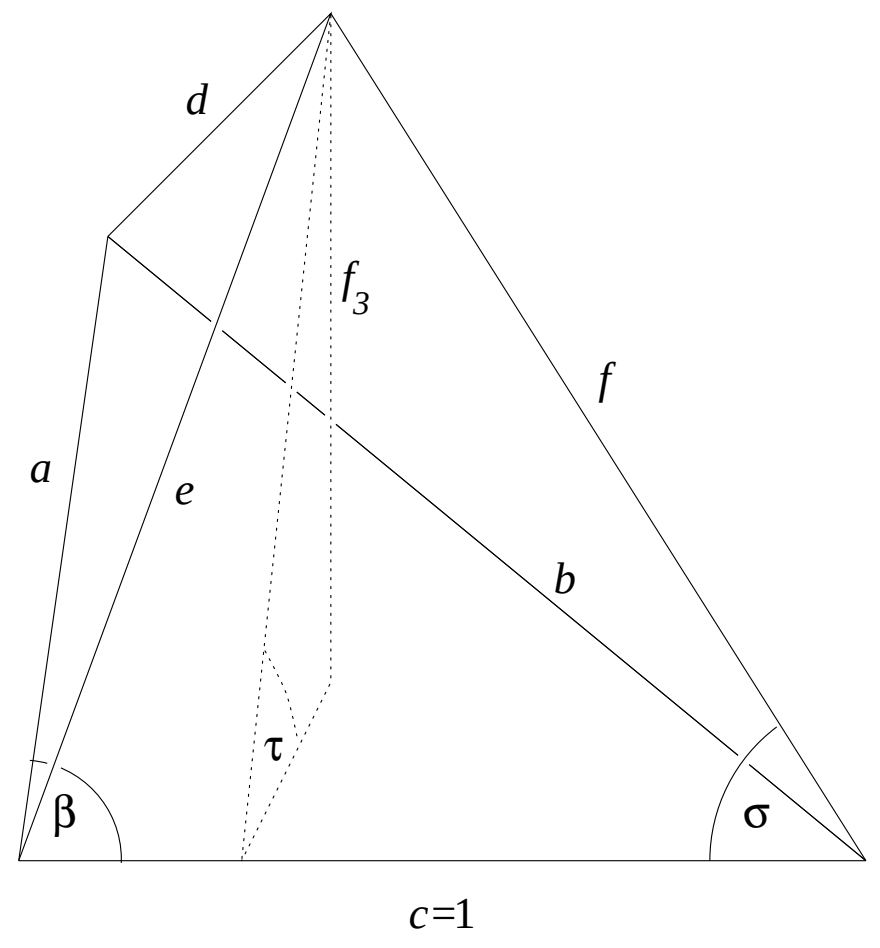

Figure 6.2: Parametrization of the tetrahedron

$$e = \sqrt{f^2 + 1 - 2f\cos\sigma} \tag{6.9}$$
$$d = \sqrt{(a\cos\beta - 1 + f\cos\sigma)^2 + (a\sin\beta - f\sin\sigma\cos\tau)^2 + (f\sin\tau\sin\sigma)^2}$$

The integrals over $a$ and $f$ run from 0 to $\infty$, the integrals over $\beta$, $\sigma$ and $\tau$ over the interval $[0,\pi]$. As we do not want to map all the points which are far away, we have to find a reparametrization of the measure which behaves for $a \to 0$ like $a^\gamma$ and for $a \to \infty$ like $a^\omega$, by this way eliminating the principle divergences. If $u$ is equally distributed we can use

$$a = u^{\frac{1}{D-\gamma}}(1-u)^{\frac{1}{D-\omega}} \ . \tag{6.10}$$

The integral over $\beta$ will be parametrized as

$$\beta = \begin{cases} \frac{\pi}{2}(2\alpha)^{\frac{1}{D-1}} & \alpha \leq 0.5 \\ \pi - \frac{\pi}{2}(2-2\alpha)^{\frac{1}{D-1}} & \alpha > 0.5 \end{cases} \tag{6.11}$$

which differs from (3.31) and following by the different range of integration: $[0,\pi]$ instead of $[0,\pi/2]$.

The integral over $f$ (remind the factor $f_3^2$ came from the partial integration with respect to $f_3$)

$$\int df_1\, df_2\, df_3\, f_3^{D-3}(f_3^2) \tag{6.12}$$

can be written as

$$\int df\, f^{D-1} \int d\sigma\,(\sin\sigma)^D \int d\tau\,(\sin\tau)^{D-1}(f^2) \ . \tag{6.13}$$

Again it will be transformed

$$f = v^{\frac{1}{D-\gamma}}(1-v)^{\frac{1}{D-\omega}} \tag{6.14}$$

Furthermore we choose in the same spirit as for $\beta$

$$\sigma = \begin{cases} \frac{\pi}{2}(2\eta)^{\frac{1}{D+1}} & \eta \leq 0.5 \\ \pi - \frac{\pi}{2}(2-2\eta)^{\frac{1}{D+1}} & \eta > 0.5 \end{cases} \tag{6.15}$$

and

$$\tau = \begin{cases} \frac{\pi}{2}(2\zeta)^{\frac{1}{D}} & \zeta \leq 0.5 \\ \pi - \frac{\pi}{2}(2-2\zeta)^{\frac{1}{D}} & \zeta > 0.5 \end{cases} \tag{6.16}$$

So the complete integral over four points is:

$$\begin{aligned} &\frac{S_{D-1}S_{D-2}}{S_D^2}\frac{\pi^3}{(2-D)(D-1)D(D+1)} \\ &\quad \int_0^1 du\, a^D \left(\frac{1}{D-\gamma}\frac{1}{u} + \frac{1}{\omega - D}\frac{1}{1-u}\right) \int_0^1 d\alpha\, \min(2\alpha, 2-2\alpha)^{\frac{2-D}{D-1}} (\sin\beta)^{D-2} \\ &\quad \int_0^1 dv\, f^D \left(\frac{1}{D-\gamma}\frac{1}{v} + \frac{1}{\omega - D}\frac{1}{1-v}\right) \int_0^1 d\eta\, \min(2\eta, 2-2\eta)^{\frac{-D}{D+1}} (\sin\sigma)^{D} \\ &\quad \int_0^1 d\zeta\, \min(2\zeta, 2-2\zeta)^{\frac{1-D}{D}} (\sin\tau)^{D-1} f^2 \mathbf{R} F(a,b,1,d,e,f) \end{aligned} \tag{6.17}$$

Another way of parametrizing consists in replacing the integral over the vectors $a$ and $f$ by the integral over the vectors $a$ and $d$. This parametrization is especially useful to eliminate divergences, when $a$ and $f$ simultaneously go to infinity. It will be used for the integration of one of the sectors in the next section. The new formulas are given here, a prime indicating new angles and distances as can be deduced from the figure. $\sigma'$ and $\tau'$

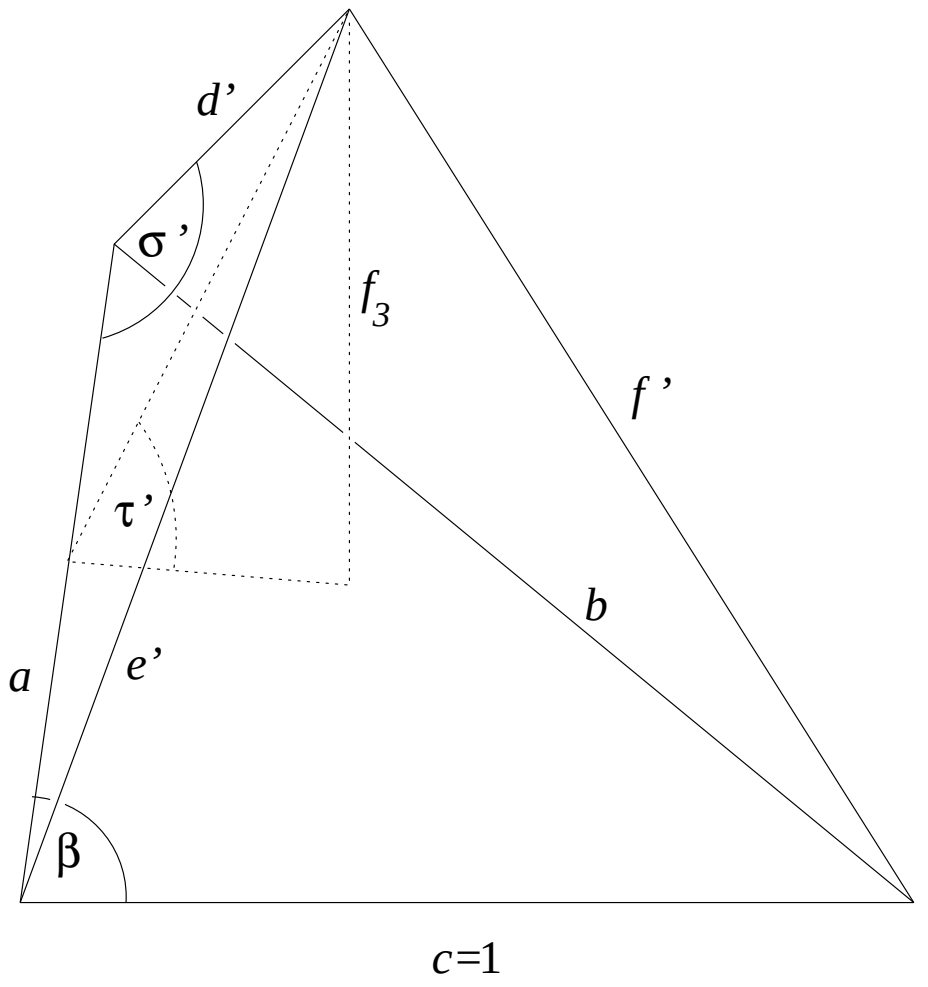

Figure 6.3: Alternative parametrization of the tetrahedron

obey the same relations as $\sigma$ and $\tau$. As before $f$, now

$$d' = v^{\frac{1}{D-\gamma}}(1-v)^{\frac{1}{D-\omega}} \tag{6.18}$$

The other new distances are:

$$e' = \sqrt{a^2 + d'^2 - 2ad'\cos\sigma'} \tag{6.19}$$

$$f' = \sqrt{(d'\sin\sigma'\sin\tau')^2 + (d'\cos\sigma' - a + \cos\beta)^2 + (\sin\beta - d'\sin\sigma'\cos\tau')^2} \tag{6.20}$$

For the integrand (6.4), the exponents are:

$$\gamma = \frac{D}{2} \tag{6.21}$$

$$\omega = \frac{3D}{2} \tag{6.22}$$

## 6.4 Decomposition into sectors

Although the measure absorbs the principal singularities it cannot handle all of them. There remains e.g. a singularity for $b \to 0$ and $e \to 0$. Two methods may be applied to handle the remaining integrable singularities. The first consists in using the second measure of section 6.3. The second is to map again some parts of the domain of integration. Thereby, we face the problem that the measure is no longer symmetric in the distances, as we have changed it in order to eliminate the relevant singularity for $f_3 \to 0$. We will therefore modify the measure (6.17) to

$$\begin{aligned}
&\frac{S_{D-1}S_{D-2}}{S_D^2}\frac{\pi^3}{(2-D)(D-1)D(D+1)} \\
&\quad\int du\, a^{D+2}\left(\frac{1}{D-\gamma}\frac{1}{u}+\frac{1}{\omega-D}\frac{1}{1-u}\right)\int d\alpha\ \min(2\alpha,2-2\alpha)^{\frac{2-D}{D-1}}(\sin\beta)^D \\
&\quad\int dv\, f^{D+2}\left(\frac{1}{D-\gamma}\frac{1}{v}+\frac{1}{\omega-D}\frac{1}{1-v}\right)\int d\eta\ \min(2\eta,2-2\eta)^{\frac{-D}{D+1}}(\sin\sigma)^D \\
&\quad\int d\zeta\ \min(2\zeta,2-2\zeta)^{\frac{1-D}{D}}(\sin\tau)^{D-1}\ ,
\end{aligned} \tag{6.23}$$

which is, except for the geometric prefactor, the invariant measure in $D+2$ dimensions. The integrand accordingly has to be modified to

$$T = \frac{1}{(2\Delta(a,b,c))^2}\mathbf{R}F\ , \tag{6.24}$$

where $\Delta(a,b,c) = \frac{1}{2}ac\sin(\beta) = \frac{1}{2}\sqrt{(a+b+c)(a+b-c)(b+c-a)(c+a-b)}$ is the area of the triangle spanned by $a$, $b$ and $c$. In our case

$$\begin{aligned}
T = &\frac{4\nu d_c}{(a+b+c)(a+b-c)(b+c-a)(c+a-b)f^2} \\
&\left[\left(\frac{1}{2}f^2\left(d^{-D}-e^{-D}\right)\left(c^{2\nu}-e^{2\nu}-b^{2\nu}+d^{2\nu}\right)-a^{2\nu}f^{2\nu}\right)\right. \\
&\left.\left(a^{2\nu}f^{2\nu}-\frac{1}{4}\left(c^{2\nu}-e^{2\nu}-b^{2\nu}+d^{2\nu}\right)^2\right)^{-d_c/2-1}+a^{-\nu d_c}f^{-\nu d_c}\right]
\end{aligned} \tag{6.25}$$

The integrand now is conformal invariant, as follows directly from equation (4.15).

The sectors are decomposed as follows:

1. ($a < 2$ or $f < 2$) and ($b > \frac{1}{2}$ or $e > \frac{1}{2}$)
   This sector is convergent: $F(a,b,c,d,e,f)$ is integrated directly, using the simple measure (6.17).

2. $a > 2$ and $f > 2$
   The measure (6.17) does not eliminate the singularity, when both $a$ and $f$ simultaneously go to $\infty$. The easiest way to integrate this sector is to use the second measure (6.18) ff. of section 6.3.

   The divergences of the integrand could also be eliminated by a mapping. This however induces new singularities due to the measure (the term $1/(2\Delta(a,b,c))^2$ in $T$, equation (6.24)). This would not be the case, if we had not been forced to use the trick of integrating by parts the measure.

3. $b < \frac{1}{2}$ and $e < \frac{1}{2}$
   In this sector the mapping can be used successfully: $a$ has to be exchanged with $b$ and $e$ with $f$. We get
   $T(b,a,c,d,f,e), \qquad a < \frac{1}{2}$ and $f < \frac{1}{2}$,
   using the measure (6.23)

## 6.5 Numerical integration

### 6.5.1 General remarks

In this section we want to discuss the algorithm used to perform the numerical integration. Let us recall that we have to treat integrals like (6.4) with the measure (6.17) and some modifications due to mappings of the sectors. These integrals run over the unit-volume $[0,1]^n$ with $n = 5$. Such high dimensional integrals cannot be done directly by means of Simpson's rule or derived methods as the integration time diverges like $(1/precision)^{n\alpha}$, $\alpha \leq 1$ and $\alpha = 1$ if no special analytical behavior is given. Normally Monte Carlo (MC) integration is a good choice, as the integration time grows like $(1/precision)^2$ independent of the dimension $n$. However in its original form it is not applicable to our integral since in large domains of integration the integrand is nearly 0 and in small domains much larger than 100, always assuming that the integral is normalized to 1. Therefore we implemented an adaptative Monte Carlo integration (AMC) procedure suggested to us by J. M. Drouffe [22].

The algorithm will be described schematically for $n = 2$:

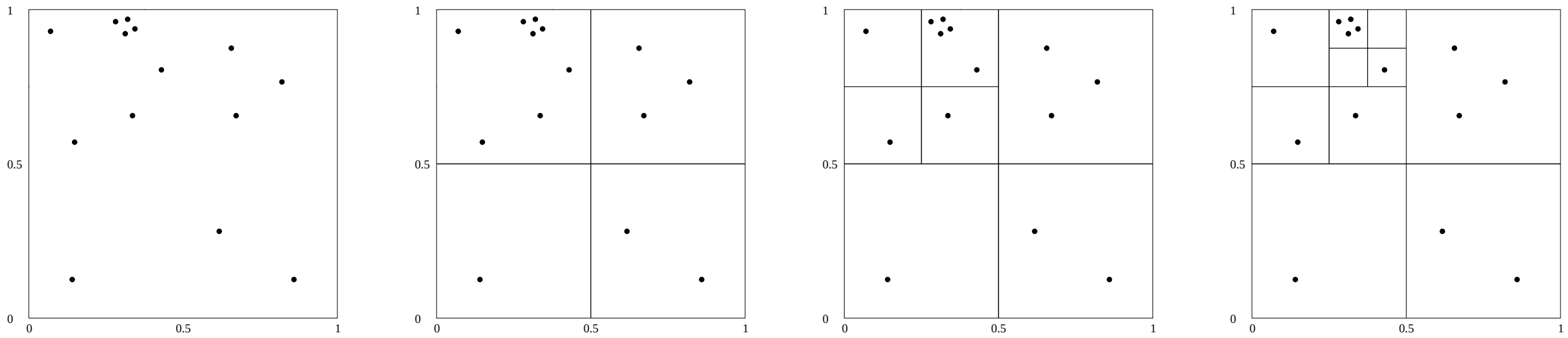

Figure 6.4: Schematical description of the AMC-algorithm

We have an integrand $f(x)$ where the main contribution to the integral

$$I = \int_0^1 d^n x \, f(x) \tag{6.26}$$

comes from a small domain, in which the integrand is large. In the figure this is indicated through a high density of points. We now try to integrate the square $[0,1]^n$ whose area is $A = 1$ with, say, 100 MC-sample points. The routine returns us the first moment $\bar{f}$, $A\bar{f}$ being an estimate of the integral and the connected second moment $\overline{f^2} - \bar{f}^2$, $A\sqrt{\overline{f^2} - \bar{f}^2}$ being an estimate of the deviation of $A\bar{f}$ from the integral $I$, henceforth called error estimate. A large error estimate indicates that the integral has not been found with a sufficient precision due to the fact that the integral is located in a small domain. Therefore we divide the domain of integration in $2^n$ subdomains (subboxes) as indicated in the figure and perform again a MC-integration with the same number of sample points in every subbox. The estimate of the integral now is

$$I \approx \sum_i \bar{f}_i A_i \tag{6.27}$$

where the sum runs over all boxes and $f_i$ and $A_i$ are the mean value of $f$ in and $A_i$ the area of the $i$-th subbox respectively. The new error estimate for the $i$-th subbox is:

$$\delta I_i = A_i \sqrt{\overline{f_i^2} - \bar{f}_i^2} \tag{6.28}$$

This determines the total (statistical) error estimate:

$$\delta I_{stat} = \sqrt{\sum_i \delta I_i^2} \tag{6.29}$$

This time only in the upper left subbox the error estimate will be large and this box has to be subdivided again. We repeat the procedure recursively until the error estimate of every subbox is smaller than a given $\epsilon$. In our example this process stops after 3 recursive subdivisions.

It should be emphasized that we did not divide the error estimate $\delta I$ in a subbox by $\sqrt{\textit{number of samplepoints}}$ as can be done for Gaussian distributions. This is due to the fact that we are normally far from the domain where the central limit theorem is valid (Gaussian domain).

### 6.5.2 Capabilities and performances

The algorithm is able to integrate singular integrands. In low dimensions ($n = 1, 2$) this works with appropriate accuracy in a very short time (some seconds). In higher dimensions (e.g. $n = 5$) however singular integrals may impose severe problems: The algorithm can simply overlook divergences. Of course, one could start with a finer sublattice, i.e. demand a minimal number of subdivisions in every direction, or one can use a smaller $\epsilon$. In practical calculations however, the execution time will explode.

As a general rule the algorithm seems to work well for functions whose second moment $\int f^2(x)$ exists. For such well-behaved integrands the integration time scales slightly better than inverse proportional to the demanded precision.

### 6.5.3 Implementation of the algorithm

The algorithm works recursively, so the appropriate language is C. It was implemented on a SUN-workstation. A portation to FORTRAN is possible, but tedious because of its missing capabilities to use structures and recursions.

### 6.5.4 Numerics for the 4-point integral

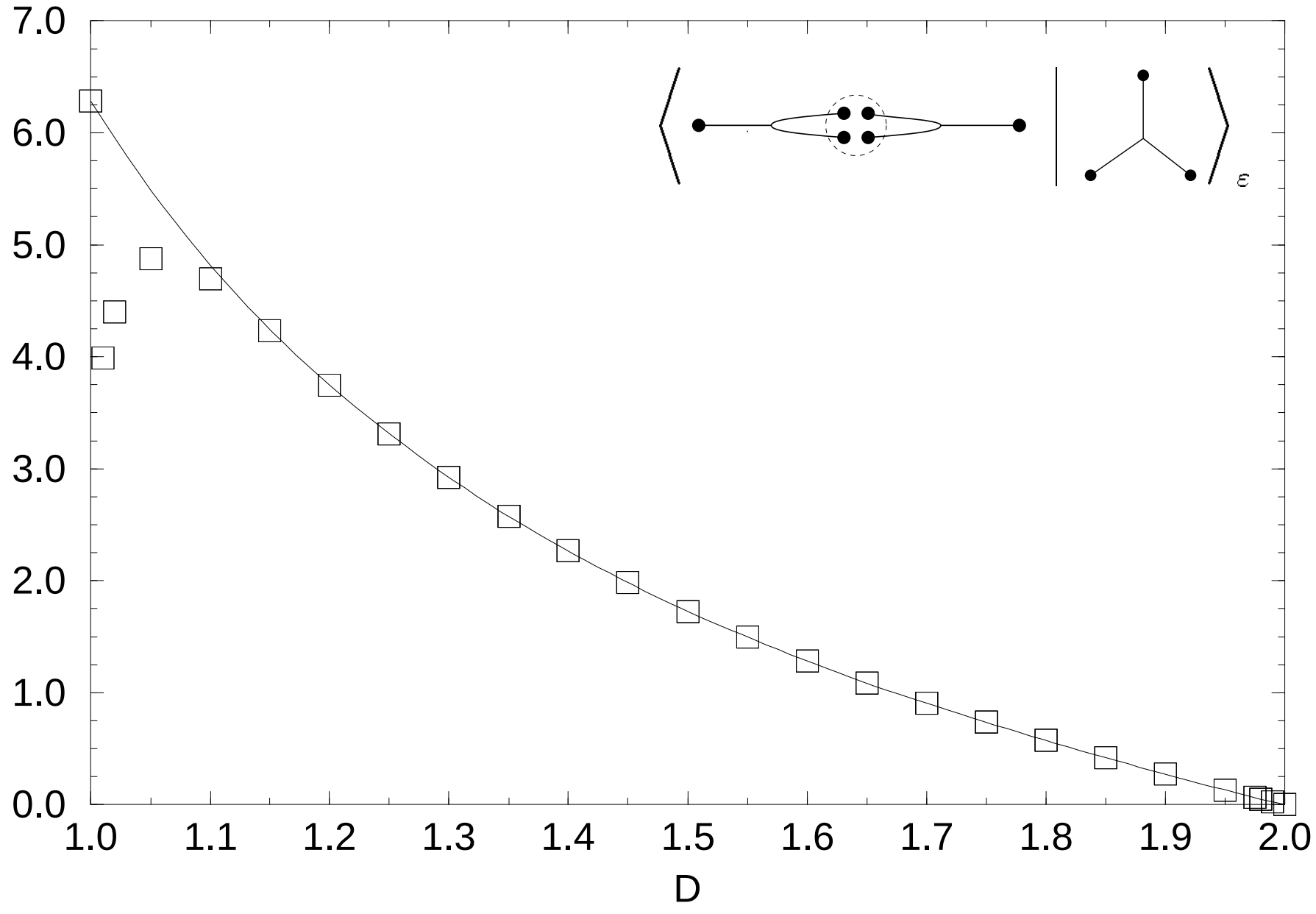

| $D$ | | $D$ | | $D$ | |
|---|---|---|---|---|---|
| 1.00 | 6.283 (exact: $2\pi$) | 1.30 | $2.92 \pm 0.03 \pm 0.004$ | 1.70 | $0.91 \pm 0.02$ |
| 1.01 | $3.99 \pm 0.03 \pm 2.0$ | 1.35 | $2.57 \pm 0.03$ | 1.75 | $0.74 \pm 0.02$ |
| 1.02 | $4.40 \pm 0.03 \pm 1.6$ | 1.40 | $2.27 \pm 0.03$ | 1.80 | $0.57 \pm 0.02$ |
| 1.05 | $4.87 \pm 0.03 \pm 1.3$ | 1.45 | $1.98 \pm 0.03$ | 1.85 | $0.42 \pm 0.02$ |
| 1.10 | $4.70 \pm 0.03 \pm 0.4$ | 1.50 | $1.73 \pm 0.03$ | 1.90 | $0.27 \pm 0.02$ |
| 1.15 | $4.23 \pm 0.03 \pm 0.08$ | 1.55 | $1.50 \pm 0.03$ | 1.95 | $0.13 \pm 0.01$ |
| 1.20 | $3.75 \pm 0.03 \pm 0.03$ | 1.60 | $1.28 \pm 0.03$ | 1.99 | $0.03 \pm 0.01$ |
| 1.25 | $3.31 \pm 0.03 \pm 0.01$ | 1.65 | $1.08 \pm 0.02$ | 2.00 | 0 |

Figure 6.5: Numerical results for the diagram (6.2). The first error in the table is the statistical error (6.29), the second the systematic error (6.30).

We recall that 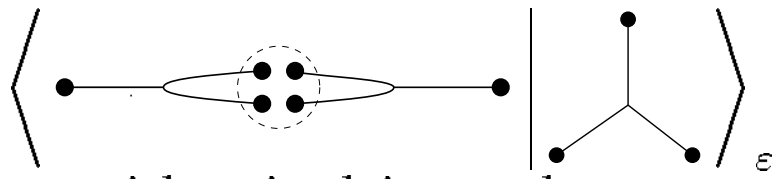 was decomposed into 3 sectors. The main numerical problem, identical in each sector, is the concentration of the integral in small domains. This problem is solved by the AMC-algorithm discussed above. Additional numerical problems arise for $D \leq 1.3$ and become dominant for $D \leq 1.1$. The floating point routines of the workstation are implemented in double-precision (64 Bits, 52 Bits fraction $\approx$ 15 digits). In this range the routines more and more often fail to calculate the integrand, i.e. $\mathbf{R}F$ (6.8) times the measure term, compare (6.17). These problems reflect the fact that for $D \to 1$ the (modified) measure becomes a distibution. They either return a wrong result or a so-called nonalgebraic number (NaN), $+\infty$ or $-\infty$. Only the latter events can be excluded from the integration. They are counted and serve as an error estimate via the formula

$$\delta I_{sys} \approx I\rho \frac{\#\{\mathrm{NaN}, \pm\infty\}}{\#\{\text{integration points}\}} , \tag{6.30}$$

where $\rho$ was determined from similar, but analytically known integrals to be

$$\rho \approx 2\ . \tag{6.31}$$

These estimates have to be taken with some precautions as $\rho$ depends on the actual integral and on the machine. The systematic errors in figure 6.5 were calculated by this method and seem to give a reasonable estimate for $D \to 1$, the limiting case for which the analytical result is $2\pi$ (see next section).

This line of arguments could be confirmed by working in quadruple precision (128 Bits, 112 Bits fraction $\approx$ 33 digits). The systematic error for $D = 1.05$ was reduced by a factor 3. No further analysis however was undertaken as the performance of the workstation drops by a factor 300. A reasonable calculation which before took one hour now needs two weeks.

## 6.6 $D \to 1$

In the limit $D \to 1$, 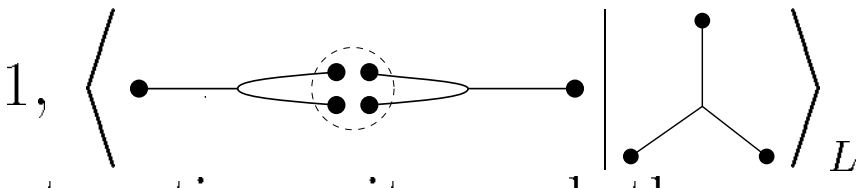 can again be calculated analytically. This calculation is interesting as it reveals the connection to standard polymer theory and the fact that 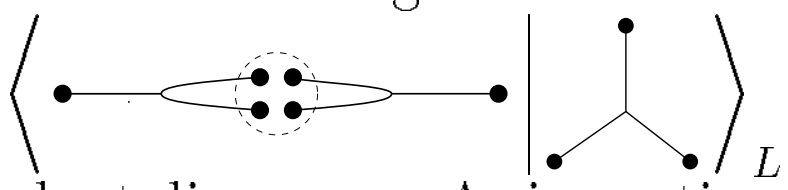 decomposes into three topologically different and non-equivalent diagrams. As in section 6.1 we keep $c = 1$ fixed. By a direct calculation it can be verified that the measure indeed reduces to an integral over a line. On this line two points, the endpoints of $c$, are already fixed. Then there are 12 different possibilities to distribute the last two points. They still can be separated into 3 topological inequivalent classes A, B and C, cf. figure 6.6. These are the three standard diagrams, arising in polymer theory. In each of these classes the line with $c = 1$ may be chosen to be the line connecting (12), (14), (23) or (34). Readers more familiar with Feynman diagrams arising in the framework of a scalar field theory may recover the three corresponding diagrams after a de Gennes transformation [23]. They contribute to the renormalization of the $\varphi^6$-interaction at $d = 3$ and are represented on the r.h.s.. Diagrams in one class can be mapped onto each other by the now well-known mapping of sectors.

As an example, we calculate the diagram A, where $c = 1$ is chosen to be the line (23). Counterterms are neglected:

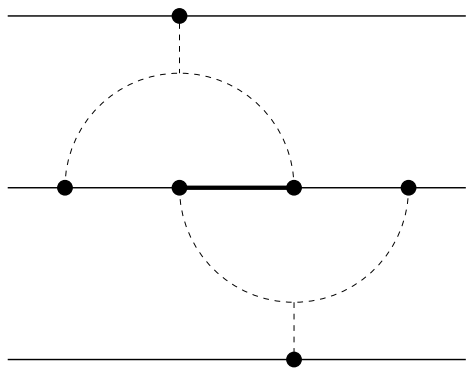

$$\begin{aligned} &= \frac{1}{4}\int_1^\infty da \int_1^\infty db\,(ab-1)^{-3/2} \\ &= \frac{1}{2}\int_1^\infty \frac{da}{a}(a-1)^{-1/2} \\ &= \frac{\Gamma^2(\frac{1}{2})}{\Gamma(1)} = \frac{1}{2}\pi \end{aligned} \tag{6.32}$$

The other three diagrams of this class were explicitly checked to give $\frac{\pi}{2}$ too. In the classes B and C all diagrams equal 0. So together, the contribution is $2\pi$ confirming the numerical results.

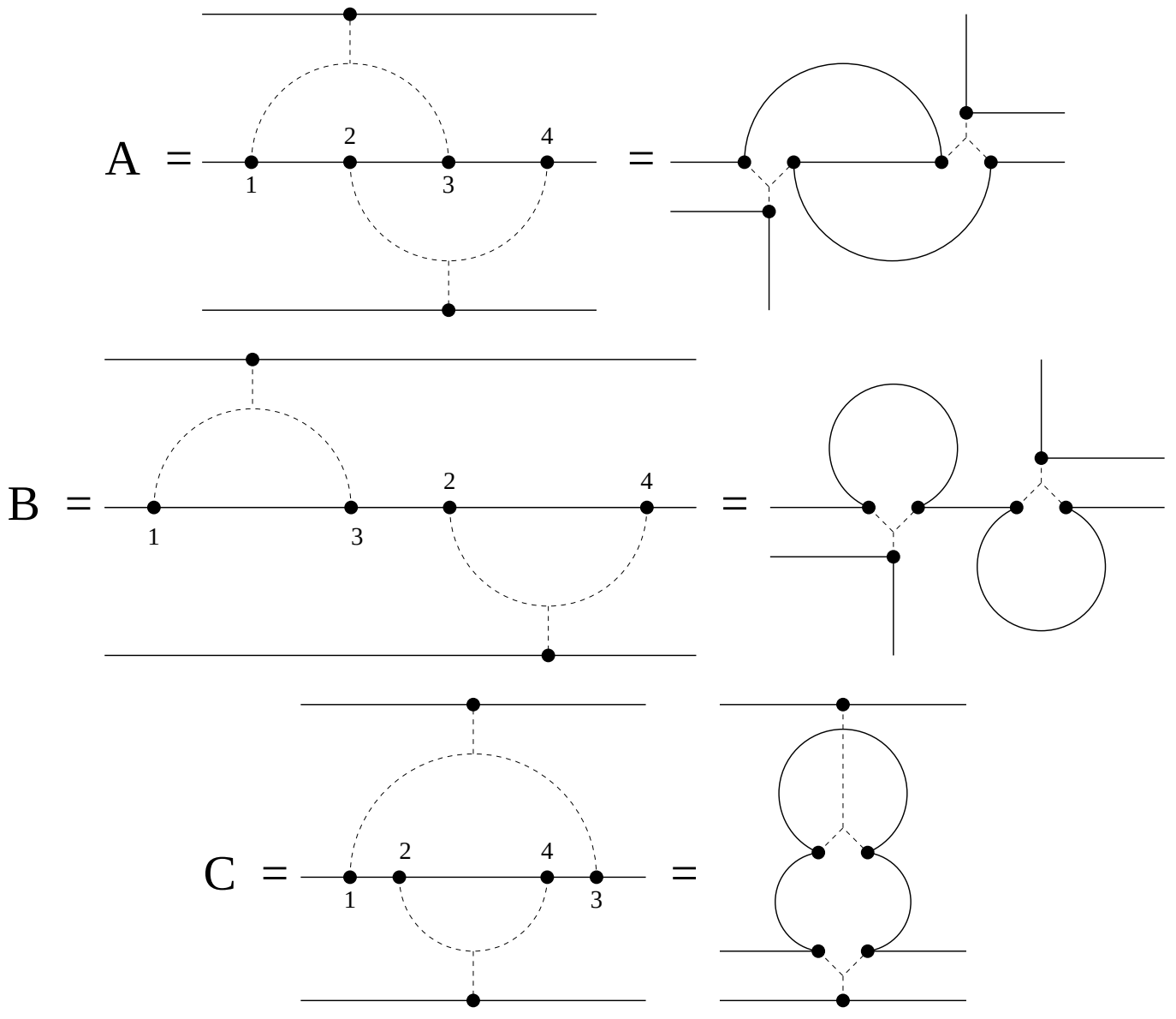

Figure 6.6: The 3 topological inequivalent classes A, B and C

This result was already obtained in a different parametrization in [24].

# 7 Results for the 3-body Hamiltonian

## 7.1 Renormalization and critical exponents

In this chapter we detail the renormalization of the model at 1-loop order, the derivation of the renormalization group equations and the determination of the critical exponents, along the line of section 2.3 and section 2.4. This is nothing but an application of the general procedure of [15, 16]. The purpose of renormalization is the following. We start from the bare Hamiltonian (2.1)

$$\mathcal{H}^0_{g_0}[r_0] = \frac{1}{2-D}\int_x \frac{1}{2}(\nabla r_0(x))^2 + g_0 \int_x\int_y\int_z \tilde{\delta}^d(r_0(x)-r_0(y))\tilde{\delta}^d(r_0(x)-r_0(z)) \ . \qquad (7.1)$$

Expectation values of "bare" observables $\mathcal{O}_0[r_0]$, i.e. functionals of the bare field $r_0$, are defined as

$$\langle \mathcal{O}_0[r_0]\rangle_{g_0} \ = \ \frac{1}{\mathcal{Z}_{g_0}}\int \mathcal{D}[r_0]\,\mathcal{O}_0[r_0]\,\mathrm{e}^{-\mathcal{H}^0_{g_0}[r_0]} \quad ; \quad \mathcal{Z}_{g_0} \ = \ \int \mathcal{D}[r_0]\,\mathrm{e}^{-\mathcal{H}^0_{g_0}[r_0]} \ . \qquad (7.2)$$

Expressed as perturbative series in the bare coupling constant $g_0$, they are UV finite for $\varepsilon > 0$, where we recall that

$$\varepsilon \ = \ 3D - 2\nu d \ , \qquad (7.3)$$

if the relevant UV divergences due to the 2-point operator are subtracted according to the prescription of section 3.2. They suffer from UV divergences for $\varepsilon = 0$, which appear at 1-loop order as single poles in $\varepsilon$. Renormalization consists in reexpressing the theory in terms of renormalized quantities – the renormalized field $r$ and the renormalized coupling constant $g$ – defined as

$$r(x) = Z^{-1/2} r_0(x) \ , \quad g = Z^{-d} Z_g^{-1} \mu^{-\varepsilon} g_0 \tag{7.4}$$

(with $\mu$ the renormalization momentum scale) so that the expectation values of physical observables, expressed in terms of the renormalized field $r$ and of the renormalized coupling constant $g$, are UV finite as $\varepsilon \to 0$. In renormalized quantities, the bare Hamiltonian is rewritten as the renormalized Hamiltonian (see equation (2.21))

$$\mathcal{H}_g^R[r] \ \equiv\ \mathcal{H}_{g_0}^0[r_0] \ =\ \frac{Z}{2-D} \int_x \frac{1}{2}\big(\nabla r(x)\big)^2 + g Z_g \mu^\varepsilon \int_x \int_y \int_z \delta^d(r(x)-r(y))\delta^d(r(x)-r(z)) \tag{7.5}$$

and the expectation values of renormalized quantities are

$$\langle \mathcal{O}[r] \rangle_g^R \ \equiv\ \langle \mathcal{O}[r] \rangle_{g_0} \ =\ \frac{\int \mathcal{D}[r]\, \mathcal{O}[r]\, \mathrm{e}^{-\mathcal{H}_g^R[r]}}{\int \mathcal{D}[r]\, \mathrm{e}^{-\mathcal{H}_g^R[r]}} \tag{7.6}$$

At 1-loop order, i.e. to first order in $g$, the counterterms are, in the spirit of the minimal subtraction scheme, chosen to have pure poles in $\varepsilon$ of the form (2.28), (2.29):

$$Z \ =\ 1 + g\,\frac{a}{\varepsilon} + \ldots \quad ; \quad Z_g \ =\ 1 + g\,\frac{b}{\varepsilon} + \ldots \tag{7.7}$$

Expanding (7.6) in terms of the renormalized coupling constant $g$ we obtain

$$\langle \mathcal{O}[r] \rangle_g^R \ =\ \Big\langle \mathcal{O} \Big\rangle_0 - g\mu^\varepsilon \iiint \Big\langle \mathcal{O}$$ 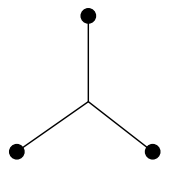 $$\Big\rangle_0^{conn} + \frac{1}{2} g^2 \mu^{2\varepsilon} \iiint \iiint \Big\langle \mathcal{O}$$ 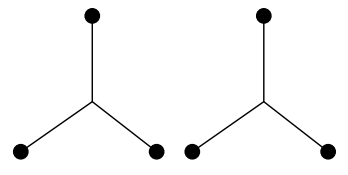 $$\Big\rangle_0^{conn}$$

$$- g\,\frac{a}{\varepsilon}\,\frac{1}{2-D} \int \Big\langle \mathcal{O}$$ 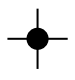 $$\Big\rangle_0^{conn} - g^2 \mu^\varepsilon \frac{b}{\varepsilon} \iiint \Big\langle \mathcal{O}$$ 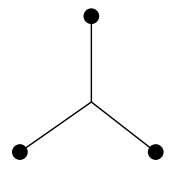 $$\Big\rangle_0^{conn}$$

$$+ \ldots \tag{7.8}$$

where we made use of the standard abbreviations for the expectation value in the free theory

$$\langle \mathcal{O}[r] \rangle_0 \ =\ \frac{\int \mathcal{D}[r]\, \mathcal{O}[r]\, \mathrm{e}^{-\frac{1}{2-D}\int_x \frac{1}{2}(\nabla r)^2}}{\int \mathcal{D}[r]\, \mathrm{e}^{-\frac{1}{2-D}\int_x \frac{1}{2}(\nabla r)^2}} \tag{7.9}$$

and for the connected correlators

$$\begin{aligned} \langle \mathcal{A}\mathcal{B} \rangle^{conn} \ &=\ \langle \mathcal{A}\mathcal{B} \rangle - \langle \mathcal{A} \rangle \langle \mathcal{B} \rangle \\ \langle \mathcal{A}\mathcal{B}\mathcal{C} \rangle^{conn} \ &=\ \langle \mathcal{A}\mathcal{B}\mathcal{C} \rangle - \langle \mathcal{A} \rangle \langle \mathcal{B}\mathcal{C} \rangle - \langle \mathcal{B} \rangle \langle \mathcal{C}\mathcal{A} \rangle - \langle \mathcal{C} \rangle \langle \mathcal{A}\mathcal{B} \rangle + 2 \langle \mathcal{A} \rangle \langle \mathcal{B} \rangle \langle \mathcal{C} \rangle \ . \end{aligned} \tag{7.10}$$

The poles in $\varepsilon$ of the bare expectation values are given by the MOPE and we have for instance

$$\iiint \Big\langle \mathcal{O}$$ 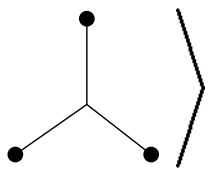 $$\Big\rangle_0^{conn} \ =\ \int \Big\langle$$ 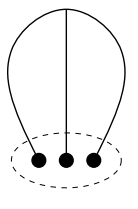 $$\Big|$$ 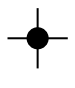 $$\Big\rangle_L \Big\langle \mathcal{O}$$ 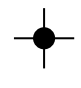 $$\Big\rangle_0^{conn} + \text{finite terms}$$

$$=\ \frac{L^\varepsilon}{\varepsilon} \Big\langle$$ 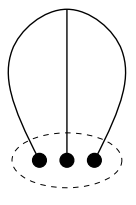 $$\Big|$$ 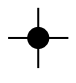 $$\Big\rangle_\varepsilon \int \Big\langle \mathcal{O}$$ 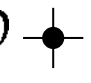 $$\Big\rangle_0^{conn} + \text{finite terms} \tag{7.11}$$

Similarly, the third term on the r.h.s. of (7.8) contains a single pole which comes from the contraction of two 3-point operators into one 3-point operator: 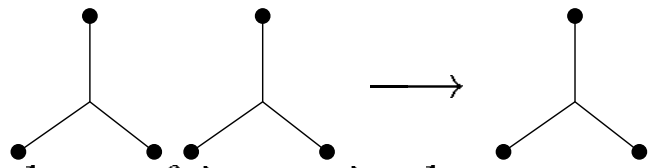 .
There are three different kind of contractions. Including the number of inequivalent ways of contracting the vertices, one has

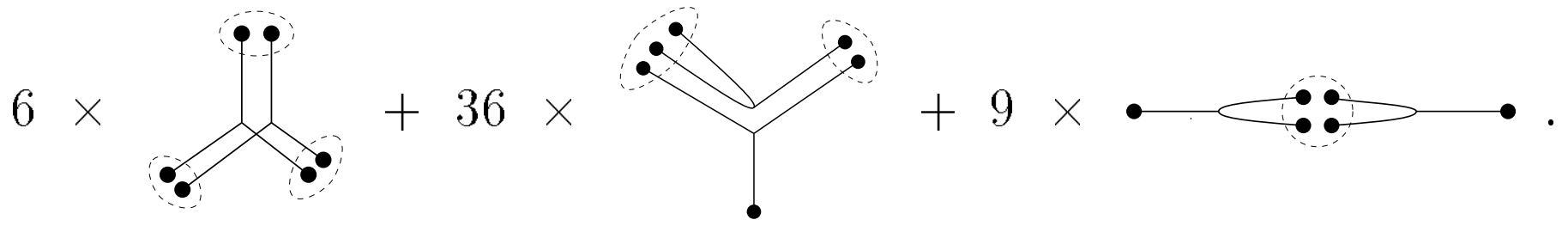

(7.12)

This term also contains poles which come from the contractions of 3-point operators into the 1-point operator 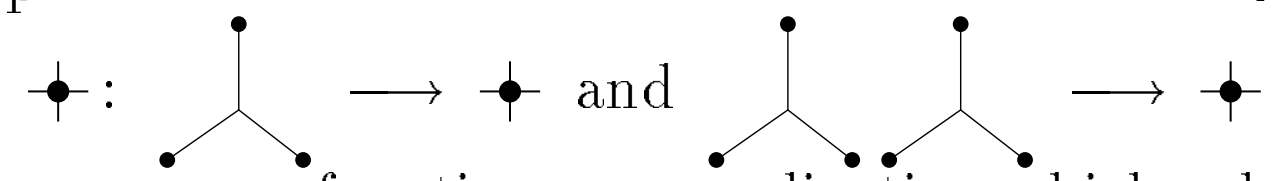. The first poles will be subtracted by the same wave-function renormalization which subtracts the 1-loop divergence (7.11). The second pole corresponds to a 2-loop wave-function divergence and need not to be considered here. Thus we have

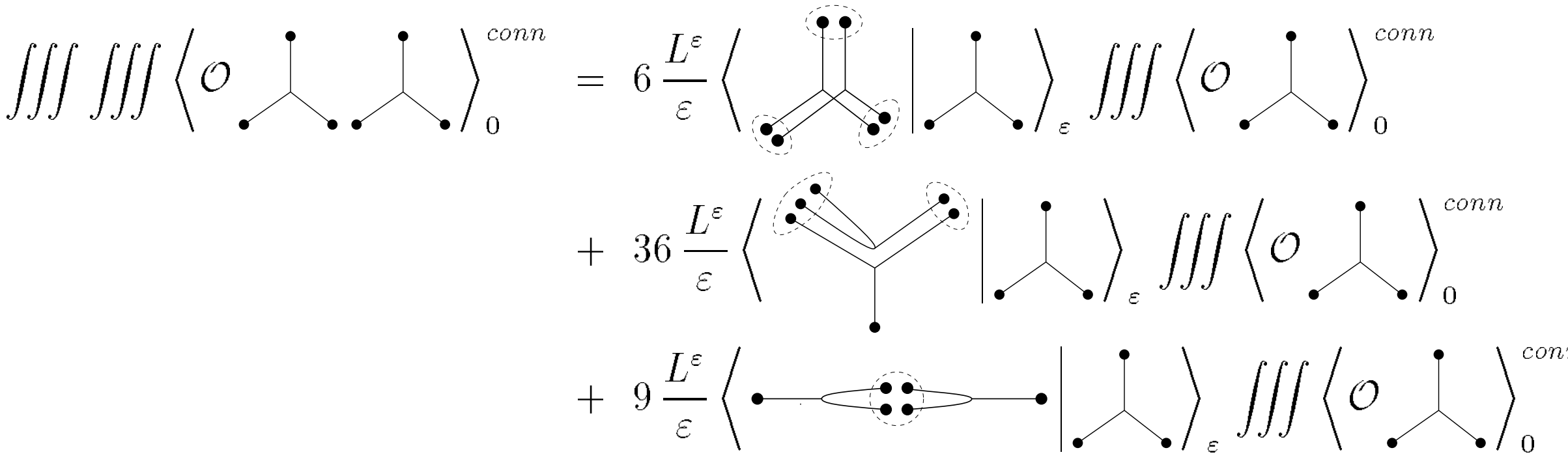

$+$ higher order wave-function divergences

$+$ finite terms (7.13)

From (7.8), (7.11) and (7.13), one sees that the single poles cancel and that the renormalized theory is finite at 1-loop order if the counterterms $Z$ and $Z_g$ are given by (7.7) with

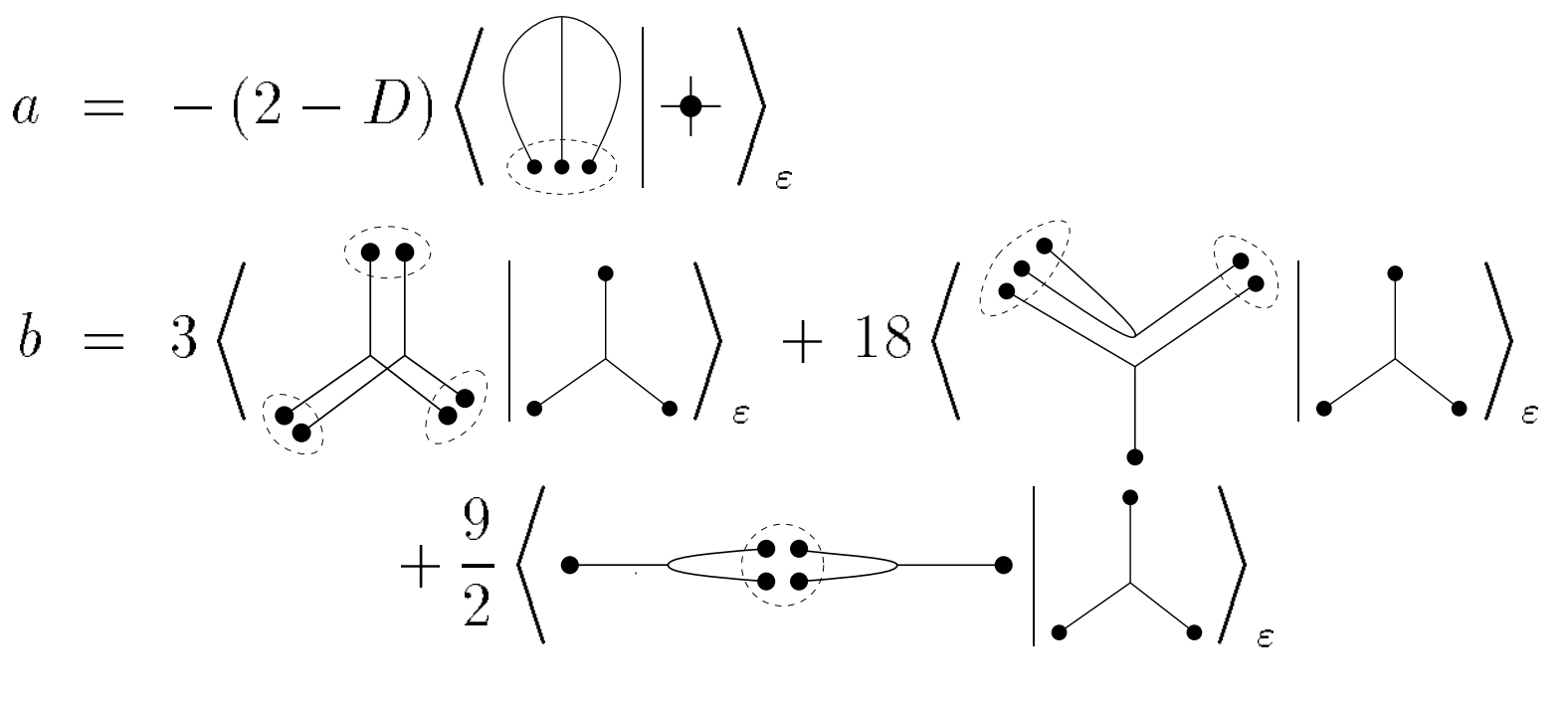

(7.14)

The renormalized observables satisfy the renormalization group equations. They are easily obtained by calculating the variation of the renormalized quantities with respect to the renormalization scale $\mu$, which corresponds to the scale at which the theory is probed, keeping the bare Hamiltonian, which represents the microscopic theory, fixed. The flow of the coupling constant is given by the $\beta$-function:

$$\beta(g) \;=\; \mu\frac{\partial}{\partial\mu}\bigg|_{g_0} g \;=\; \frac{-\varepsilon g}{1+g\frac{\partial}{\partial g}\ln Z_g + d_c g\frac{\partial}{\partial g}\ln Z}$$
$$=\; -\varepsilon g \,+\, g^2(d_c a + b) \,+\, \mathcal{O}(g^2)\;, \tag{7.15}$$

with $d_c = 3D/(2-D)$. The scaling dimension of the field, $\nu(g)$, is similarly obtained from the flow of the *dimensionless* renormalized field $\tilde r = \mu^{(2-D)/2}\, r$:

$$\nu(g) \;=\; \mu\frac{\partial}{\partial\mu}\bigg|_{g_0,\,r_0} \ln(\tilde r) \;=\; \frac{2-D}{2} \,-\, \frac{1}{2}\,\mu\frac{\partial}{\partial\mu}\bigg|_{g_0} \ln(Z)$$
$$=\; \frac{2-D}{2} \,+\, g\,\frac{a}{2} \,+\, \mathcal{O}(g^2) \tag{7.16}$$

$\nu(g)$ is related to the "fractal dimension" $d_F(g)$ of the membrane through $\nu(g) = D/d_F(g)$.

The analytical and numerical calculations of the previous sections show that the coefficients $a$ and $b$ (7.14) are both strictly positive for $1 < D < 4/3$. This implies that for positive (and at least small) $\varepsilon$, the $\beta$-function has a non-trivial IR-attractive fixed point at

$$g^\star \;=\; \varepsilon\frac{1}{a d_c + b} \,+\, \mathcal{O}(\varepsilon^2) \tag{7.17}$$

This fixed point governs the large distance (small $\mu$) behavior of the membrane at the $\Theta$-point. The scaling dimension of the membrane at the $\Theta$-point is given by

$$\nu^\star \;=\; \frac{D}{d_F^\star} \;=\; \nu(g^\star) \;=\; \frac{2-D}{2} \,+\, \varepsilon\frac{a}{2(a d_c + b)} \,+\, \mathcal{O}(\varepsilon^2) \tag{7.18}$$

Finally let us recall that, while the $\beta$-function and the anomalous dimension $\nu(g)$ depend on the normalization of the Hamiltonian and on the choice of the renormalization scheme, the result for the anomalous dimension $\nu^\star$ is universal. In particular, it does not depend on the normalization introduced in section 2.5.

## 7.2 Results for the anomalous dimension at 1-loop order

One possibility to make the expansion is to consider the dimension $D$ of the membrane as fixed and to vary the dimension of the external space $d$ around the critical dimension $d_c = 3D/(2-D)$. This means to expand $\nu^\star$ in $\bar\varepsilon = d_c - d$:

$$\nu^\star \;=\; \frac{2-D}{2} \,+\, \bar\varepsilon\, A(D) \,+\, \mathcal{O}(\bar\varepsilon^2)\;, \quad A(D) \;=\; \frac{a(2-D)}{2(a d_c + b)} \tag{7.19}$$

The value of the coefficient $A(D)$, which determines the scaling dimension of the membrane at order $\bar\varepsilon$, is plotted in figure 7.1 for $1 < D < 4/3$.

In this region, $A(D)$ is positive and the scaling dimension $\nu^\star$ is larger than the scaling dimension of the free "phantom" membrane $\nu = (2-D)/2$, as expected since the interaction should swell the membrane. The correction vanishes for $D \to 1$. This is in agreement with the known results [21] for polymers at the Θ-point: corrections to the scaling dimension first appear at two loops, i.e. at order $\bar{\varepsilon}^2$. Interestingly, although individual diagrams diverge at $D = 4/3$, the correction for the anomalous dimension seems to be regular as $D \to 4/3$. We shall prove in the next section that this is indeed the case and that the correction can be calculated analytically at $D = 4/3$.

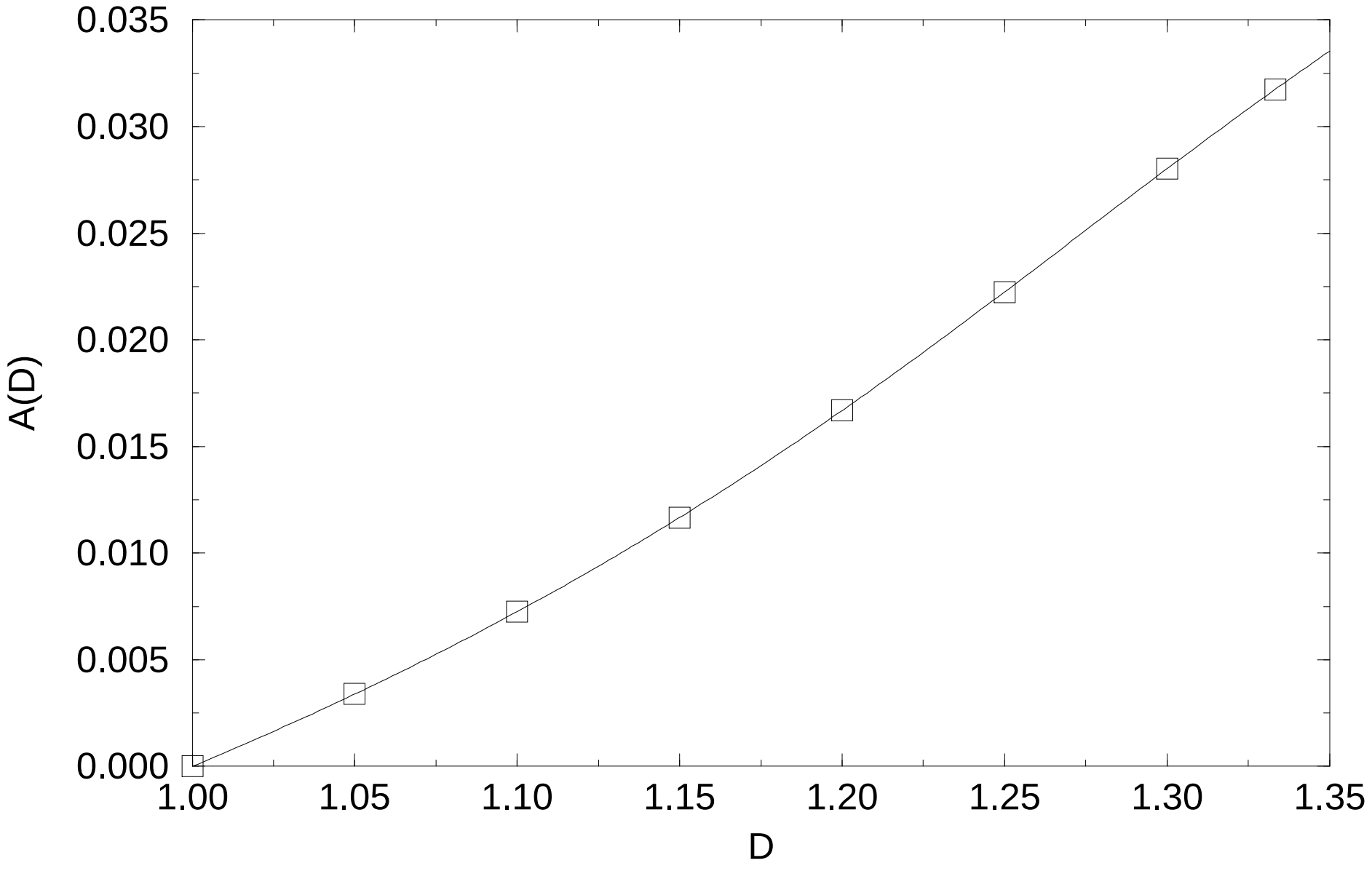

| $D$ | $A(D)$ |
|---|---|
| 1.00 | 0 |
| 1.05 | 0.003390 |
| 1.10 | 0.007252 |
| 1.15 | 0.01167 |
| 1.20 | 0.01669 |
| 1.25 | 0.02225 |
| 1.30 | 0.02802 |
| 4/3 | 0.03175 |

Figure 7.1: Anomalous correction to $\nu^\star$ at 1-loop order (3-point interaction)

# 8 $D = 4/3$

As we remarked in the calculation of some of the diagrams, new singularities appear at $D = 4/3$. They come from the fact that the modified 2-body operator •—++—•, defined by (2.41), which is irrelevant for $D < 4/3$ and $\varepsilon = 0$, becomes marginal at $D = 4/3$ and is relevant for $D > 4/3$. As argued in [15], it is this last interaction term which is relevant to describe the Θ point for $D > 4/3$ and $d$ close to the critical dimension, which is now given by $d'_c = 2(3D-2)/(2-D)$.

In this section we first discuss the limit $D \to 4/3$ of the 1-loop results obtained for $D < 4/3$ by the 3-point interaction only and show how the modified 2-point interaction emerges from the 3-point interaction as $D \to 4/3$. This allows to prove that this limit is regular as far as physical quantities such as the anomalous dimensions are concerned.

In the second subsection we briefly recall the 1-loop results for the modified 2-point interaction, which are a priori valid for $D > 4/3$. The limit $D \to 4/3$ also turns out to be regular and the 1-loop correction for the anomalous dimension can be continued to $D < 4/3$.

Finally in the third subsection we study the domain around $D = 4/3$ and $d = 6$, which is characterized by a crossover between the 3-point interaction and the modified 2-point interaction, described by the full Hamiltonian (2.20). We show that at 1-loop order this case can be treated analytically and that the crossover between the two interactions is understood.

## 8.1 The limit $D \to 4/3$ for the 3-body interaction

Regarding the analytical and numerical results for the integrals involved in the calculations for the 3-body interaction (figure 3.4, 5.2, 5.4 and 6.5), one remarks that the integrals 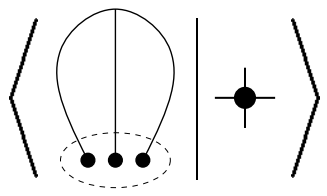 $_{\varepsilon}$ and 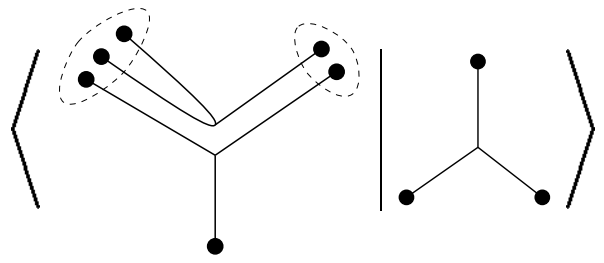 $_{\varepsilon}$ are diverging for $D \to 4/3$ whereas the two others, 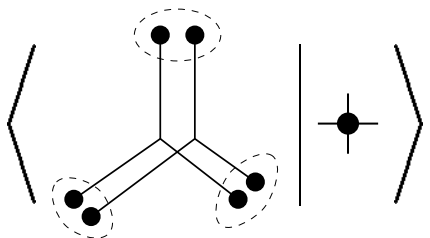 $_{\varepsilon}$ and 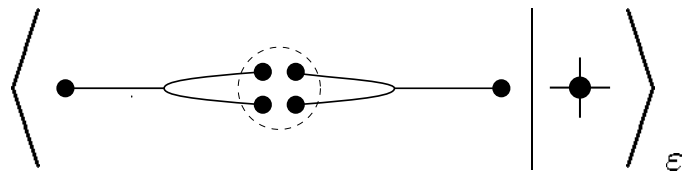 $_{\varepsilon}$, stay finite. Therefore the first two are expected to dominate in the limit $D \to 4/3$.

Let us start with the integral 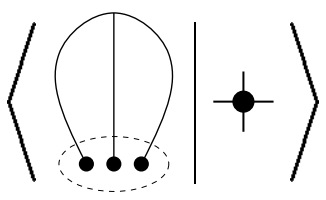 $_{L}$, given by (3.17) and (3.9). This integral has a global divergence as $\varepsilon \to 0$. The additional subdivergence as $D \to 4/3$ is due to the subcontraction of two out of the three points: 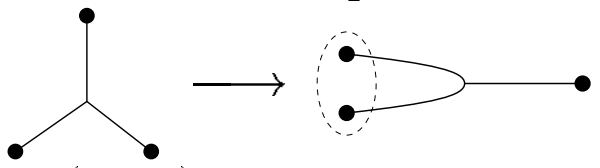. This subcontraction is given by the MOPE (2.13). If we denote by $(y, z)$ the pair of subcontracted points, it takes the explicit form:

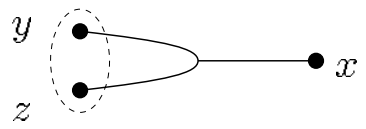 $= C(\{y,z\})$ 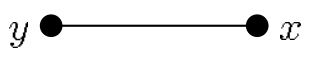 $+ D(\{y,z\})$ 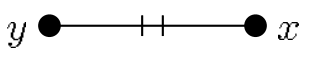 $+ \ldots$

$$C(\{y,z\}) = |y-z|^{-\nu d}\ , \quad D(\{y,z\}) = \frac{1}{4}|y-z|^{-\nu(d-2)} \tag{8.1}$$

We remind the notation of points and distances represented in figure 3.1. The first term is the most singular one and is subtracted by the finite part prescription described in section 3.2. The second gives the singularity at $D = 4/3$. Remind, that the factorization of the integrations is valid only if $b = |y-z|$ is (much) smaller than $a = |x-y|$. So an effective IR cutoff $L'$ has to be introduced. It will be specified later. Integrating in the domain $b = |y-z| < L'$, we get

$\int_{b=|y-z|<L'}$ 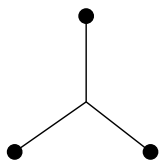 $=$ 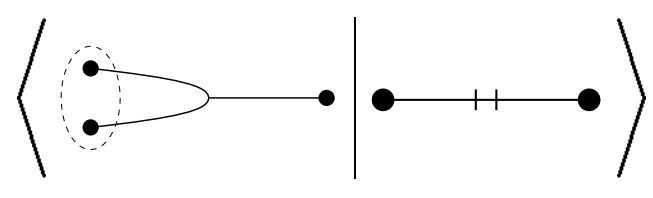 $_{L'}$  $+$ finite terms (8.2)

where we used the notation, similar to that of section 3.3, for the singular coefficient

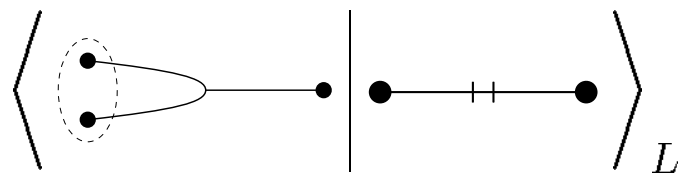 $_{L'} = \int_{b=|y-z|<L'} D(\{y,z\}) = \int_{b<L'} \frac{1}{4} b^{-\nu(d-2)}$

$$= \frac{1}{4(2-\nu d)} L'^{2-\nu d} \tag{8.3}$$

which diverges as $\nu d - 2 \to 0$. Since we are interested in the pole given by the global divergence for $\varepsilon = 0$, we can replace $\nu d$ by $3D/2$ and the pole term $1/(2-\nu d)$ by $\frac{2}{3}\frac{1}{4/3-D}$. It is this term, associated to the subcontraction 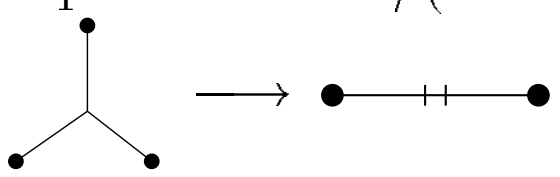, which gives the dominant contribution as $D \to 4/3$.

In this limit, another interesting thing happens: (8.3) depends only logarithmically on $L'$. This solves the delicate problem of how to choose this regulator when one integrates over the other distances. One can simply take $L'$ to be the distance $a$ between the two points in . This is natural since this distance is the only one available. It is equivalent to integrate in the sector $b < a$.

In the integration over the remaining distance, the dominant contribution comes from the contraction  $\longrightarrow$ 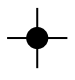, given by a MOPE of the form

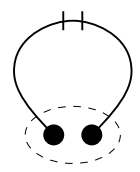 $= H(\{x,y\})\bullet \ + \ I(\{x,y\})$ 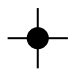 $+ \ \ldots$

$$H(\{x,y\}) = \frac{d}{2}|x-y|^{-\nu(d+2)}\ , \quad I(\{x,y\}) = -\frac{d+2}{4D}|x-y|^{2D-2-\nu d} \tag{8.4}$$

The integration over $a = |x-y|$ has to be performed for $a < L$. Since there are 3 inequivalent sectors we end up with

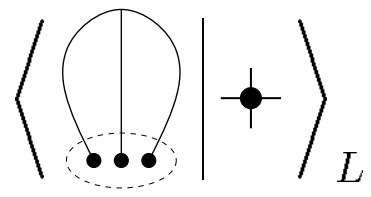

$$\begin{aligned} &= \int_{a,b,c<L} B(\{x,y,z\}) \\ &= 3\int_{b=|y-z|<a=|x-y|<L} D(\{y,z\})\, I(\{x,y\}) + \text{less singular terms} \\ &= -\frac{3}{4}\frac{1}{4/3-D}\frac{1}{\varepsilon}L^{\varepsilon} + O\left(\varepsilon^0, \left(\frac{4}{3}-D\right)^0\right) \end{aligned} \tag{8.5}$$

This result agrees with the numerical predictions.

A similar analysis shows that

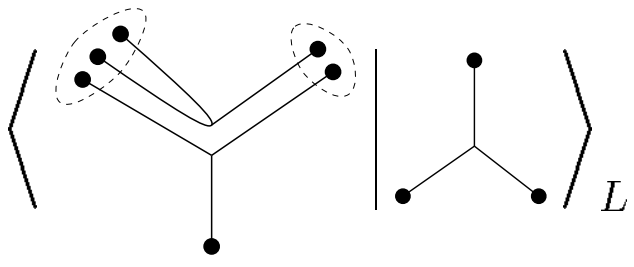

$$= \frac{1}{8}\frac{1}{4/3-D}\frac{1}{\varepsilon}L^{\varepsilon} + O\left(\varepsilon^0, \left(\frac{4}{3}-D\right)^0\right)\ . \tag{8.6}$$

To perform the limit $D \to 4/3$ it is convenient to reexpress $g$ in (7.5) in terms of

$$\bar{g} \ = \ \frac{1}{2\,(4/3-D)}\,g \tag{8.7}$$

and analogously for bare quantities. The renormalization group functions become in this limit

$$\bar{\beta}(\bar{g}) = \mu\frac{\partial}{\partial\mu}\bigg|_{\bar{g}_0}\bar{g} = -\varepsilon\bar{g} + \frac{21}{2}\bar{g}^2 + \mathcal{O}(\bar{g}^3) \tag{8.8}$$

$$\nu(\bar{g}) = \frac{2-D}{2} + \frac{1}{2}\bar{g} + \mathcal{O}(\bar{g}^2) \tag{8.9}$$

These functions are used to analytically calculate the function $A(D)$ plotted in figure 7.1 in the limit $D \to 4/3$.

In addition an interesting and striking property can be remarked: $\bar{g}$ is the appropriate variable to analytically continue to $D > 4/3$. For $\varepsilon > 0$ the fixed point $\bar{g}^*$ stays positive whereas in terms of the original coupling $g$ the associated $g^*$ becomes negative. This fact will be further clarified in section 8.3.

## 8.2 Comparison with the modified 2-body Hamiltonian

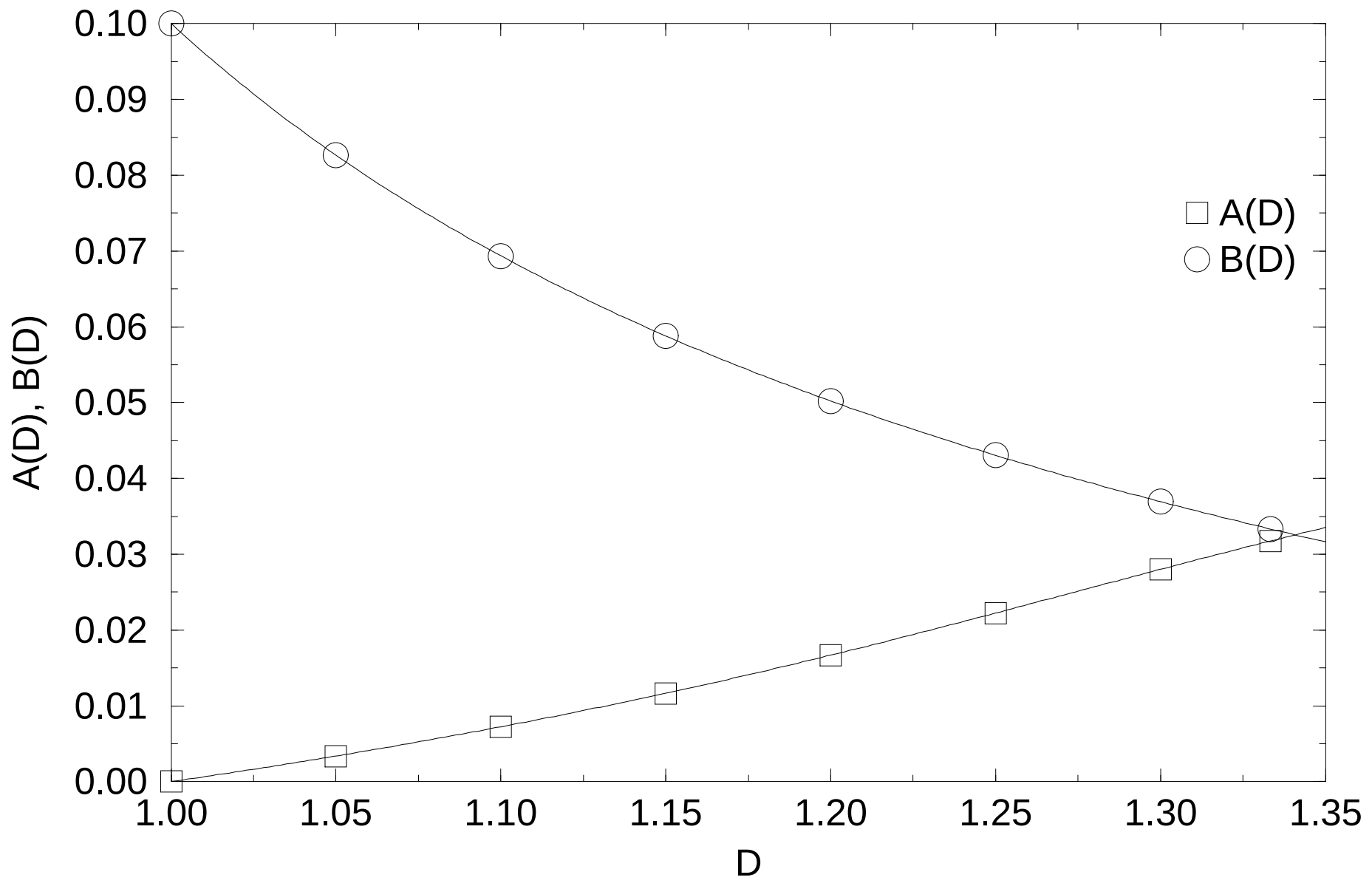

| $D$ | $B(D)$ |
|---|---|
| 1.00 | 0.10000 |
| 1.05 | 0.08266 |
| 1.10 | 0.06937 |
| 1.15 | 0.05882 |
| 1.20 | 0.05023 |
| 1.25 | 0.04306 |
| 1.30 | 0.03695 |
| 4/3 | 0.03333 |

Figure 8.1: Anomalous corrections to $\nu^\star$ at 1-loop order ($A(D)$ from the 3-point interaction and $B(D)$ from the modified 2-point interaction)

Since the modified 2-point interaction ●—⧺—● becomes relevant for $D > 4/3$, it is interesting to compare the 1-loop renormalization group calculations for this interaction with the results for the 3-point interaction. These calculations are very similar to the original calculation for the 2-point interaction and can be performed analytically at 1-loop order [25]. Here we recall the principle of the calculation and the results.

Let us start from the bare Hamiltonian

$$\mathcal{H}^0_{b_0}[r_0] \;=\; \frac{1}{2-D}\int_x \frac{1}{2}(\nabla r_0(x))^2 \,+\, b_0 \int_x\int_y (-\Delta_r)\tilde{\delta}^d(r_0(x)-r_0(y)) \qquad (8.10)$$

The dimension of the coupling constant $b_0$ is

$$\varepsilon' = 3D - 2 - \nu d \qquad (8.11)$$

and the model has UV divergences, i.e. poles in $\varepsilon'$, for $\varepsilon' = 0$, that is for $d = d'_c = 2(3D-2)/(2-D)$. The model is made finite by defining a renormalized field $r$ and a renormalized coupling constant $b$:

$$r(x) = Z^{-1/2} r_0(x) \qquad (8.12)$$
$$b = Z^{-d/2-1} Z_b^{-1} \mu^{-\varepsilon'} b_0 \qquad (8.13)$$

At 1-loop order, the counterterms are found to be

$$Z \;=\; 1 \,-\, \frac{b}{\varepsilon'}(2-D)$$ 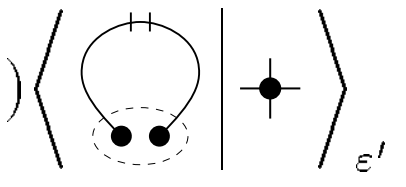 $$+\, \mathcal{O}(b^2)\,, \qquad (8.14)$$

$$Z_b \;=\; 1 \,+\, \frac{b}{\varepsilon'}$$ 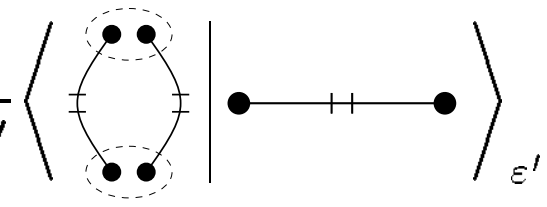 $$+\, \mathcal{O}(b^2)\,. \qquad (8.15)$$

The residues 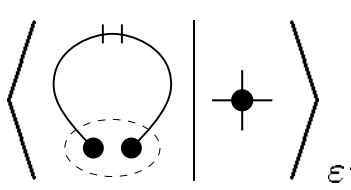 and 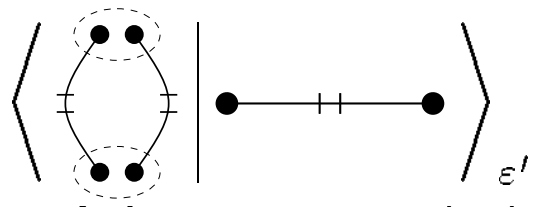 are analytic functions of $D$ for $0 < D < 2$, listed in appendix B. The $\beta$-function and the scaling dimension $\nu(b)$ for the field $r$ are:

$$\beta(b) \;=\; \mu \frac{\partial}{\partial \mu}\bigg|_{b_0} b\,, \quad \nu(b) \;=\; \nu \;-\; \frac{1}{2}\mu \frac{\partial}{\partial \mu}\bigg|_{b_0} \ln Z \tag{8.16}$$

The new $\beta$-function has a non-trivial IR fixed point $b^\star > 0$ for $\varepsilon' > 0$ and the scaling dimension of the membrane at the $\Theta$-point, $\nu'^\star$, can be expanded in $\bar{\varepsilon}' = d_c' - d$ for fixed $D$:

$$\nu'^\star \;=\; \frac{2-D}{2} \;+\; \bar{\varepsilon}'\, B(D) \;+\; \mathcal{O}(\bar{\varepsilon}'^2) \tag{8.17}$$

The 1-loop coefficient $B(D)$ has no singularity at $D = 4/3$. Its value for $1 \le D \le 4/3$ is shown in figure 8.1, where it is compared to the 1-loop coefficient $A(D)$ from the 3-point interaction. At $D = 4/3$, the expansion is around the same critical dimension $d_c = 6$ and it is interesting to note that, although the expansion parameters $\varepsilon$ and $\varepsilon'$ are different, the two coefficients $A$ and $B$ are very close, but not identical at $D = 4/3$.

## 8.3 Mixing of 2- and 3-body interaction at $D = 4/3$

At $D = 4/3$ the two operators  and 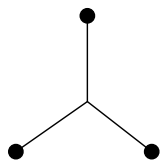 interchange their role. Below 4/3 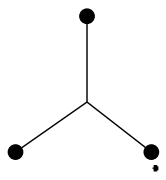 is more relevant, above 4/3 . At $D = 4/3$ and $d = 6$ both operators are marginal: We have the interesting situation of a system with two coupling constants. The associated bare Hamiltonian, one wants to renormalize, is:

$$H^0_{b_0,g_0}[r_0] = \frac{1}{2-D}\int_x \frac{1}{2}(\nabla r_0(x))^2 + b_0 \int_x\int_y (-\Delta_r)\tilde{\delta}^d(r_0(x)-r_0(y))$$
$$+\,g_0 \int_x\int_y\int_z \tilde{\delta}^d(r_0(x)-r_0(y))\tilde{\delta}^d(r_0(x)-r_0(z)) \tag{8.18}$$

In contrast to standard perturbation theory, where in first order of the coupling constant the divergences are single poles, the leading singularity of 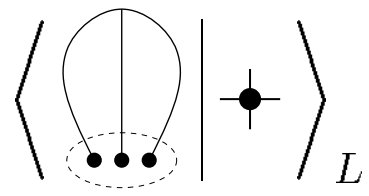 is a double pole, due to the sequence of divergent contractions

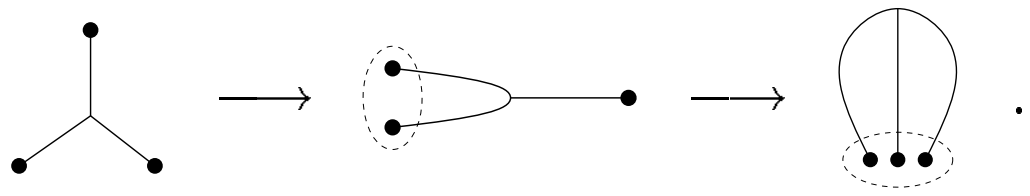

This prevents us from performing the renormalization in the standard way. Let us look at the problem from another point of view. As the modified 2-point interaction  renormalizes the elastic energy $\frac{1}{2}(\nabla r)^2$, perturbations in this operator have to be controlled by a small coupling $b$ as is done in the Hamiltonian (8.18). On the other hand, the 3-point interaction 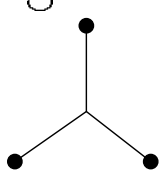 renormalizes the modified 2-point interaction  via the

contraction

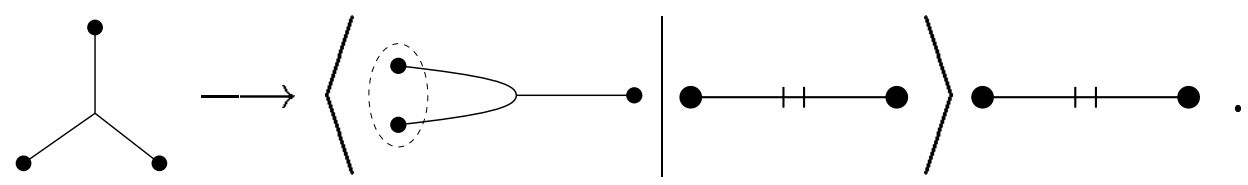

Therefore there has to be a small parameter controlling the ratio of 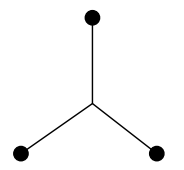 and .
These demands are satisfied by redefining the 3-point coupling constant $g_0$ as $b_0 g_0$. The bare Hamiltonian becomes

$$H^0_{b_0,g_0}[r_0] = \frac{1}{2-D}\int_x \frac{1}{2}(\nabla r_0(x))^2 + b_0 \int_x\int_y (-\Delta_r)\tilde{\delta}^d(r_0(x)-r_0(y))$$

$$+ b_0 g_0 \int_x\int_y\int_z \tilde{\delta}^d(r_0(x)-r_0(y))\tilde{\delta}^d(r_0(x)-r_0(z)) \qquad (8.19)$$

Orders in $b_0$ and $g_0$ are counted equivalently. The canonical dimensions of the coupling constants $b_0$ and $g_0$ are:

$$\varepsilon_b = 3D - 2 - \nu d\ , \quad \varepsilon_g = 2 - \nu d \qquad (8.20)$$

As the perturbative expansion is valid in the vicinity of the critical point $d_c = 6$ and $D_c = 4/3$, the two small parameters $\varepsilon_b$ and $\varepsilon_g$ are independent. The theory will be finite in terms of the renormalized Hamiltonian

$$H^R_{b,g}[r] = \frac{Z}{2-D}\int_x \frac{1}{2}(\nabla r(x))^2 + bZ_b\mu^{\varepsilon_b}\int_x\int_y (-\Delta_r)\tilde{\delta}^d(r(x)-r(y))$$

$$+ bgZ_bZ_g\mu^{\varepsilon_b+\varepsilon_g}\int_x\int_y\int_z \tilde{\delta}^d(r(x)-r(y))\tilde{\delta}^d(r(x)-r(z)) \qquad (8.21)$$

involving the renormalized quantities

$$\begin{aligned} r(x) &= Z^{-1/2} r_0(x) \\ b &= Z^{-d/2-1} Z_b^{-1}\mu^{-\varepsilon_b} b_0 \\ g &= Z^{-d/2+1} Z_g^{-1}\mu^{-\varepsilon_g} g_0 \end{aligned} \qquad (8.22)$$

where the renormalization factors have up to first order in $b$ and $g$ the form

$$Z = 1 + p\frac{g}{\varepsilon_g} + q\frac{b}{\varepsilon_b}$$

with constants $p$ and $q$. We draw the attention to the important point that pole terms in $\varepsilon_b$ are always proportional to $b$ as should be evident from dimensional arguments. The same is true for $\varepsilon_g$ and $g$. It is also important to note that this would not be the case for the parametrization (8.18), nor if one there replaces $g_0$ by $g_0^2$, what one might be tempted to do.

In order to explicitly perform the calculations, we have as in section 7.1 to expand $\langle \mathcal{O}\rangle^R_{g,b}$ in $g$ and $b$:

$\langle \mathcal{O}\rangle^R_{b,g}[r] = \Big\langle \mathcal{O}\Big\rangle_0 + \frac{1-Z}{2-D}\int \Big\langle \mathcal{O}$ 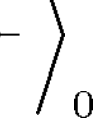 $\Big\rangle_0^{conn} - Z_b b\mu^{\varepsilon_b}\iint \Big\langle \mathcal{O}$  $\Big\rangle_0^{conn}$

$$- Z_b Z_g b g \mu^{\varepsilon_b+\varepsilon_g} \iiint$$ 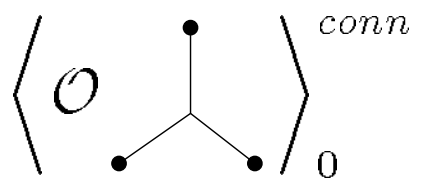

$$+ \frac{1}{2} Z_b^2 b^2 \mu^{2\varepsilon_b} \iint \iint$$ 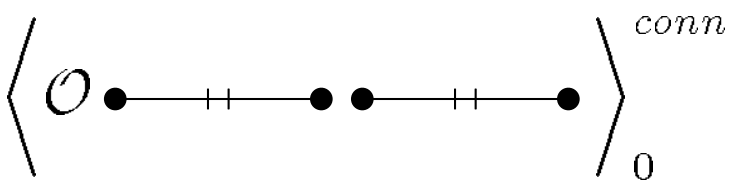

$$+ Z_b^2 Z_g b^2 g \mu^{2\varepsilon_b+\varepsilon_g} \iint \iiint$$ 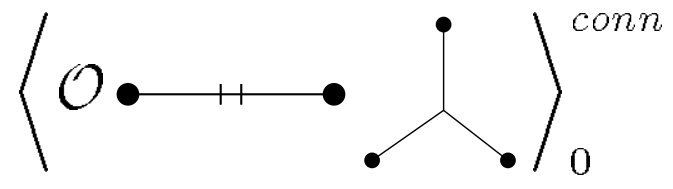

$$+ \text{higher order terms} \tag{8.23}$$

The following diagrams contribute in first order of $g$ and $b$ at $D = 4/3$ (for more details cf. appendix B):

$$\frac{2}{3}$$ 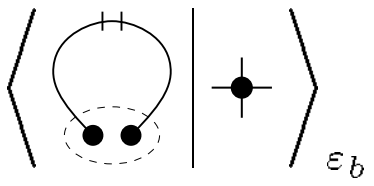 $$= -1 \tag{8.24}$$

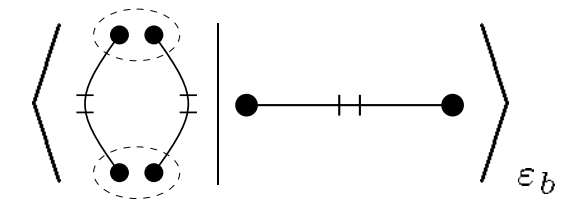 $$= 1 \tag{8.25}$$

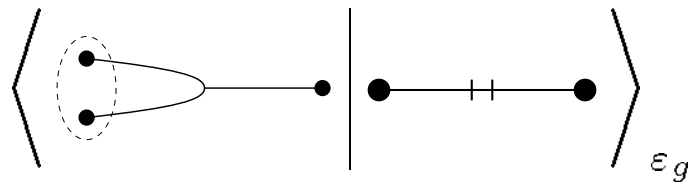 $$= \frac{1}{4} \tag{8.26}$$

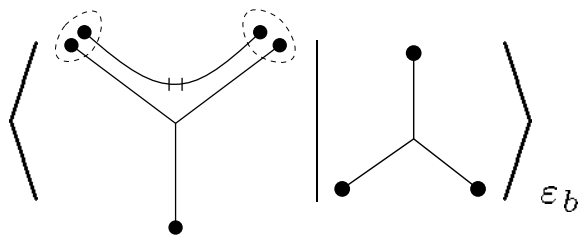 $$= \frac{3}{4} \tag{8.27}$$

They determine the renormalization factors at 1-loop order:

$$Z = 1 + \frac{b}{\varepsilon_b} \tag{8.28}$$

$$Z_g = 1 + \frac{3}{4}\frac{g}{\varepsilon_g} + \frac{7}{2}\frac{b}{\varepsilon_b} \tag{8.29}$$

$$Z_b = 1 - \frac{3}{4}\frac{g}{\varepsilon_g} + \frac{b}{\varepsilon_b} \tag{8.30}$$

As usually we define the renormalization group $\beta$-functions of the two couplings $b$ and $g$ as their variation with respect to the renormalization scale $\mu$ at fixed bare parameters:

$$\beta_b(b,g) = \mu \frac{\partial}{\partial \mu}\bigg|_{b_0\,,\,g_0} b \tag{8.31}$$

$$\beta_g(b,g) = \mu \frac{\partial}{\partial \mu}\bigg|_{b_0\,,\,g_0} g \tag{8.32}$$

Plugging in the definitions of $b$ and $g$, we get two coupled linear equations in $\beta_b$ and $\beta_g$, which can be solved after some algebra. They lead to the $\beta$-functions at 1-loop order:

$$\beta_b(b,g) = -\varepsilon_b b - \frac{3}{4} b g + 5 b^2 \tag{8.33}$$

$$\beta_g(b,g) = -\varepsilon_g g + \frac{11}{2} b g + \frac{3}{4} g^2 \tag{8.34}$$

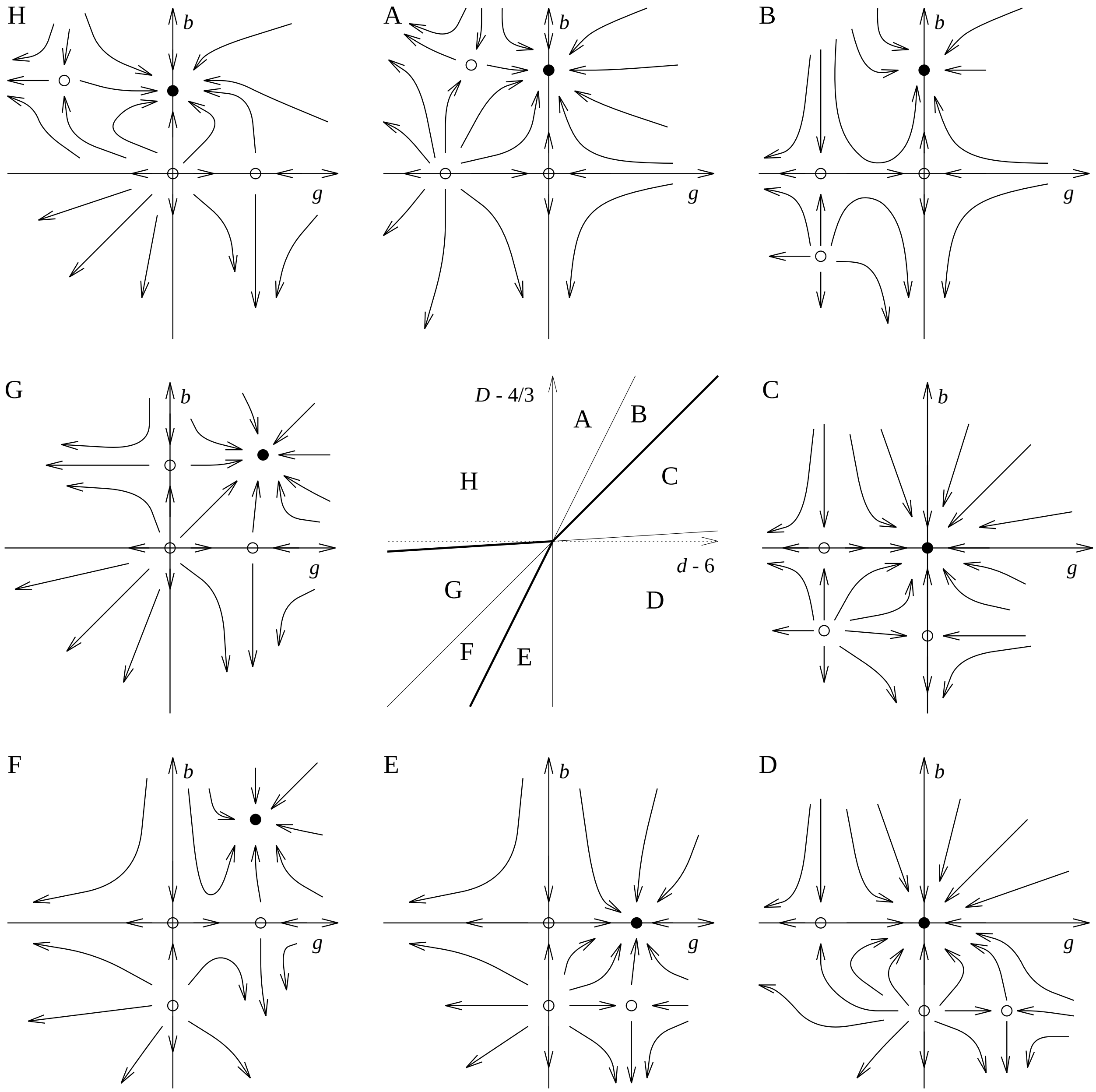

Figure 8.2: The different domains in the $d$-$D$-plane (middle). The corresponding flowdiagrams are drawn around.

The scaling function of the field $\nu(b, g)$ becomes:

$$\begin{aligned}\nu(b,g) &= \nu - \frac{1}{2}\mu\frac{\partial}{\partial\mu}\ln Z \\ &= \frac{1}{3} + \frac{1}{2}b \qquad (8.35)\end{aligned}$$

The system of equations (8.33), (8.34) determines four fixed points in the $(g, b)$ plane. The physical couplings must correspond to a repulsive interaction at short distance, hence to the domain $(b \geq 0,\ g \geq 0)$. One of the fixed points is IR-attractive, one IR-repulsive and the other two have one attractive and one repulsive direction. For special values of the parameters $\varepsilon_b$ and $\varepsilon_g$, fixed points may coincide. Passing through these special values describes the transition from one fixed point to another, resulting in an eventual non-analyticity of the critical exponent $\nu(b,g)$. We first list the different critical points visualized in figure 8.2.

$P_1$: The Gaussian fixed point $b_c = 0$ and $g_c = 0$: it is stable for $\varepsilon_b < 0$ and $\varepsilon_g < 0$.

$P_2$: The fixed point $b_c = 0$ and $g_c = \frac{4}{3}\varepsilon_g$ describes also a trivial theory, although $g_c$ has a nontrivial value. Indeed, regarding the action (8.19), we see that both interactions are renormalized to 0. Also the critical exponent $\nu(b,g)$ equals that of the free (Gaussian) theory. The stability condition is $\varepsilon_g > 0$ and $\varepsilon_b + \varepsilon_g < 0$.

$P_3$: The fixed point $b_c = \frac{1}{5}\varepsilon_b$ and $g_c = 0$: for this non-trivial fixed point only the modified 2-point interaction plays a role. It is stable for $\varepsilon_b > 0$ and $11\varepsilon_b > 10\varepsilon_g$.

$P_4$: The fixed point $b_c = \frac{2}{21}(\varepsilon_b + \varepsilon_g)$ and $g_c = \frac{4}{63}(-11\varepsilon_b + 10\varepsilon_g)$ is the most interesting. Both couplings flow to a finite non-zero value. This point is stable for $\varepsilon_b + \varepsilon_g > 0$ and $11\varepsilon_b < 10\varepsilon_g$. It corresponds to the earlier derived fixed point for the case of a 3-point interaction only in the limit $D \to 4/3$ from below. We will explain that in more detail below.

Let us discuss the graphics of figure 8.2: We can distinguish 8 different regions in the $(d,D)$ plane around the critical point $(d_c = 6, D_c = 4/3)$, named A to H. The separating lines are:

1. $\varepsilon_g = 0$ separating D,E and A,H
2. $\varepsilon_b + \varepsilon_g = 0$ between E,F and A,B
3. $\varepsilon_b = 0$ separating F,G and B,C
4. $11\varepsilon_b = 10\varepsilon_g$ between C,D and G,H

The flowgraphs in figure 8.2 correspond to these regions A to H, starting with region H in the upper left corner.

The situation encountered in the discussion of section 8.1 corresponds to an expansion in $d-6$ for $D = 4/3$ fixed. This direction lies in the sector H and is near to the transition line $11\varepsilon_b = 10\varepsilon_g$, separating the domain of attraction of the fixed points $b_c = \frac{1}{5}\varepsilon_b$, $g_c = 0$ and $b_c = \frac{2}{21}(\varepsilon_b + \varepsilon_g)$, $g_c = \frac{4}{63}(-11\varepsilon_b + 10\varepsilon_b)$. The correct value for the anomalous scaling dimension of the field is thus given by the calculation with the modified 2-point interaction only. The fixed point described by the calculations for a 3-point interaction only before taking the limit $D \to 4/3$ is equivalent to the non-trivial but unstable fixed point $b_c = \frac{2}{21}(\varepsilon_b + \varepsilon_g)$, $g_c = \frac{4}{63}(-11\varepsilon_b + 10\varepsilon_b)$. This can be seen by calculating the flow equation for $bg$, which has the same fixed point as the flow equation ($\beta$-function) for $\bar{g}$ in section 8.1, if $b$ here and $\bar{g}$ there are identified. In this parametrization also $\nu(b,g)$ here and $\nu(\bar{g})$ there are equivalent. The fact that the direction of expansion is near to the transition line, at which the 2 fixed points coincide, explains why the anomalous corrections for the scaling of the membrane differ so little. The flow graph representing this situation is drawn in figure 8.3.

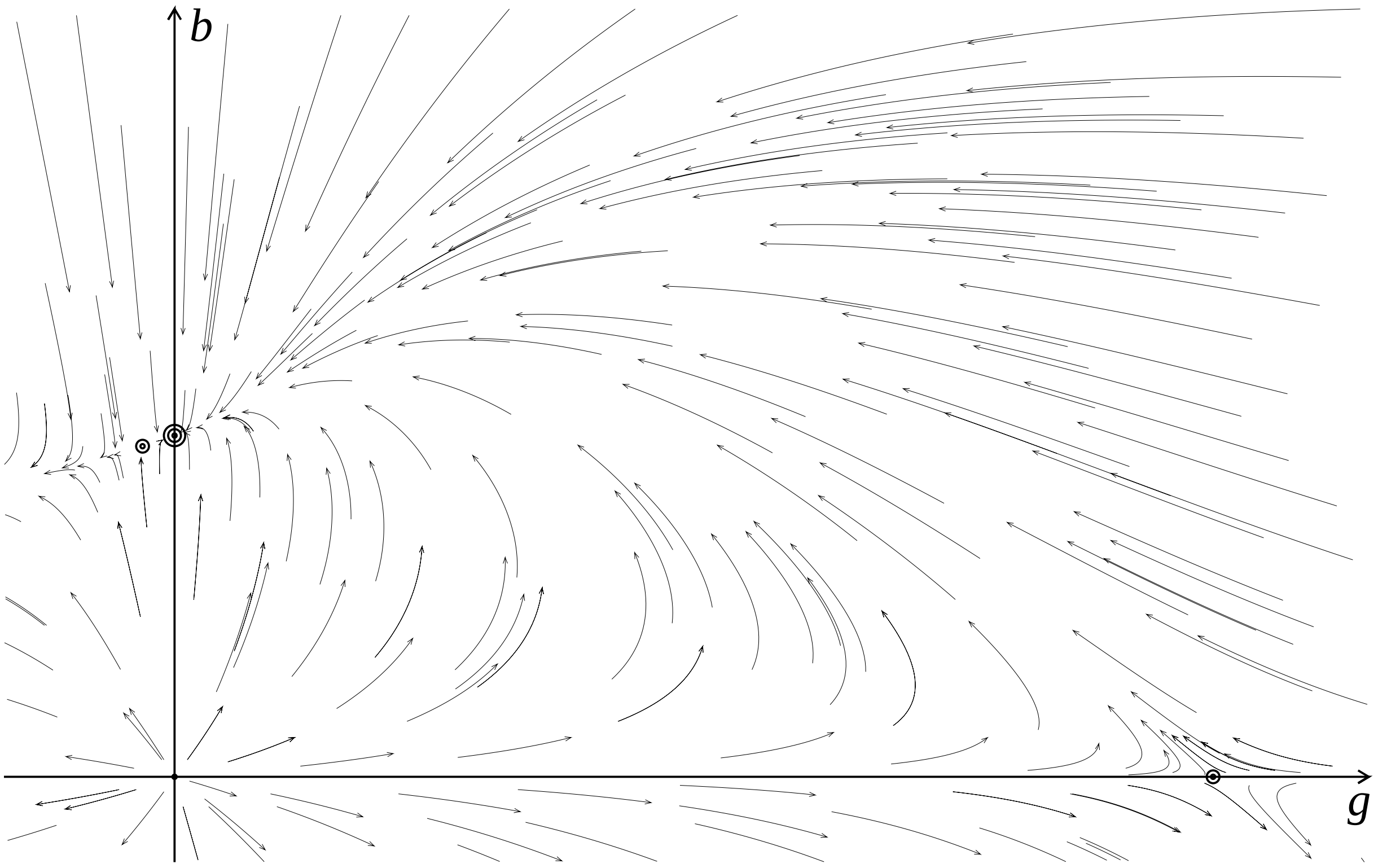

Figure 8.3: The renormalization flow of the expansion for constant $D$. The fixed points are marked with points. An additional circle indicates an attractive direction. The graph was obtained through a numerical study of (8.33) and (8.34). The length of the arrows is scaled as the square root of the speed of the renormalization flow.

Coming back to the general situation depicted in figure 8.2, the flows are such that:

- In regions C, D and E, the Gaussian fixed point $P_1$ or the pseudo-Gaussian fixed point $P_2$ are IR stable. The modified 2-point and 3-point interactions are irrelevant and the large distance properties of the manifold at the Θ-point are those of a free Gaussian manifold

- In regions A, B and H, the fixed point $P_3$, described by the modified 2-point interaction only, is IR stable. The 3-point interaction is irrelevant and the modified 2-point Hamiltonian (8.10), discussed in section 8.2, is sufficient to describe the large distance properties of the manifold at the Θ-point through an $\varepsilon_b$-expansion.

- Finally, in regions F and G the fixed point $P_4$, which contains a mixture of 3-point and modified 2-point interactions, is IR stable. As discussed above, this fixed point corresponds to the limit $D \to 4/3$ for the 3-point Hamiltonian (2.1). Therefore the pure 3-point Hamiltonian is sufficient to describe the Θ-point in an $\varepsilon_g$ expansion.

If we extrapolate these 1-loop results, we obtain the picture already summarized in figure 1.1 for the Θ-point as a function of the external dimension of space $d$ and of the internal dimension of the membrane $D$: The $(d, D)$ plane is separated into three regions:

- For $D < 2$ and $d$ large enough, both the 3-point interaction and the modified 2-point interaction are irrelevant. The $\Theta$-point is described by the Gaussian model.
- For $d < d_c = 3D/(2-D)$ and $D$ small enough, the 3-point interaction is more relevant than the modified 2-point interaction and governs the $\Theta$-point.
- For $d < d'_c = 2(3D-2)/(2-D)$ and $D$ large enough, the modified 2-point interaction is more relevant than the 3-point interaction and governs the $\Theta$-point.

At 1-loop order, the separatrix between these two domains is given by line number 4. ($11\varepsilon_b = 10\varepsilon_g$, with $\varepsilon_b$ and $\varepsilon_g$ given by (8.20)), i.e. by the line

$$d \;=\; 108\,D \;-\; 138 \tag{8.36}$$

Thus, if we trust this picture far from the critical point ($d = 6,\ D = 4/3$), we expect that for 2-dimensional membranes ($D = 2$), the modified 2-point interaction will always be the most relevant to describe the $\Theta$-point, even for $d < 6$. One also checks that the modified 2-point interaction is less relevant than the standard 3-point interaction to describe polymers ($D = 1$) in two dimensions ($d = 2$) at the $\Theta$-point.

Finally, let us stress that our analysis of the relevance of the two interaction terms leads to results drastically different from naive power counting or approximate schemes. Naive power counting predicts a separating line given by

$$d \;=\; \frac{4}{2-D} \tag{8.37}$$

and that for $D = 2$ the 3-body interaction is always more relevant than the modified 2-body interaction. Flory-type arguments give a separatrix

$$d \;=\; 3D + 2 \tag{8.38}$$

while a Gaussian variational approximation leads to

$$d \;=\; 6\ . \tag{8.39}$$

Both approximations predict that for $D = 2$ the 3-body interaction is relevant for low dimensions $d$ ($d < 8$ and $d < 6$ respectively).

# 9 Conclusions

In this paper we discussed the renormalization of self-avoiding tethered membranes at the tricritical point. The 3-body repulsive interaction has been considered first. From a technical point of vue the calculations at 1-loop order are difficult and led us to the developement of new technical tools. These tools have successfully been applied to membranes with intrinsic dimension $1 \leq D \leq 4/3$ at the $\Theta$-point and gave the critical exponent $\nu^*$ at 1-loop order. They equally should apply for the 2-loop calculations in the case of pure self-avoidance, equation (1.1).

The limit $D \to 4/3$, $d \simeq 6$ is especially interesting, since in this situation a crossover between the 3-body repulsive interaction and a modified 2-body interaction takes place. For $D \to 4/3$ we were able to analytically calculate $\nu^*$. The expansion around the point $D = 4/3$ and $d = 6$ revealed new and interesting phenomena. Thanks to a new double $\varepsilon$-expansion we were able to completely describe the structure of the renormalization group flow and of the fixed points. The crossover is understood as the passing from one IR attractive fixed point to another. This approach also settles the question which interaction is expected to be relevant to describe the $\Theta$-point for the "physical" systems: While for polymers it is the 3-body interaction, it is the modified 2-body interaction which is relevant for 2-dimensional tethered surfaces.

Although the models considered in this paper seem to be rather complicated and cumbersome they possess a great mathematical and physical richness, which we hope will further be explored in the future.

## Acknowledgement

We thank B. Duplantier and E. Guitter for discussions and E. Guitter for useful remarks. We would like to thank J.M. Drouffe for the idea to implement the AMC-algorithm. It is also a pleasure to thank J. Zinn-Justin for useful remarks concerning the double $\varepsilon$-expansion and J.-B. Zuber for his interest and a careful reading of the manuscript.

# Appendix

## A About the Finite Part Prescription

In order to eliminate the relevant UV divergences which appear in subcontractions 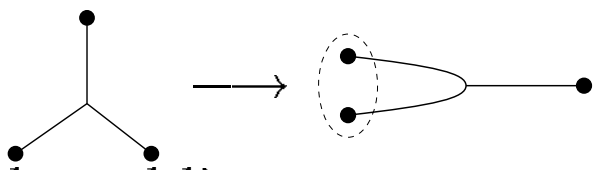, we had in equation (3.13) defined a subtracted 3-point interaction by adding a counterterm proportional to the 2-point interaction. Another possibility is not to add such a counterterm, but to subtract the relevant divergences by a finite part prescription on the level of diagrams. Let us demonstrate that for the diagram involved in the elastic term renormalization (section 3). We start from expression (3.9)

$$B(\{x,y,z\}) \;=\; -\frac{2^d}{D}\,\frac{(c^2+a^2-b^2)b^{2\nu}+(a^2+b^2-c^2)c^{2\nu}+(b^2+c^2-a^2)a^{2\nu}}{\Big[(a^\nu+b^\nu+c^\nu)(a^\nu+b^\nu-c^\nu)(b^\nu+c^\nu-a^\nu)(c^\nu+a^\nu-b^\nu)\Big]^{1+\frac{d}{2}}}\,, \tag{A.1}$$

which shall be integrated over the sector $\mathcal{A}$, defined in section 4 and characterized by $L = a > b, c$. Divergences occur only for $b \to 0$ and $c \to 0$. The finite part prescription amounts to:

$$\text{f.p.}\int_{\mathcal{A}} B(\{x,y,z\}) \;=\; \int_{\mathcal{A}} B(\{x,y,z\}) \;-\; \int_{\mathcal{A},\mathcal{B},\mathcal{C}} \frac{1}{4D}(b^{-\nu d}+c^{-\nu d})a^{D-\nu d} \tag{A.2}$$

In the counterterm the integration is not restricted by $b, c < a = L$. Mapping the sectors $\mathcal{B}$ and $\mathcal{C}$ onto $\mathcal{A}$ yields a new expression $c'(\varepsilon)$ for the residue in (3.18). It differs from the one obtained in section 3, given in (3.21), by

$$\begin{aligned} c'(\varepsilon) \;=\; c(\varepsilon) - \frac{1}{4D}\int_{\mathcal{A}} & (a^{-\nu d}b^{D-\nu d} + c^{-\nu d}b^{D-\nu d})(b^{-\varepsilon}-1) \\ & +(a^{-\nu d}c^{D-\nu d} + b^{-\nu d}c^{D-\nu d})(c^{-\varepsilon}-1) \end{aligned} \tag{A.3}$$

This difference is of order $\varepsilon$ as $\varepsilon \to 0$ and so does not change the residue $c(0)$ at $\varepsilon = 0$.

The same statement holds for the other diagrams involved in the 1-loop renormalization of the 3-point interaction. Different subtraction prescriptions of the relevant divergences (associated to the 2-point interaction) thus do not change the pole terms at $\varepsilon = 0$, which determine the renormalization group functions at 1-loop order, but they may change the finite parts at that order. This implies that the renormalization group functions at 2-loop order will depend on the choice of the subtraction prescriptions.

Another point, which has to be discussed, is whether the limit $d \to d_c$ in (3.21) or with the modified counterterm discussed above can be performed. It is clear that problems may only arise for small $b$ or $c$. In this limit, the integrals may be expanded in the form

$$\int_b b^{-\lambda} \tag{A.4}$$

According to the definition of the finite part prescription it is sufficient to check that no pole term in $1/(D-\lambda)$ may occur for $D - \lambda = 0$. The reader may easily verify that this is indeed the case.

# B Some Diagrams

In this appendix the MOPE is given for all contractions not calculated in the main text. In addition the residue of some more diagrams can be found, which was used but not calculated up to now.

The labeling of distances is either unambiguous or follows the notations in figures 3.1, 5.1, 5.3 or 6.1.

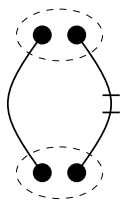 $= \frac{d}{2}\left(\frac{1}{a^{2\nu}+b^{2\nu}}\right)^{d/2+1}$  $+\, \frac{d+2}{8}\left(\frac{1}{a^{2\nu}+b^{2\nu}}\right)^{d/2}$  $+ \dots$ (B.1)

$\Big\langle$ 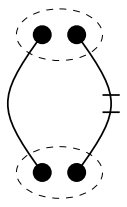 $\Big|$  $\Big\rangle_{3D-2-\nu d} = \frac{3D-2}{(2-D)^2}\frac{\Gamma\left(\frac{D}{2-D}\right)^2}{\Gamma\left(\frac{2D}{2-D}\right)}$ (B.2)

$\Big\langle$ 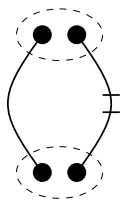 $\Big|$  $\Big\rangle_{2D-\nu d} = \frac{D+2}{4(2-D)^2}\frac{\Gamma\left(\frac{D}{2-D}\right)^2}{\Gamma\left(\frac{2D}{2-D}\right)}$ (B.3)

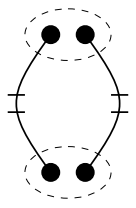 $= \frac{d(d+2)}{4}\left(\frac{1}{a^{2\nu}+b^{2\nu}}\right)^{d/2+2}$ 

$+\frac{d^2+6d-8}{16}\left(\frac{1}{a^{2\nu}+b^{2\nu}}\right)^{d/2+1}$  $+\ldots$ (B.4)

$\Big\langle$ 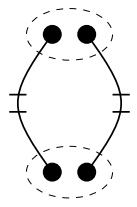 $\Big|$  $\Big\rangle_{3D-2-\nu d} = \frac{10D-D^2-8}{2(2-D)^3}\frac{\Gamma\left(\frac{D}{2-D}\right)^2}{\Gamma\left(\frac{2D}{2-D}\right)}$ (B.5)

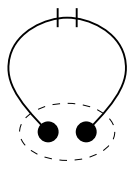 $= \frac{d}{2}a^{-\nu(d+2)}$  $- \frac{d+2}{4D}a^{2D-2-\nu d}$ 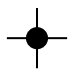 $+\ldots$ (B.6)

$\Big\langle$ 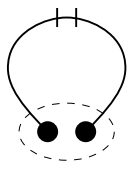 $\Big|$ 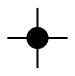 $\Big\rangle_{3D-2-\nu d} = -\frac{1}{2-D}$ (B.7)

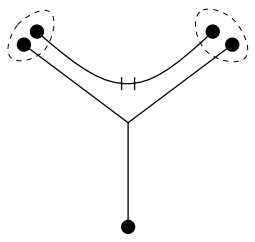 $= \frac{d}{2}\left(\frac{1}{a^{2\nu}+b^{2\nu}}\right)^{d/2+1}$ 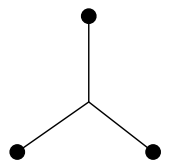 $+\ldots$ (B.8)

$\Big\langle$ 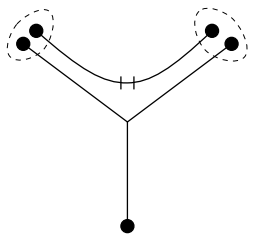 $\Big|$ 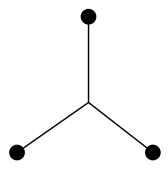 $\Big\rangle_{3D-2-\nu d} = \frac{3D-2}{(2-D)^2}\frac{\Gamma\left(\frac{D}{2-D}\right)^2}{\Gamma\left(\frac{2D}{2-D}\right)}$ (B.9)

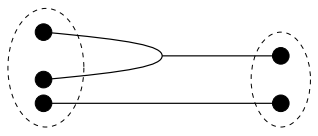 (B.10)

$$= \left(\frac{4}{4b^{2\nu}e^{2\nu}+(a^\nu+c^\nu+b^\nu)(a^\nu+c^\nu-b^\nu)(a^\nu-c^\nu+b^\nu)(c^\nu+b^\nu-a^\nu)}\right)^{d/2}$$  $+\ldots$

$\Big\langle$ 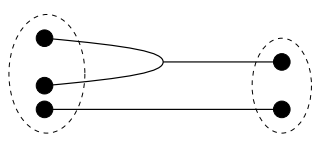 $\Big|$  $\Big\rangle_{3D-2-\nu d} = \Big\langle$ 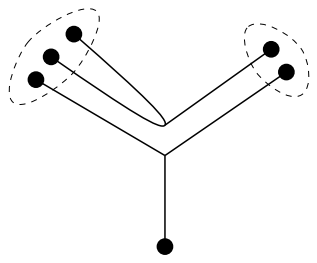 $\Big|$ 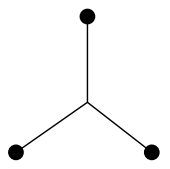 $\Big\rangle_{3D-2-\nu d}$ (B.11)

# C Renormalization of Relevant Operators: Away from the Tricritical Point

The tricritical point ($\Theta$-point) was obtained by demanding the *renormalized* coupling $t$ for the 2-point interaction to vanish. It is possible to study the general situation with $t \neq 0$ by regarding the renormalized Hamiltonian:

$$\mathcal{H}^R_{g,t}[r] = \frac{Z}{2-D}\int_x \frac{1}{2}\big(\nabla r(x)\big)^2 + tZ_t\mu^{(D+\varepsilon)/2}\int_x\int_y \delta^d(r(x)-r(y))$$
$$+gZ_g\mu^\varepsilon\int_x\int_y\int_z \delta^d(r(x)-r(y))\delta^d(r(x)-r(z)) \qquad \text{(C.1)}$$

The dimensions of the renormalized couplings $t$ and $g$ are adjusted to vanish. $g$ was defined in (7.4), $t$ analogously is

$$t = Z^{-d/2} Z_t^{-1} \mu^{-(D+\varepsilon)/2} t_0 \tag{C.2}$$

From general arguments it is known that $t$ as the coupling of a relevant operator does not appear in the renormalization factors $Z$, $Z_t$ and $Z_g$ as long as we stay within the frame of the minimal subtraction scheme. So $Z$ and $Z_g$ are unchanged. In addition to (7.8) there is to first order in $t$:

$$\langle \mathcal{O}[r]\rangle^R_{g,t} - \langle \mathcal{O}[r]\rangle^R_g = -tZ_t\mu^{D/2+\varepsilon/2} \;[\ldots]\; + tg\mu^{D/2+3\varepsilon/2}\;[\ldots]\; + \mathcal{O}(g^2)$$

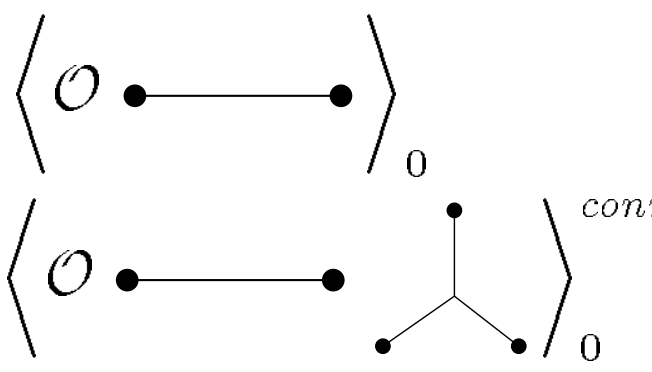

(C.3)

This determines

$$Z_t = 1 + 6 \;[\ldots]\; \frac{g}{\varepsilon} + \mathcal{O}(g^2)\ .$$

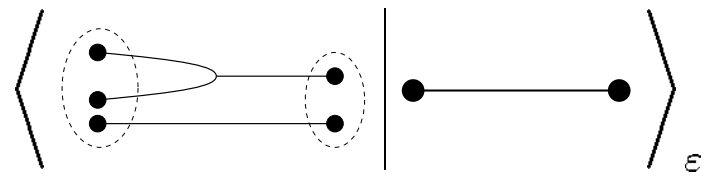

(C.4)

The scaling of the 2-point interaction is thus described by the renormalization group $\gamma_t$-function:

$$\begin{aligned}\gamma_t &= \mu \frac{\partial}{\partial \mu}\bigg|_{t_0,\, g_0} t \\ &= -\frac{D+\varepsilon}{2} - \beta(g)\left[\frac{d}{2}\frac{\partial}{\partial g}\ln Z + \frac{\partial}{\partial g}\ln Z_t\right] \\ &= -\frac{D+\varepsilon}{2} + g_R \left[-\nu d_c \;[\ldots]\; + 6 \;[\ldots]\right] + \mathcal{O}(g^2)\end{aligned}$$

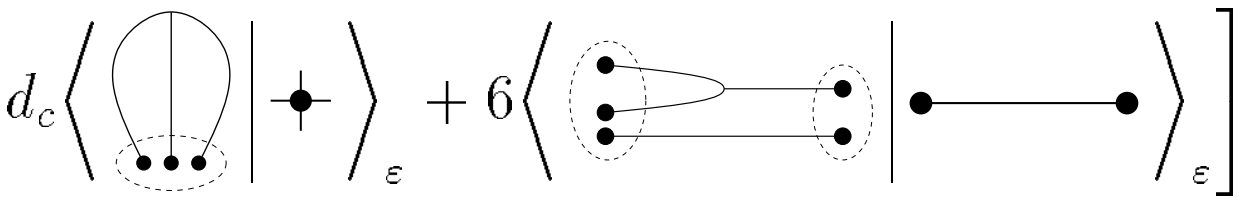

(C.5)

The new diagram involved in (C.5) has already been calculated:

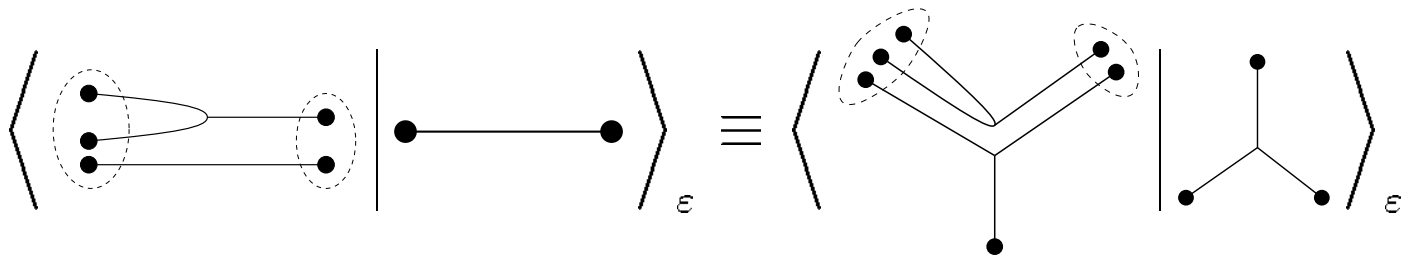

(C.6)

This is understood from the fact that the additional exterior leg on the r.h.s. of (C.6) does not appear in the calculation of the MOPE coefficient.

By now it should be clear that the limit $D \to 4/3$ is regular if one uses the rescaled coupling $\bar{g}$ in (8.7). It is equally possible to calculate around $D = 4/3$ and $d = 6$ following the discussion in section 8.3. The reader willing to perform this exercise will find the necessary diagrams in appendix B.